\documentclass[trackchanges]{aastex701}
\usepackage{amsmath}

\begin{document}

\title{Optical--NIR Multi-band Photometric Analysis and Characterization of Giant Exoplanets with CPI-C}

\author[0000-0002-9106-8718]{Yiming Zhu}
\affiliation{Nanjing Institute of Astronomical Optics $\&$ Technology, Chinese Academy of Sciences, \\Nanjing 210042, China}
\affiliation{CAS Key Laboratory of Astronomical Optics $\&$ Technology, Nanjing Institute of Astronomical Optics $\&$ Technology, \\Nanjing 210042, China}
\email{ymzhu@niaot.ac.cn}

\author[0000-0001-8544-0280]{Gang Zhao}
\affiliation{Nanjing Institute of Astronomical Optics $\&$ Technology, Chinese Academy of Sciences, \\Nanjing 210042, China}
\affiliation{CAS Key Laboratory of Astronomical Optics $\&$ Technology, Nanjing Institute of Astronomical Optics $\&$ Technology, \\Nanjing 210042, China}
\email{gzhao@niaot.ac.cn}

\author{Xi Zhang}
\affiliation{Nanjing Institute of Astronomical Optics $\&$ Technology, Chinese Academy of Sciences, \\Nanjing 210042, China}
\affiliation{CAS Key Laboratory of Astronomical Optics $\&$ Technology, Nanjing Institute of Astronomical Optics $\&$ Technology, \\Nanjing 210042, China}
\email{xzhang@niaot.ac.cn}

\author{Gang Wang}
\affiliation{Nanjing Institute of Astronomical Optics $\&$ Technology, Chinese Academy of Sciences, \\Nanjing 210042, China}
\affiliation{CAS Key Laboratory of Astronomical Optics $\&$ Technology, Nanjing Institute of Astronomical Optics $\&$ Technology, \\Nanjing 210042, China}
\email{gwang@niaot.ac.cn}

\author[0000-0002-3346-4659]{Bingli Niu}
\affiliation{Nanjing Institute of Astronomical Optics $\&$ Technology, Chinese Academy of Sciences, \\Nanjing 210042, China}
\affiliation{CAS Key Laboratory of Astronomical Optics $\&$ Technology, Nanjing Institute of Astronomical Optics $\&$ Technology, \\Nanjing 210042, China}
\email{blniu@niaot.ac.cn}

\author[0009-0006-4561-3697]{Zhonghua Lv}
\affiliation{Nanjing Institute of Astronomical Optics $\&$ Technology, Chinese Academy of Sciences, \\Nanjing 210042, China}
\affiliation{CAS Key Laboratory of Astronomical Optics $\&$ Technology, Nanjing Institute of Astronomical Optics $\&$ Technology, \\Nanjing 210042, China}
\email{zhhlv@niaot.ac.cn}

\author[0000-0002-7612-6377]{Jiangpei Dou}
\affiliation{Nanjing Institute of Astronomical Optics $\&$ Technology, Chinese Academy of Sciences, \\Nanjing 210042, China}
\affiliation{CAS Key Laboratory of Astronomical Optics $\&$ Technology, Nanjing Institute of Astronomical Optics $\&$ Technology, \\Nanjing 210042, China}
\email{jpdou@niaot.ac.cn}

\correspondingauthor{Jiangpei Dou; Gang Zhao}
\email{jpdou@niaot.ac.cn; gzhao@niaot.ac.cn}



\begin{abstract}

We present a multi-band photometric approach to characterize giant exoplanets, which represents one of the anticipated core scientific outcomes of Cool Planet Imaging Coronagraph (CPI-C). CPI-C operates with two observational channels covering visible and near-infrared wavelengths, each equipped with four broadband filters. The planet--star flux ratio integrated over each filter bandpass is calculated for photometric analysis. For cool planets observed in the visible bands, the data are primarily used to fit the overall spectral shape and methane-induced modulation, providing sensitivity to metallicity- and cloud-dependent spectral variations while constraining the reflected-light spectral shape and the combined scaling involving planet radius, orbital separation, and orbital phase. In the near-infrared bands, which probe thermal emission, the data help to better constrain fundamental planetary parameters including the effective temperature, radius, surface gravity and mass. For a synthetic giant planet with measurable reflected-light and thermal-emission components, the combined VIS4+NIR4 data provide tighter same-target constraints than either filter set alone, especially for the planet radius and cloud sedimentation parameter. Our simulations incorporate realistic instrument throughput, detector noise, and residual speckle noise. The results demonstrate that the eight-band design spanning visible to near-infrared wavelengths supports reflected-light diagnostics, thermal-emission characterization, and joint optical--NIR analysis of giant exoplanets within CPI-C science observations.

\end{abstract}



\section{Introduction} \label{sec:intro}

Direct imaging has emerged as a powerful approach to investigate the atmospheres, orbits, and bulk properties of exoplanets by spatially separating the planetary signal from the overwhelming glare of the host star~\citep{2009ARA&A..47..253O,2025ARA&A..63..179K,2025Natur.642..905L,2025Natur.643..938H}. Achieving such detections requires extreme starlight suppression at small angular separations: ground-based facilities typically target a contrast of $10^{-7}$ at 0.4--0.5$\arcsec$  \citep{2006ApJS..167...81G,2009A&A...495..363M,2010RAA....10..189D,2010PASP..122..590R,2024AstTI...1..166D} with extreme adaptive optics and advanced coronagraphy, whereas spaceborne coronagraphs with precise wavefront control aim for a higher contrast approaching $10^{-8}$ $\sim$ $10^{-10}$ at $\sim0.1\arcsec$ \citep{2007Natur.446..771T,2016ApJ...832...84D}. 

Despite recent progress, the direct imaging of mature exoplanets in reflected starlight remains a highly anticipated goal rather than a routine capability. To date, most directly imaged exoplanets are young, widely separated, and self-luminous companions detected primarily via their thermal emission. In contrast, detecting reflected light from mature planets is far more difficult, requiring substantially deeper starlight suppression at much smaller angular separations. For mature giant planets, the expected planet-to-star flux ratios approach the $10^{-8}$ to $10^{-10}$ regime, depending on the planet's radius, orbital distance, albedo, and phase. Overcoming this extreme contrast challenge has therefore become a primary science and technology driver for the next generation of high-contrast instruments and space missions. These include ELT--PCS \citep{2021Msngr.182...38K}, GMagAO-X \citep{2024SPIE13096E..0YM}, Roman CGI \citep{2022SPIE12180E..1WM}, the Habitable Worlds Observatory \citep{2026arXiv260106233B}, and the Lazuli Space Observatory \citep{2026arXiv260102556R}.

In parallel, recent progress in high-contrast imaging via thermal emission is increasingly extending direct imaging toward colder and more mature giant planets. Representative examples include JWST coronagraphic observations of $\epsilon$ Indi Ab, a temperate super-Jupiter with an effective temperature of approximately 275 K in a system with an estimated age of approximately 3.5 Gyr \citep{2024Natur.633..789M}, and 14 Her c, a nearby mature cold giant planet with an estimated system age of approximately 4.6 Gyr, directly imaged with JWST/NIRCam \citep{2025ApJ...988L..18B}. These detections indicate that direct imaging is moving beyond the predominantly young directly imaged population, thereby motivating stable multi-band photometric characterization across wavelengths sensitive to both reflected light and thermal emission.

Cool giant exoplanets are expected to exhibit rich atmospheric spectra in the visible (VIS) and near-infrared (NIR), with the dominant source of planetary flux depending on the target temperature and age. For cool and mature giant planets, the visible-light signal is mainly reflected starlight, dominated by molecular absorption and scattering. The reflected-light albedo spectrum of a cool giant is shaped by Rayleigh scattering and clouds (setting the continuum level) and by broad molecular bands (in particular CH$_4$). Classic models of cold Jupiters and Neptunes \citep[e.g.,][]{1999ApJ...513..879M,2000ApJ...538..885S,2004ApJ...609..407B,2010ApJ...724..189C} show that methane is a dominant absorber across optical wavelengths and into the NIR, while reflective clouds and hazes (NH$_3$, H$_2$O or possibly silicates in warmer cases) strongly control the overall brightness. For example, \citet{2016AJ....152..217L} note that all cool giant models have methane-dominated optical reflection spectra. Likewise, \citet{2010ApJ...724..189C} demonstrate that the presence and vertical structure of clouds strongly influence the albedo spectra of giant exoplanets. 

In practice, CH$_4$ produces broad absorption bands near 500-900 nm and beyond, whereas cloud particles and Rayleigh scattering create a bright continuum at shorter wavelengths. Well-placed optical photometric bands can therefore sample CH$_4$ features and adjacent continuum, yielding leverage on the methane abundance and cloud properties. Previous forward-modeling and retrieval studies have shown that multiple methane bands of varying strength must be detected to constrain CH$_4$ mixing ratios, because band depths and widths depend on both composition and cloud opacity \citep{2010ApJ...724..189C,2016AJ....152..217L}. Previous reflected-light studies provide useful context for this approach. \citet{2018AJ....156..158B} investigated color classification of extrasolar giant planets using a large grid of theoretical reflected-light spectra, demonstrating both the diagnostic value and the limitations of broadband color measurements for giant-planet characterization. \citet{2018ApJ...858...69M} examined the detectability of H$_2$O features in reflected-light spectra of cool giant planets, showing that molecular-band constraints depend sensitively on atmospheric parameters such as temperature, gravity, metallicity, and cloud sedimentation. These studies motivate the use of band-integrated reflected-light observables as compact diagnostics of atmospheric composition and cloud properties, while also emphasizing the degeneracies inherent in sparse photometric measurements.

A key limitation of optical reflected-light direct imaging is that it measures only the photometric contrast (planet–star flux ratio) in each band, rather than the planet’s albedo or radius independently. Thus an observed contrast may result from different combinations of planet size, phase, albedo, and thermal emission. Uncertainties in the planet’s orbital phase and radius thus translate into degeneracies when inferring atmospheric properties. In the visible reflected-light regime, these degeneracies directly affect the interpretation of methane-sensitive colors and broadband albedo variations. 
Retrieval studies have explored these issues: \citet{2016AJ....152..217L} performed MCMC retrievals on synthetic reflected-light data and showed that, if the planet radius and phase are assumed known, one can infer methane abundance and cloud optical depth. However, they caution that in real observations the planet’s radius and phase would not be directly measured, which can strongly affect the results. Similarly, \citet{2017PASP..129c4401N} showed that unknown planetary radius and uncertain observer--planet--star phase angle can affect reflected-light retrievals of gas giants, including constraints on methane abundance, cloud properties, and surface gravity. More recent reflected-light retrieval studies have further emphasized that prior knowledge of orbital distance, phase angle, mass, and radius can influence the inferred planetary radius and atmospheric interpretation \citep{2024ApJ...969L..22S,2025AJ....169...97D}. \citet{2013ApJ...775..137L} and \citet{2014ApJ...786..154B} have similarly emphasized that direct imaging spectra have low information content and inherent degeneracies in temperature, composition and cloud properties. 
In addition, unresolved circumplanetary material or companions can introduce further ambiguity: rings or moons may contribute reflected light and modify the apparent phase- or color-dependent contrast of the system, thereby complicating the interpretation of radius, albedo, and atmospheric properties 
\citep[e.g.,][]{2004A&A...420.1153A,2018A&A...618A.162B}. In practice, then, it is common to use comprehensive forward models coupled to Bayesian inference to interpret photometric contrasts and to quantify degeneracies. These methods allow simultaneous fits for methane abundance, cloud properties, and planet radius/phase, or suitable parametrizations thereof, and thereby identify which spectral bands most effectively constrain each quantity.

In contrast, for young and/or moderately warm giant planets, self-luminosity can be strong in the near-infrared, especially in the 1000--2000~nm windows, so NIR photometry probes thermal emission. In this regime, NIR photometry enables constraints on fundamental properties such as effective temperature, radius, and surface gravity. In such thermal-emission fits, effective temperature primarily shapes the overall spectral energy distribution; the planet radius enters primarily through its absolute flux normalization, while surface gravity affects the atmospheric pressure--temperature structure and the relative depths of broad molecular absorption features and opacity windows~\citep{2015A&A...582A..83B}. 
However, NIR thermal-emission photometry also has its own degeneracies, because effective temperature, radius, surface gravity, metallicity, and cloud treatment can produce partially similar broadband spectral energy distributions. 

In this framework, the VIS and NIR channels play distinct roles: the optical bands target methane and cloud-controlled reflectance in cool giants, while the NIR bands provide complementary sensitivity to the thermal state of young, self-luminous planets. When $R_p$ and $\log g$ can be constrained from the NIR thermal-emission fit, the planet mass can also be estimated as a derived quantity through $M_p = gR_p^2/G$. The combination of VIS and NIR measurements can therefore mitigate characterization degeneracies by linking reflected-light spectral shape and phase-dependent brightness to thermal-emission constraints on the planetary spectral energy distribution.

The Chinese Space-station Survey Telescope (CSST) is a planned space-based survey telescope developed as part of China's manned space program. Utilizing a 2~m primary mirror, it is designed to provide a wide field of view, high image quality, and multi-band observing capability~\citep{2026SCPMA..6939501C}. The telescope will operate in the same orbital environment as the China Space Station. Its scientific payload includes five instruments: the Survey Camera (SC), Multi-Channel Imager (MCI), Integral Field Spectrograph (IFS), Cool Planet Imaging Coronagraph (CPI-C), and THz Spectrometer (TS)~\citep{2026SCPMA..6939501C}. Within this suite, CPI-C \citep{Dou2026CPIC} is the dedicated high-contrast imaging instrument for exoplanet observations. It adopts a dual-channel architecture with VIS and NIR cameras operating simultaneously. CPI-C aims to survey planets from Neptune to Jupiter size at separations of roughly 0.5--5~AU around nearby ($\lesssim 40$~pc) solar-type stars, targeting contrasts of $\sim10^{-8}$ in the 500--900~nm band and $\sim10^{-6}$ in the 900--1600~nm band. In this design, the visible channel is naturally suited to reflected-light observations of cool or mature giant planets, while the near-infrared channel provides complementary access to thermal emission from warmer companions. CPI-C therefore provides a concrete instrumental context for studying the atmospheric composition, orbital parameters, and physical properties of giant exoplanets over a parameter space that allows access to cool reflected light targets, warm self-luminous targets, and probes of the same target in both regimes simultaneously.

In this work, we adopt a CPI-C-like configuration as a representative case and compute band-integrated planet--star contrasts and signal-to-noise ratios (SNRs) using an explicit forward model under stated throughput, detector-noise, and residual-speckle assumptions. These calculations are implemented within our simulation program \citep{2026RAA....26b4010Z,2026RAA....26b4011Z}, which provides the bandpass throughputs and a consistent noise model. The conclusions are therefore framed as an evaluation of the fundamental information content accessible through the CPI-C optical--NIR bandpasses, rather than as a prediction tied to a single target or observing realization.

The overarching goal of this work is to evaluate the information content of the CPI-C eight-band photometric system: namely, which atmospheric, orbital, and fundamental planetary parameters can be constrained from sparse VIS+NIR photometry, and under what observational limitations. We first study the two limiting regimes separately and then test a same-target case in which reflected light and thermal emission both contribute to the observed spectrum. This paper is organized as follows. Section~\ref{sec:physics_diagnostics} defines the adopted eight-band set (VIS4+NIR4) and the corresponding diagnostic observables, treating band-integrated planet--star flux ratios as the primary measurements and using atmospheric models only as an interpretable mapping from wavelength dependence to physical parameters. Section~\ref{sec:optical_param} focuses on the optical reflected-light regime: it specifies the reflected-light atmosphere model used for inference, introduces the forward-model and noise prescription for band-integrated photometry, and presents scenario-based results that illustrate how phase, atmospheric state, and residual-speckle assumptions shape the information content of sparse optical measurements. Section~\ref{sec:near_infrared_regime} turns to the thermal-emission regime relevant to self-luminous giant planets. In this section, we use atmosphere-model spectra to evaluate CPI-C optical--NIR thermal-emission diagnostics through representative benchmark and synthetic cases. Section~\ref{sec:joint__analysis} presents a same-target test for synthetic giant planet whose optical–NIR spectrum contains measurable contributions from both reflected light and thermal emission. This test compares VIS4-only, NIR4-only, and combined VIS4+NIR4 retrievals under the same model assumptions. Section~\ref{sec:dis} summarizes the main scientific implications and limitations, emphasizing the mapping between band placement and retrievable information content.

\section{Physical basis and diagnostic observables}
\label{sec:physics_diagnostics}

This section first defines the eight-band photometric set used throughout this work and then summarizes the physical diagnostics enabled by its placement. The adopted bandpass definitions are summarized in Table \ref{tab:wave}. The primary measured quantities are band-integrated planet-to-star flux ratios, while geometric-albedo spectra serve as an interpretable intermediate output of forward models. In direct imaging, a geometric-albedo spectrum is generally not observed directly because accessible phase angles are limited by angular resolution and the planetary radius is often unknown, leading to an intrinsic degeneracy between reflectivity and the $(R_p/r)^2$ scale factor \citep{2016AJ....152..217L}. Accordingly, we treat band-integrated observables as the core diagnostics and use atmospheric models to interpret their wavelength dependence.

For a given bandpass with transmission $T(\lambda)$, we define the band-averaged planet-to-star contrast C as
\begin{equation}
C \equiv
  \left\langle \frac{f_p}{f_\star} \right\rangle
  =
  \frac{\int f_p(\lambda)\,T(\lambda)\,d\lambda}{\int f_\star(\lambda)\,T(\lambda)\,d\lambda},
  \label{eq:band_flux_ratio_def}
\end{equation}
which is the fundamental observable used throughout this paper.

The adopted eight-band set is designed to sample this wavelength dependence across two regimes in which the dominant source of planetary flux differs. As discussed in Section~\ref{sec:intro}, cool giant planets are expected to be observed primarily in reflected starlight at visible wavelengths, so relative contrasts among bands encode information about scattering, cloud reflectivity, molecular absorption, and viewing geometry. At near-infrared wavelengths, young or moderately warm giant planets can instead be dominated by intrinsic thermal emission, making the same band-integrated contrast measurements sensitive to the shape of the thermal spectral energy distribution. Throughout this work, the planet-to-star flux ratio therefore serves as a common observable whose physical interpretation depends on whether reflected-light or thermal-emission processes dominate in a given wavelength range.

\subsection{Adopted eight-band filter set}
\label{subsec:bands_def}

\begin{figure}
  \centering
  \plotone{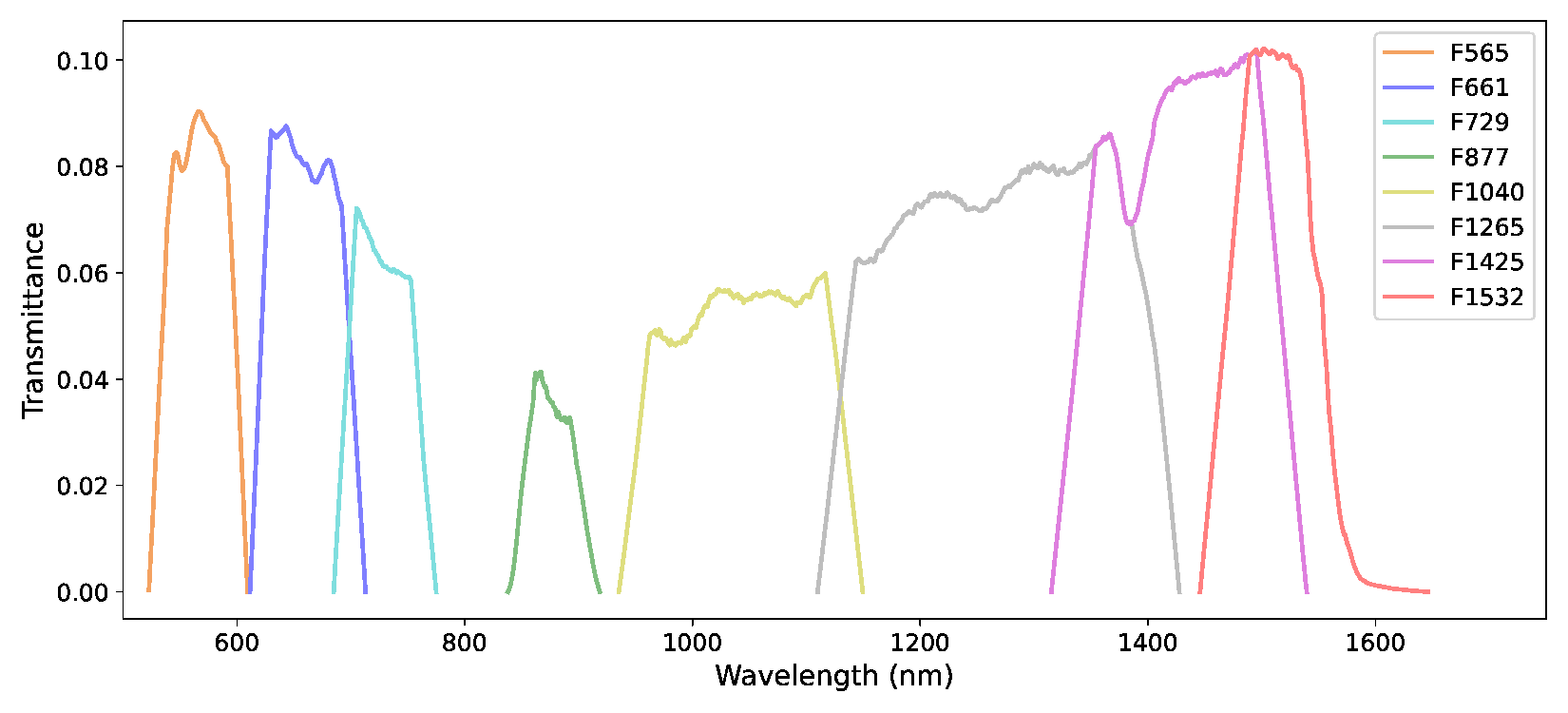}
  \caption{
  Adopted bandpass transmission curves. These curves represent the system throughput, incorporating the transmittance of the CSST main optical assembly, the CPI-C instrument optical path, and the quantum efficiency of the detector.}
  \label{fig:bandpass}
\end{figure}

We adopt a fixed set of eight bands spanning the optical and near-infrared. We refer to the four visible-channel filters F565, F661, F729, and F877 collectively as VIS4, and to the four near-infrared filters F1040, F1265, F1425, and F1532 collectively as NIR4. Each band is defined by a central wavelength and a nominal width. We quantify the effective bandwidth using the 90\% in-band transmission points on the blue and red sides ($\lambda_{\rm L90}$ and $\lambda_{\rm R90}$). The adopted band definitions are listed in Table~\ref{tab:wave}. The transmission curves for these eight bands are illustrated in Figure~\ref{fig:bandpass}. Throughout the paper, these curves enter the photometric simulations through the transmission function $T(\lambda)$ used in Equation~(\ref{eq:band_flux_ratio_def}).
The corresponding band definitions for this eight-filter set are also described in the CPI-C instrument paper~\citep{Dou2026CPIC}. The adopted filter widths are taken from the baseline CPI-C instrument design rather than re-optimized in this study. They reflect the practical passbands adopted for the optical system and detectors, together with the intended placement on continuum and molecular-feature regions. In the following simulations, we use the full adopted throughput curves and characterize their high-transmission portions by the $\lambda_{\rm L90}$ and $\lambda_{\rm R90}$ values listed in Table \ref{tab:wave}.

\begin{deluxetable}{cccccc}
\tablecaption{Design of eight broadband filters. \label{tab:wave}}
\tablewidth{0pt}
\tablehead{
\colhead{Channel} & \colhead{Band} &
\colhead{$\lambda_c$} & \colhead{$\Delta\lambda$} &
\colhead{$\lambda_{\mathrm{L90}}$} & \colhead{$\lambda_{\mathrm{R90}}$} \\
\colhead{} & \colhead{} &
\colhead{(nm)} & \colhead{(nm)} &
\colhead{(nm)} & \colhead{(nm)}
}
\startdata
VIS & F565  & 565.0  & 56.5  & 536.8  & 593.3  \\
VIS & F661  & 661.0  & 66.0  & 628.0  & 694.0  \\
VIS & F729  & 729.0  & 52.0  & 703.0  & 755.0  \\
VIS & F877  & 877.5  & 35.0  & 860.0  & 895.0  \\
NIR & F1040 & 1040.0 & 160.0 & 960.0  & 1120.0 \\
NIR & F1265 & 1265.0 & 250.0 & 1140.0 & 1390.0 \\
NIR & F1425 & 1425.0 & 150.0 & 1350.0 & 1500.0 \\
NIR & F1532 & 1532.5 & 95.0  & 1485.0 & 1580.0 \\
\enddata
\tablecomments{
$\lambda_{\mathrm{L90}}$ and $\lambda_{\mathrm{R90}}$ are defined at 90\% of the peak in-band transmission on the blue and red sides, measured from the designed bandpass curves.}
\end{deluxetable}

The following two subsections describe the physical diagnostic motivation for the optical and NIR components of this fixed filter set. The visible bands are interpreted primarily as reflected-light diagnostics of cool giant planets, whereas the NIR bands are interpreted primarily as thermal-emission diagnostics for young or moderately warm self-luminous companions.

\subsection{Optical reflected-light diagnostics: continuum anchors and methane modulation}
\label{subsec:optical_reflected}

For cool exoplanets with negligible thermal emission in the optical, the emergent planetary flux is dominated by reflected starlight \citep[e.g.,][]{2015ARA&A..53..279M,2019ApJ...878...70B,2021ApJ...910..158M}. To identify the atmospheric constituents that dominate the spectra within our visible bands, we use the Planetary Spectrum Generator (PSG) \citep{2018JQSRT.217...86V,2022fpsg.book.....V} as a radiative-transfer suite to compute model albedo spectra for controlled perturbations around a baseline template.

\begin{figure}
  \centering
  \plotone{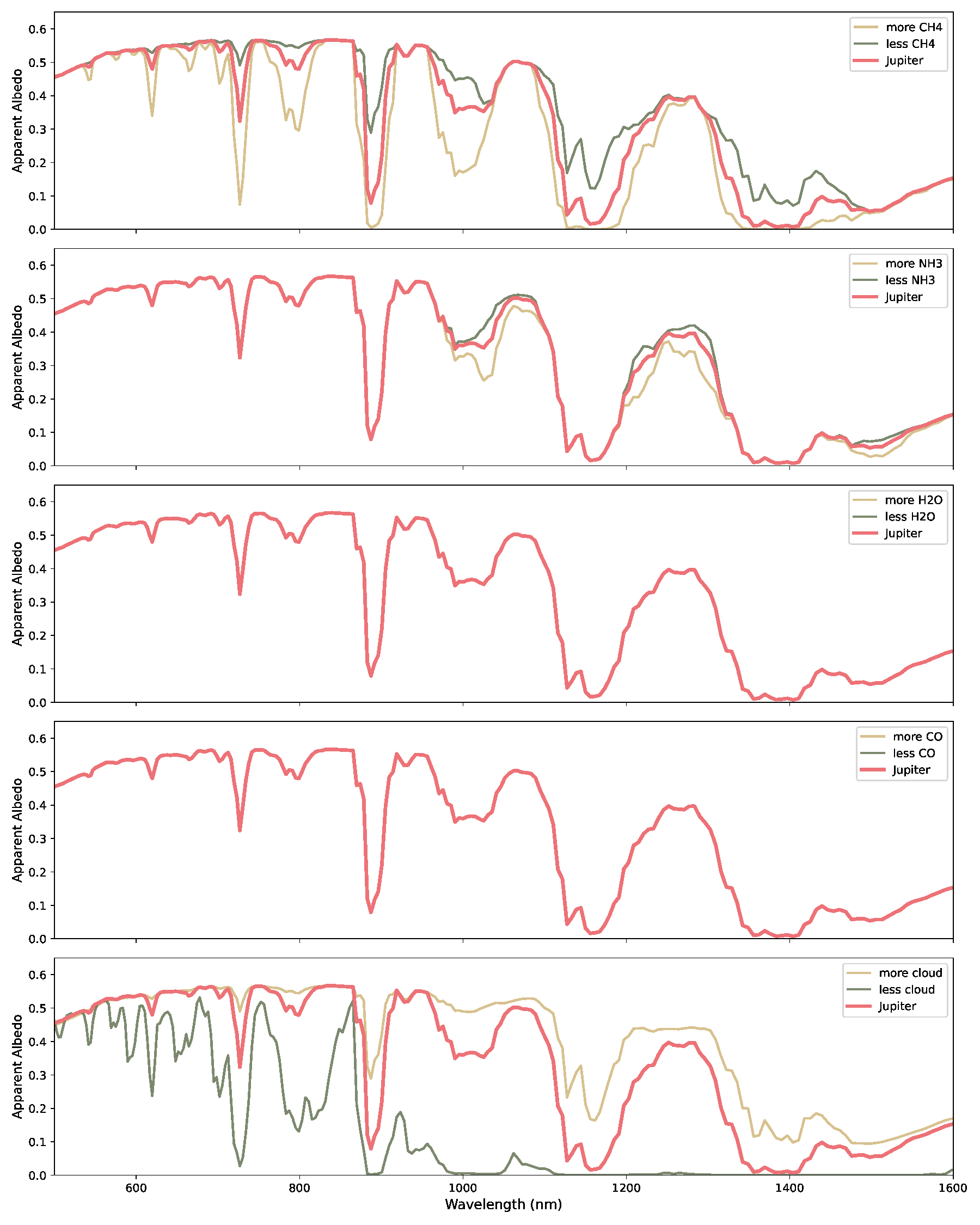}
  \caption{
  PSG-based albedo spectra adopting an HST Jupiter template as a baseline, illustrating the impact of controlled perturbations in key atmospheric parameters. The examples shown include (1) a standard Jupiter-like atmosphere, (2) methane abundance increased by a factor of 5, (3) ammonia abundance increased by a factor of 5, (4) water-vapor abundance increased by a factor of 5, (5) carbon-monoxide abundance increased by a factor of 5, and (6) water-cloud abundance increased by a factor of 100. Within this setup, methane produces the strongest modulation of the reflectance at longer optical wavelengths and motivates
  methane-sensitive band placement in our optical diagnostics.}
  \label{fig:composition}
\end{figure}

We adopt the built-in “Jupiter from HST” template provided by PSG as the baseline configuration \citep{2018JQSRT.217...86V,2022fpsg.book.....V} and apply separate abundance perturbations (relative to the baseline) for methane, water vapor, carbon monoxide, and ammonia. We also explore an increased cloud-content case (water cloud abundance) consistent with our baseline setup. Figure~\ref{fig:composition} illustrates the resulting albedo spectra for these controlled perturbations. Within this Jupiter-template experiment, the resulting spectra show that increasing methane produces the largest modulation of the reflectance at longer optical wavelengths ($\sim$700--1000~nm) among the tested species, strengthening absorption features, whereas ammonia primarily affects longer wavelengths ($\sim$1000--1500~nm); variations in water vapor and carbon monoxide induce comparatively smaller changes in the optical range. These trends are consistent with the general picture of cold-giant reflected-light spectra in which methane shapes much of the longer-wavelength optical structure \citep[e.g.,][]{2015ARA&A..53..279M,2019ApJ...878...70B,2021ApJ...910..158M,2023ApJ...942...71M,2026ApJ..1000...98M}. This motivates a baseline optical photometric strategy: at least one band placed on an optical continuum anchor, and one or more bands placed where methane absorption modulates the reflectance.

In the reflected-light regime, a convenient physical interpretation is obtained by expressing the monochromatic
flux ratio in terms of the geometric albedo and a phase function,
\begin{equation}
  \frac{f_p}{f_\star}
  =
  A_g(\lambda)\left(\frac{R_p}{r}\right)^2 \Phi(\alpha),
  \label{eq:reflected_contrast_general}
\end{equation}
where $A_g(\lambda)$ is the geometric albedo, $R_p$ is the planet radius, $r$ is the planet--star separation,
$\alpha$ is the star--planet--observer phase angle, and $\Phi(\alpha)$ is the phase function.
Equation~\ref{eq:reflected_contrast_general} is used here as the standard analytical form for interpreting reflected-light contrast. For a simple first-order description of the overall phase-dependent brightness, one commonly adopts the Lambert-sphere phase function, $\Phi(\alpha)=\left[\sin\alpha+(\pi-\alpha)\cos\alpha\right]/\pi$
\citep{2012ApJ...747...25M}. This approximation is useful for illustrating the basic coupling between phase angle and reflected-light normalization. However, realistic giant-planet atmospheres can exhibit non-Lambertian and wavelength-dependent phase behavior because scattering by clouds, hazes, and gas molecules is generally anisotropic.
Therefore, in the simulations below, we distinguish this analytical expression from the phase-dependent reflected-light spectra supplied by the adopted atmospheric model grid.

\begin{figure}
  \centering
  \plotone{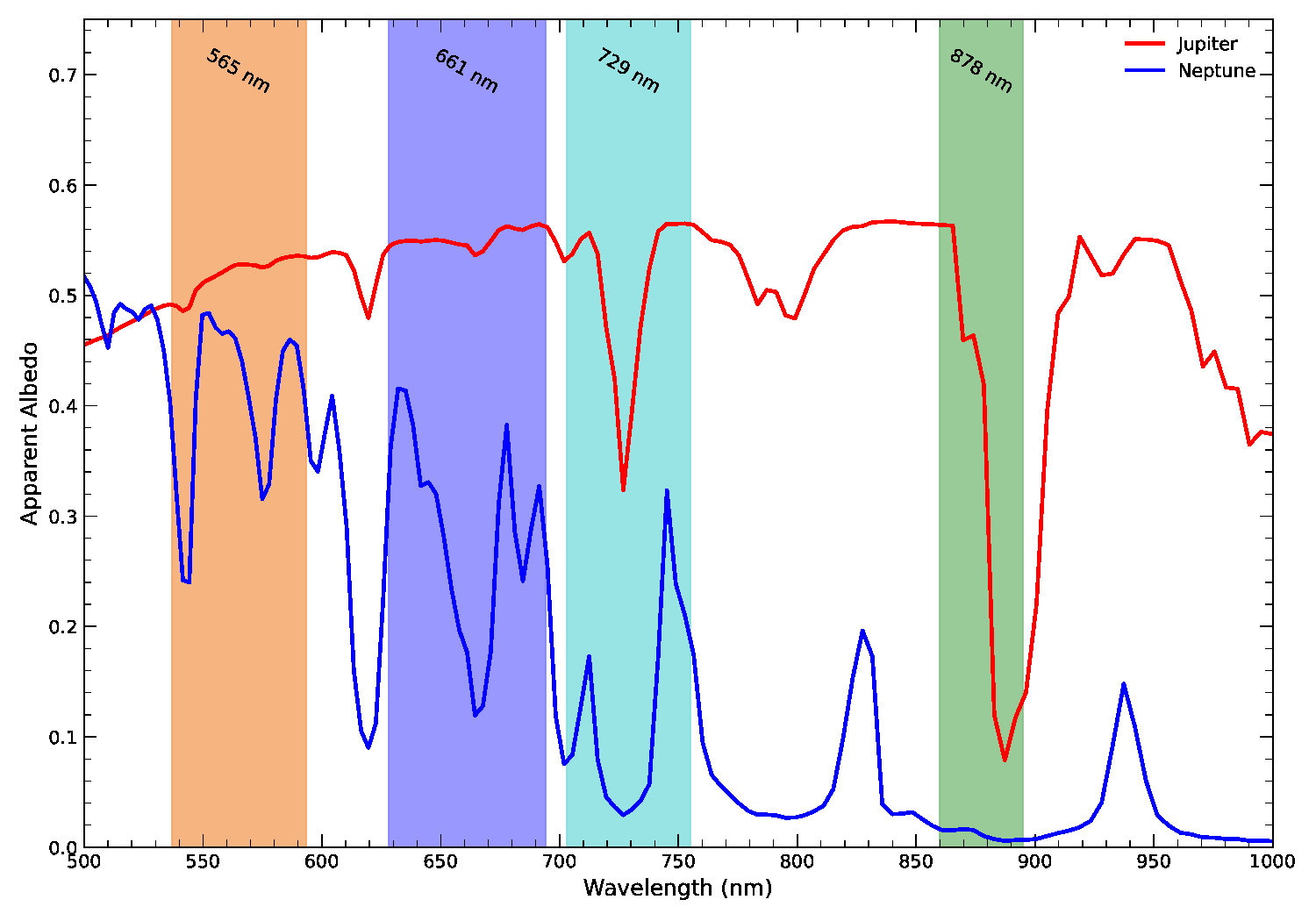}
  \caption{
  Representative reflected-light spectra overlaid with the adopted optical bandpasses (transmissions
  normalized for display). The four optical bands sample an optical continuum anchor and methane-sensitive
  regions, enabling color- and index-based diagnostics of methane modulation versus broadband albedo level.}
  \label{fig:visible}
\end{figure}

In this context, the optical filters are not intended to resolve individual molecular bands spectroscopically, but to preserve the broad color information most relevant to reflected-light interpretation. Figure~\ref{fig:visible} shows representative Jupiter and Neptune reflected-light spectra overlaid with the adopted optical bandpasses, illustrating how the four filters sample both continuum-dominated and methane-sensitive regions. For the optical subset introduced above, the four bands are centered at 565, 661, 729, and 877 nm (F565, F661, F729, F877). The placement is designed so that at least one band samples a relatively smooth optical continuum region, while additional bands probe wavelengths where methane absorption can suppress the reflected flux. F565 serves as a continuum anchor in the 500--600~nm range, where scattering and cloud reflectivity contribute strongly to the broadband albedo level. F661 provides a redder reference point that remains comparatively less affected by the deepest methane absorption in typical cold-giant spectra. F729 and F877 are positioned near prominent methane absorption structures around $\sim$720--750~nm and $\sim$880~nm, respectively, and therefore sample progressively stronger methane absorption at longer optical wavelengths. The resulting four-band sequence is sensitive to both the overall reflected-light normalization and the coarse spectral curvature produced by methane absorption and cloud-controlled continuum variations \citep[e.g.,][]{2010ApJ...724..189C,2016AJ....152..217L,2018AJ....156..158B,2021ApJ...910..158M}.

\subsection{Near-infrared diagnostics: opacity windows and broad molecular absorption}
\label{subsec:nir_emission}

While optical reflected light is central for cool targets, an optical--near-infrared photometric set can also
deliver significant science return for moderately warm and young companions for which self-luminosity dominates
in the near-infrared. Here we consider a representative thermal-emission regime spanning
$T_{\rm eff}\sim 1000$--$2000$~K, motivated by benchmark directly imaged planets such as $\beta$~Pictoris~b and
HR~8799~bcde \citep{2008Sci...322.1348M,2010Natur.468.1080M,2009A&A...493L..21L,2010Sci...329...57L}. In this
regime, broadband NIR measurements provide leverage on $T_{\rm eff}$, radius $R$, and surface gravity
$\log g$ by sampling a combination of opacity windows and broad molecular absorption features.

Thermal-emission spectra of young giant planets can be predicted using self-consistent atmosphere grids developed for substellar objects and directly imaged planets, such as \texttt{Exo-REM} and the \texttt{Sonora} model series \citep[e.g.,][]{2015ARA&A..53..279M,2015A&A...582A..83B,2024ApJ...975...59M}. These models compute pressure--temperature structures and emergent spectra as functions of parameters such as $T_{\rm eff}$, $\log g$, atmospheric composition, and cloud properties. A key design principle is to combine (i) a window in the 900--1100~nm region, where molecular opacity is relatively reduced and the emergent flux traces deeper atmospheric layers, with (ii) a band placed on the broad $\sim$1400~nm H$_2$O/CH$_4$ absorption complex, bracketed by nearby quasi-continuum anchors.

For the NIR subset introduced above, the four bands are centered at 1040, 1265, 1425, and 1532 nm (F1040, F1265, F1425, F1532). In this configuration, F1040 samples the short-wavelength window and provides leverage on the overall NIR continuum slope/normalization, F1425 targets the molecular absorption complex, and F1265/F1532 act as adjacent anchors that help separate band-depth variations from broadband SED changes. In this sense, the NIR4 set is designed to preserve the relative shape of the thermal spectrum rather than to provide four independent brightness measurements.

\begin{figure}
  \centering
  \plotone{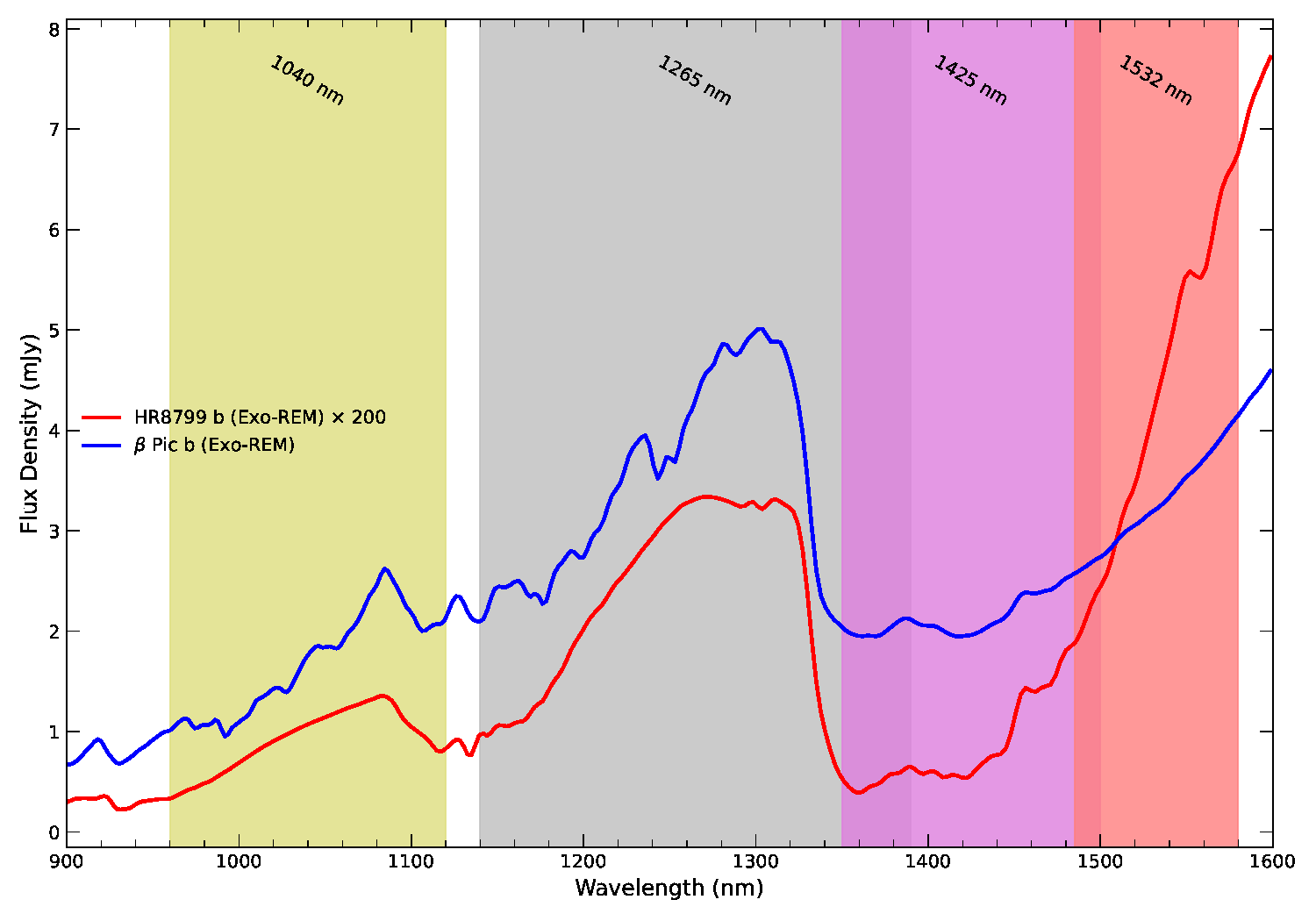}
  \caption{Near-infrared model spectra (900--1600~nm) generated from the self-consistent Exo-REM grid for HR~8799~b (red; multiplied by a factor of 200 for visibility) and $\beta$~Pic~b (blue). The spectra were produced with the \texttt{species} framework \citep{2023ascl.soft07057S} by interpolating the pre-computed Exo-REM model grid \citep{2015A&A...582A..83B}. Shaded regions mark the four NIR bandpasses
  adopted in this work (1040, 1265, 1425, and 1532~nm), illustrating how the filter set samples (i) the
  900--1100~nm window/continuum slope and (ii) the $\sim$1400~nm H$_2$O/CH$_4$ absorption complex.}
  \label{fig:infrared}
\end{figure}

To visualize how the four NIR bands map onto the dominant spectral structures of young, self-luminous
giant planets, Figure~\ref{fig:infrared} shows representative 900--1600~nm thermal-emission spectra
for two benchmark companions, HR~8799~b and $\beta$~Pic~b, generated from the self-consistent Exo-REM model grid
\citep{2015A&A...582A..83B} using the \texttt{species} framework \citep{2023ascl.soft07057S}. For HR~8799~b, we adopted the Exo-REM single best-fit grid parameters reported by \citet{2024A&A...687A.298N}: $T_{\rm eff}=850$~K, $\log g=3.5$, [M/H]$=0.5$, C/O$=0.55$, and $R=1.05\,R_{\rm Jup}$; the spectrum is plotted with an additional $\times200$ scaling to facilitate comparison across the same axis range. For $\beta$~Pic~b, we used the recent multi-modal Exo-REM constraints from \citet{2025A&A...704A.325R}: $T_{\rm eff}=1502.74$~K, $\log g=4.00$, [M/H]$=0.50$, and C/O$=0.552$; we adopted $R=1.73\,R_{\rm Jup}$ for the flux scaling, consistent with earlier Exo-REM-based radius estimates \citep{2015A&A...582A..83B}. The shaded CPI-C bandpasses highlight how the selected NIR filters sample the continuum slope and molecular bands in these young, self-luminous atmospheres.

\section{Optical reflected-light regime}	\label{sec:optical_param}

\subsection{Reference target and baseline geometry}
\label{subsec:ref_target}

We adopt $\pi^3$~Ori as a reference target to define a representative observing scenario. As a nearby, bright F-type main-sequence star, it provides a convenient benchmark for illustrating the expected performance of our simulated observations. Its parameters are taken as stellar inputs: $V=3.19$, spectral type F6V, and distance $d=8.024$~pc \citep{2024A&A...682A.145S}. We simulate a synthetic planet characterized by an instantaneous star-planet distance $r=2.0$~au and a baseline radius $R_p = 1.2\,R_{\rm J}$. The apparent separation is approximated as
\begin{equation}
\rho \;=\; \frac{r}{d} \;\;{\rm arcsec},
\end{equation}
which serves as a proxy for projected separation, neglecting orbital inclination and true anomaly for simplicity. This reference configuration yields $\rho\simeq 0.25^{\prime\prime}$. At the baseline phase angle of $\alpha=60^\circ$, the corresponding projected separation is approximately $\rho\sin\alpha \simeq 0.22^{\prime\prime}$, close to the inner working angle of CPI-C, and therefore provides a stringent but informative baseline for comparing scenarios.

\subsection{Atmospheric Model}	\label{sub:atmospheric_model}
For performance simulations and retrieval-style demonstrations with the optical bands,
we adopt the reflected-light model of \citet{2018AJ....156..158B} (hereafter B18), available via the \texttt{colorcolor} package. In Section \ref{sec:physics_diagnostics}, PSG was used only to illustrate the physical motivation for the filter placement using controlled perturbations of a Jupiter-like spectrum. For the sampling-based simulations in this section, we instead use the B18 grid because it provides a computationally convenient pre-computed reflected-light model grid parameterized by atmospheric metallicity and cloud sedimentation efficiency, while retaining the broad methane-driven spectral structures relevant to the VIS4 bands. From the comprehensive B18 multi-dimensional grid, we extract target spectra parameterized by two atmospheric variables: the logarithmic metallicity $[\mathrm{M/H}]\equiv \log_{10}(Z/Z_\odot)$, where $Z/Z_\odot$ is the heavy-element abundance relative to solar, and the cloud sedimentation efficiency $f_{\rm sed}$. In the representative grid calculations below, we use $[\mathrm{M/H}]=0,1,2$ and $f_{\rm sed}=0.01,1,6$, spanning $1$--$100\times$ solar metallicity and cloudy-to-relatively-clear atmospheric cases within the B18 grid. The standard B18 cloudy grid extends to $f_{\rm sed}=6$. In the continuous MCMC interpolation used below, we allow a limited extension from $f_{\rm sed}=6$ to $f_{\rm sed}=10$ in the internal $\log_{10}(f_{\rm sed})$ coordinate to avoid artificially truncating the posterior at the upper grid boundary. This interval is obtained by interpolation toward the pre-tabulated high-sedimentation endpoint and is used only to maintain numerical continuity in the fit; values between 6 and 10 are not treated as additional native B18 cloudy-grid nodes. The phase angle is treated as an additional model-grid coordinate, and the corresponding phase-dependent reflected-light spectrum is used directly in the synthetic photometry. 

For the controlled methane-abundance experiment introduced in Section \ref{sec:methane_sweep} below, we do not use the B18 reflected-light grid, because CH$_4$ abundance is not treated there as an independent model parameter. Instead, we generate a separate set of reflected-light spectra with \texttt{PICASO}~\citep{2019ApJ...878...70B}, starting from a Jupiter-like baseline atmosphere and varying the CH$_4$ abundance by multiplicative scaling while keeping the other baseline assumptions fixed. The resulting spectra are then integrated over the VIS4 bandpasses to obtain the corresponding broadband contrasts for direct comparison with the other visible-band scenarios.

\subsection{Noise model and signal-to-noise estimation}
\label{subsec:noise_snr}
To propagate instrument and detector effects into band-integrated observables, we use a custom simulation program, CPI-C Image Simulator (\texttt{CPISM}; \citealt{2026RAA....26b4010Z,2026RAA....26b4011Z}). \texttt{CPISM} models the complete optical train, including broadband PSF generation and detector response (e.g., EMCCD electron multiplication, dark current, readout noise), and provides the quantities needed for aperture photometry and noise estimation within the adopted bandpasses. The fixed baseline parameters adopted in the \texttt{CPISM} simulations are summarized in Table~\ref{tab:fixed_inst_params}.

We model broadband photometric measurements of a planet in reflected starlight as an aperture-summed counting experiment on a diffraction-limited, high-contrast image.  For a given bandpass with system throughput $T(\lambda)$, the stellar and planetary photo-electron rates at the detector are
\begin{equation}
\dot N_\star = A_{\rm tel}\int T(\lambda)\,F_\star(\lambda)\,{\rm d}\lambda,
\qquad
\dot N_p = A_{\rm tel}\int T(\lambda)\,F_p(\lambda)\,{\rm d}\lambda,
\end{equation}
where $A_{\rm tel}$ is the collecting area.  The planet spectrum $F_p(\lambda)$ used in the count-rate calculation is constructed from the adopted reflected-light model spectrum and then integrated over each bandpass.

For the B18-based calculations, the reflected-light model spectrum is taken from the B18 grid and interpolated over metallicity, cloud sedimentation efficiency, and phase angle. For the controlled methane-abundance experiment, the corresponding wavelength-dependent albedo spectrum is instead generated with \texttt{PICASO}, as described in Section \ref{sub:atmospheric_model}. This treatment allows phase angle to affect not only the overall brightness but also the relative band-to-band contrasts, since realistic giant-planet atmospheres can exhibit non-isotropic scattering and wavelength-dependent phase behavior~\citep{2010ApJ...724..189C}. The general contrast scaling follows the standard reflected-light description used in direct imaging studies (e.g., \citealt{1975lpsa.book.....S,2010exop.book..111T,2010ARA&A..48..631S,2016PASP..128b5003R}), while the simulations use the phase-dependent spectra supplied by the adopted atmospheric models.

\paragraph{Aperture photometry and photon/detector noise.} We sum the signal within a fixed photometric aperture that contains a fraction $\alpha_{\rm ap}$ of the planet point-spread function (PSF) and spans $N_{\rm pix}$ pixels.  For an exposure time $t$, the per-frame planet signal in the aperture is
\begin{equation}
S \equiv \alpha_{\rm ap}\,\dot N_p\,t .
\label{eq:signal_frame}
\end{equation}
The mean background rate per pixel is modeled as
\begin{equation}
\dot N_{\rm bg}=\dot N_{\rm sky}+\dot N_{\rm dark}+\dot N_{\rm leak},
\end{equation}
where $\dot N_{\rm sky}$ represents diffuse background (e.g., zodiacal light), $\dot N_{\rm dark}$ is the dark current, and $\dot N_{\rm leak}$ is the residual stellar leakage rate per pixel in the high-contrast region. In our implementation, $\dot N_{\rm leak}$ is parameterized through a band-dependent, dimensionless residual contrast factor multiplying $\dot N_\star$ and a PSF peak-to-region scaling, yielding a leakage rate consistent with a generic coronagraphic dark region.

For electron-multiplying CCD (EMCCD) operation, multiplication statistics introduce an excess noise factor $F$ and an effective reduction of read noise by the electron-multiplication gain $G$ \citep{2003SPIE.4796..164D,2003ITED...50.1227R}.  The per-frame variance of the aperture-summed measurement from photon and detector noise is written as \citep{2011MNRAS.411..211T}
\begin{equation}
\sigma_{\rm ph+det}^2
=
\left[\left(\dot N_{\rm bg}\,N_{\rm pix}+\alpha_{\rm ap}\,\dot N_p\right)t\right]F
+
\left[\left(\frac{\sigma_{\rm RN}}{G}\right)^2+C_{\rm CIC}F\right]N_{\rm pix},
\label{eq:ph_det_var}
\end{equation}
where $\sigma_{\rm RN}$ is the read noise at unit gain and $C_{\rm CIC}$ is the clock-induced charge (CIC) contribution expressed in electron units per pixel per frame.  

\paragraph{Residual speckle term and post-processing parameterization.}
In the high-contrast regime, quasi-static speckles produce a structured stellar residual that is not purely Poisson-distributed and may not average down as $t^{-1/2}$ once a speckle floor is reached. Following standard treatments in exoplanet direct imaging yield and detectability studies, we represent post-processed residual speckles as an additional, approximately Gaussian error term whose amplitude is a fixed
fraction of the raw stellar leakage in the photometric aperture \citep{2016AJ....152..217L}. Using the leakage rate $\dot N_{\rm leak}$ defined above, the corresponding raw leakage counts in the aperture in one frame are
\begin{equation}
N_{\rm leak}^{\rm (frame)}=\dot N_{\rm leak}\,N_{\rm pix}\,t\,F,
\end{equation}
and we define a post-processing factor $f_{\rm pp}$ such that the per-frame $1\sigma$ residual-speckle
amplitude is
\begin{equation}
\sigma_{\rm speck}\equiv f_{\rm pp}\,N_{\rm leak}^{\rm (frame)}.
\label{eq:speck_sigma}
\end{equation}
This form captures the fact that residual speckle errors behave as a systematic (or quasi-systematic) component proportional to the underlying stellar leakage, and it is the amplitude (not the mean count)
that enters the variance budget \citep{2016AJ....152..217L}.  In practice, $f_{\rm pp}$ parameterizes the effectiveness of wavefront control and subsequent PSF subtraction, serving as an explicit scenario parameter to systematically explore varying contrast limits.

The post-processing procedure motivating Equation~\ref{eq:speck_sigma} can be realized by reference differential imaging (RDI), in which each science frame is modeled as a linear combination of a reference library constructed to be free of planet signal. A standard implementation uses a low-rank basis obtained via singular value decomposition (SVD) to fit and subtract the speckle field in the region of interest \citep[e.g.,][]{2021MNRAS.502.2158R}. For completeness, alternative strategies that do not rely on field rotation, such as image-rotation subtraction, have also been proposed to attenuate quasi-static speckles \citep{2012ApJ...753...99R,2015ApJ...802...12D}. In the present work, the details of the post-processing pipeline are not simulated explicitly; instead, their net impact is represented by the tunable factor $f_{\rm pp}$.

\paragraph{Stacked-frame SNR and $3\sigma$ contrast limits.}
We consider a sequence of $n_{\rm frame}$ independent exposures, each of duration $t$.  Photon and detector
noise average down with the number of frames, while the residual speckle term is treated as fully correlated
between frames (i.e., a speckle-floor limit) and therefore does not decrease with $n_{\rm frame}$ in the
baseline calculations. With this noise model, the SNR for the stacked measurement can be written in a per-frame form as
\begin{equation}
{\rm SNR}
=
\frac{S}{\sqrt{\sigma_{\rm ph+det}^2/n_{\rm frame}+\sigma_{\rm speck}^2}},
\label{eq:snr_lupu_style}
\end{equation}
where $S$ is given by Equation~\ref{eq:signal_frame}, $\sigma_{\rm ph+det}^2$ by Equation~\ref{eq:ph_det_var}, and
$\sigma_{\rm speck}$ by Equation~\ref{eq:speck_sigma}. For each band, we additionally compute a $3\sigma$
detection (or upper-limit) contrast by solving Equation~\ref{eq:snr_lupu_style} for the required planet count rate
$\dot N_p$ at ${\rm SNR}=3$, keeping all background, detector, and speckle terms fixed for the assumed stellar
brightness, observing strategy, and $f_{\rm pp}$.  

\subsection{Scenario definitions and outputs}
\label{subsec:scenario_defs}

We define four complementary scenarios, each implemented by a dedicated script and each intended to probe
a different sensitivity axis.  Across all scenarios, the phase angle is treated as an independent input parameter. Phase-angle effects are incorporated through the phase-dependent reflected-light spectra adopted in the simulations. For clarity, we summarize the fixed instrumental and detector parameters used by the simulation software in Table~\ref{tab:fixed_inst_params}. These values serve as baseline configuration settings for the simulator to evaluate the assumed observing scenarios.

\begin{deluxetable*}{lcc}
\tablecaption{Fixed instrumental and detector parameters adopted in the noise model.\label{tab:fixed_inst_params}}
\tablewidth{0pt}
\tablehead{
\colhead{Parameter} & \colhead{Symbol} & \colhead{Value}
}
\startdata
Telescope collecting area & $A_{\rm tel}$ & $3.1416\times10^{4}\ \mathrm{cm^{2}}$ \\
Plate scale & $p$ & $0.016\ \mathrm{arcsec\ pix^{-1}}$ \\
Dark current & $D$ & $0.001\ \mathrm{e^{-}\ s^{-1}\ pix^{-1}}$ \\
Readout noise & $\sigma_{\rm r}$ & $160\ \mathrm{e^{-}\ pix^{-1}\ frame^{-1}}$ \\
EM gain & $G$ & $300$ \\
Clock-induced charge & $C_{\rm CIC}$ & $0.2\ \mathrm{e^{-}\ pix^{-1}\ frame^{-1}}$ \\
Excess noise factor & $F$ & $2.0$ \\
Sky surface brightness & $m_{\rm sky}$ & $21\ \mathrm{AB\ mag\ arcsec^{-2}}$ \\
\enddata
\tablecomments{The listed values are the baseline inputs adopted in \texttt{CPISM} for the simulations and noise calculations described in Section~\ref{subsec:noise_snr}.}
\end{deluxetable*}

\subsubsection{Scenario A: Atmospheric-parameter grid}
In this scenario we explore an atmosphere grid while holding the observing geometry fixed.  We evaluate a $3\times3$ grid in two atmospheric parameters that index the albedo catalog: logarithmic metallicity $[\mathrm{M/H}]\in\{0,1,2\}$, corresponding to $1$, $10$, and $100\times$ solar metallicity, and cloud sedimentation efficiency $f_{\rm sed}\in\{0.01,1,6\}$, representing cloudy, intermediate, and relatively clear atmospheres. The phase angle is fixed to $\alpha=60^\circ$.  The observing strategy is set to per-frame exposure time $t=120$~s with $n_{\rm frame}=30$ frames in each VIS4 band. Residual speckles are parameterized by a post-processing factor $f_{\rm pp}=0.05$, consistent with the ``speckle-amplitude fraction'' definition in Eq.~\ref{eq:speck_sigma}. This choice follows the representative value fpp = 1/20 adopted by \citet{2016AJ....152..217L}, who noted that post-processing speckle-reduction factors are expected to lie roughly between 1/10 and 1/30 based on \citet{2016JATIS...2a1020T}. For each grid point, we compute the per-band SNR and the band-integrated contrast $C$ defined in Eq.~\ref{eq:band_flux_ratio_def}, and we also solve for the $3\sigma$ contrast limits per band for the adopted reference configuration.  The primary products are (i) a contrast table over the atmosphere grid and (ii) a per-band summary of $3\sigma$ limits to be utilized as upper constraints.

For this scenario, we report the VIS4-band SNRs, band-integrated contrasts, and per-band relative uncertainties for each $([\mathrm{M/H}],f_{\rm sed})$ grid point. The per-band $3\sigma$ for the adopted reference configuration are listed in the note to Table~\ref{tab:sim_5mag}. These limits are used as band-specific upper limits in the low-SNR regime and as a consistent reference for comparing
different scenario choices.

\subsubsection{Scenario B: Methane-abundance sweep}	\label{sec:methane_sweep}
To directly test the methane sensitivity implied by the VIS4 band placement, we introduce a controlled methane-abundance experiment in which the observing geometry and instrumental assumptions are held fixed while the methane abundance is varied. Because the adopted B18 reflected-light grid does not provide CH$_4$ abundance as an independent model axis, we generate this sweep with PICASO using a Jupiter-like baseline atmosphere. We adopt the same $\pi^3$ Ori reference configuration as in Scenario~A. The CH$_4$ abundance is scaled by factors of $0.1$, $0.3$, $1.0$, $3.0$, $10.0$, and $30.0$ relative to the baseline atmosphere, while the remaining assumptions are kept fixed. For each methane-scaling case, we compute the continuous reflected-light contrast spectrum and then project it into the VIS4 bandpasses using the same throughput and noise prescription adopted elsewhere in Section~\ref{sec:optical_param}.

For this scenario, the output consists of a set of continuous reflected-light contrast spectra together with the corresponding VIS4-band integrated contrasts, SNRs, and per-band relative uncertainties for each adopted CH$_4$ scaling factor. This product is used to quantify how methane-driven spectral modulation maps onto the four-band optical measurements under the adopted baseline noise model.

We quantify the separation between two methane-scaling cases using the pairwise Mahalanobis distance
$D_{ij}=\left[(\mathbf{x}_i-\mathbf{x}_j)^{\rm T}(\Sigma_i+\Sigma_j)^{-1}(\mathbf{x}_i-\mathbf{x}_j)\right]^{1/2}$.
We evaluate this metric in the color--color plane,
$\mathbf{x}_{\rm CC}=(m_{p,\mathrm{F729}}-m_{p,\mathrm{F661}},\,m_{p,\mathrm{F877}}-m_{p,\mathrm{F661}})$,
and in the color--magnitude plane,
$\mathbf{x}_{\rm CM}=(m_{p,\mathrm{F729}},\,m_{p,\mathrm{F877}}-m_{p,\mathrm{F661}})$,
where the planet AB magnitudes are derived from the band-integrated contrasts using the adopted $\pi^3$~Ori stellar spectrum. The covariance matrices include the correlated color error introduced by the shared F661 measurement. The summed covariance provides a conservative measure of the separation between the two expected measurement distributions. We evaluate this metric only when all bands entering the diagnostic are detected; cases involving an F877 upper limit are retained as one-sided constraints.

Starting from the 3600~s per-band SNRs adopted for the baseline methane sweep, we evaluate equal total integration times per band from 30~s to 24~hr using individual exposures of at most 120~s. We also vary the residual-speckle factor over $0\leq f_{\rm pp}\leq0.05$. The $f_{\rm pp}=0$ case represents an idealized photon- and detector-noise-limited limit.

\subsubsection{Scenario C: Phase-angle sweep}
This scenario isolates the impact of viewing geometry and phase on detectability.  We fix the atmospheric parameters to a representative point $([\mathrm{M/H}],f_{\rm sed})=(1,1)$ and sweep the phase angle over a set of values: $\alpha\in\{30^\circ,60^\circ,90^\circ,120^\circ\}$. The phase angle is treated as a coordinate of the adopted phase-dependent reflected-light model grid, rather than as a separate multiplicative phase law. The observing strategy is again $t=120$~s per frame with $n_{\rm frame}=30$ in each band.  The post-processing factor is held fixed at $f_{\rm pp}=0.05$. For each phase angle, we compute the per-band stacked SNR and the band-integrated true contrast. This phase-sweep scenario produces a compact text table (one row per phase angle) and is intended to guide the interpretation of phase-dependent photometry, as well as the choice of ``representative'' phase angles in subsequent performance demonstrations.

For this scenario, the output is a four-row text table (one row per phase angle) with VIS4-band
SNRs and true contrasts.  This product is used to quantify the phase dependence of detectability and to
select representative phase angles for performance demonstrations.

\subsubsection{Scenario D: Post-processing-level sweep}
The fourth scenario explores how the assumed effectiveness of post-processing changes the inferred measurement quality in the speckle-limited regime.  We fix the atmosphere to $([\mathrm{M/H}],f_{\rm sed})=(1,1)$ and use the same target geometry as in Scenario~A, while scanning $f_{\rm pp}\in\{0.025,0.05,0.075,0.1\}$. For each $f_{\rm pp}$ we compute the four-band SNRs, contrasts, and $3\sigma$ upper limits, and we visualize the results in a four-panel comparison plot. This scenario provides a transparent mapping between an assumed post-processing performance level and the achievable photometric constraints.

For this scenario, we generate a multi-panel comparison plot that overlays (i) the model
contrast spectrum implied by the chosen phase-dependent reflected-light model and (ii) the simulated
four-band measurements with either symmetric error bars (when ${\rm SNR}\ge 3$) or downward arrows denoting
$3\sigma$ upper limits (when ${\rm SNR}<3$). This visualization demonstrates how the assumed post-processing performance translates into practical photometric constraints across the VIS4 bands.

\subsection{Scenario results}
\label{subsec:scenario_outcomes_vis}

We present the results below in the same order as the scenario definitions in Section~\ref{subsec:scenario_defs}, using the Scenario A--D nomenclature consistently throughout. The discussion focuses on the information directly supported by the simulated band-integrated contrasts, their uncertainties, and the $3\sigma$ upper limits produced by the adopted noise model and speckle-floor parameterization.

\subsubsection{Scenario A: Atmospheric-parameter grid}
\begin{figure}
  \centering
  \includegraphics[width=\linewidth]{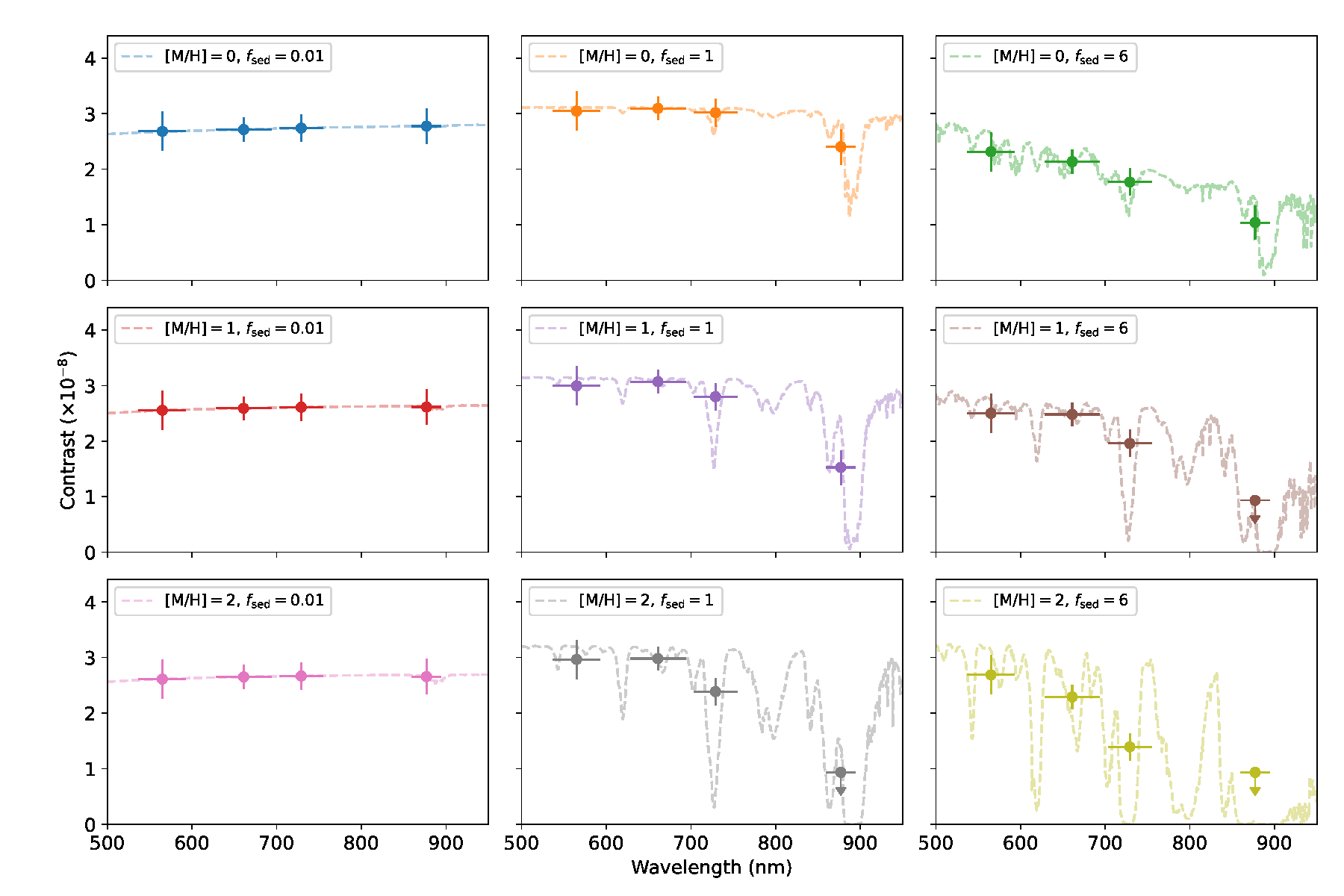}
  \caption{Visible-band scenario A: VIS4 contrasts for the $([\mathrm{M/H}],f_{\rm sed})$ grid at fixed phase and observing
  setup. Points with error bars denote ${\rm SNR}\ge3$ measurements; arrows denote per-band $3\sigma$ upper
  limits computed from Equation~\ref{eq:snr_lupu_style}.}
  \label{fig:model_pi3_ori}
\end{figure}

Figure~\ref{fig:model_pi3_ori} shows the VIS4 outputs for the $3\times3$ grid in $([\mathrm{M/H}],f_{\rm sed})$ at fixed geometry and phase. The dashed curves represent the phase-dependent model contrast spectra supplied by the B18 grid after applying only the geometric $(R_p/r)^2$ scaling, while the points show the band-integrated contrasts. No additional phase-function factor is applied at this stage because the phase dependence is already encoded in the model spectra. For bands with ${\rm SNR}\ge 3$, we plot symmetric error bars computed from the simulated SNR; for ${\rm SNR}<3$, we plot downward arrows indicating the per-band $3\sigma$ upper limits derived by solving Equation~\ref{eq:snr_lupu_style}. The corresponding band-integrated contrasts and SNR values for all nine $([\mathrm{M/H}],f_{\rm sed})$ scenarios are listed in Table~\ref{tab:sim_5mag}.

Two practical features are evident. First, within this four-band sampling, much of the atmospheric information is primarily derived from the relative flux differences between bands (i.e., coarse spectral shape) rather than through an absolute contrast normalization, which remains strongly coupled to the overall scaling $\Phi(\alpha)(R_p/r)^2$.  Second, the redder band (F877) frequently transitions into an upper-limit regime under the fiducial observing strategy, so that the effective number of informative measurements can be reduced below four for some grid points.  In such cases, apparent spectral curvature suggested by the forward-model curve is only weakly constrained by the data and is instead bounded primarily by the upper limit.  These behaviors are consistent with the general experience in high-contrast photometry that sparse-band measurements can provide strong shape constraints when multiple bands are detected, but quickly become prior- and upper-limit-dominated when one or more bands fall below the detection threshold \citep[e.g.,][]{2016AJ....152..217L}.

\begin{deluxetable*}{c|cc|cc|cc|cc}
\tablecaption{Signal-to-noise ratios and contrasts of simulated planets with different atmospheric properties in the $\pi^3$~Ori system observed by CPI-C.\label{tab:sim_5mag}}
\tablewidth{0pt}
\tablehead{
 & \multicolumn{2}{c|}{F565} & \multicolumn{2}{c|}{F661} & \multicolumn{2}{c|}{F729} & \multicolumn{2}{c}{F877} \\
 & \colhead{SNR}    & \colhead{Contrast} \vline  & \colhead{SNR}    & \colhead{Contrast} \vline  & \colhead{SNR}    & \colhead{Contrast} \vline  & \colhead{SNR}    & \colhead{Contrast} \\
 & &($\times 10^{-8}$)& &($\times 10^{-8}$)& &($\times 10^{-8}$)& &($\times 10^{-8}$)
}
\startdata
$[\mathrm{M/H}]=0$, $f_{\rm sed}=0.01$ & 7.5165 & 2.6819 & 12.4324 & 2.7149 & 10.8779 & 2.7407 & 8.5167 & 2.7757 \\
$[\mathrm{M/H}]=0$, $f_{\rm sed}=1$ & 8.5290 & 3.0477 & 14.1139 & 3.0957 & 11.9329 & 3.0184 & 7.4460 & 2.4046 \\
$[\mathrm{M/H}]=0$, $f_{\rm sed}=6$ & 6.4991 & 2.3153 & 9.8492 & 2.1363 & 7.1193 & 1.7686 & 3.3330 & 1.0390 \\
$[\mathrm{M/H}]=1$, $f_{\rm sed}=0.01$ & 7.1665 & 2.5556 & 11.8865 & 2.5920 & 10.3812 & 2.6107 & 8.0563 & 2.6153 \\
$[\mathrm{M/H}]=1$, $f_{\rm sed}=1$ & 8.3885 & 2.9969 & 14.0059 & 3.0711 & 11.1048 & 2.8002 & 4.8334 & 1.5263 \\
$[\mathrm{M/H}]=1$, $f_{\rm sed}=6$ & 7.0184 & 2.5023 & 11.3944 & 2.4815 & 7.8626 & 1.9587 & 1.5680 & 0.4814 \\
$[\mathrm{M/H}]=2$, $f_{\rm sed}=0.01$ & 7.3220 & 2.6117 & 12.1479 & 2.6508 & 10.5956 & 2.6667 & 8.1711 & 2.6552 \\
$[\mathrm{M/H}]=2$, $f_{\rm sed}=1$ & 8.3025 & 2.9658 & 13.6115 & 2.9816 & 9.5201 & 2.3864 & 2.4817 & 0.7680 \\
$[\mathrm{M/H}]=2$, $f_{\rm sed}=6$ & 7.5351 & 2.6886 & 10.5413 & 2.2906 & 5.6296 & 1.3907 & 0.2294 & 0.0696 \\
\enddata
\tablecomments{Each row corresponds to one atmospheric scenario defined by metallicity $[\mathrm{M/H}]$ and cloud sedimentation parameter $f_{\rm sed}$. For each band (F565, F661, F729, F877), we report the simulated band-integrated contrast (in units of $10^{-8}$) and the resulting SNR computed from the same noise prescription used in the visible-band
analysis. Values with ${\rm SNR}<3$ are treated as non-detections and are handled as one-sided $3\sigma$ upper limits in the likelihood. The adopted per-band $3\sigma$ upper limits are $[2.05,\,0.25,\,1.25,\,4.99] \times 10^{-8}$ for (F565, F661, F729, F877), respectively. }
\end{deluxetable*}

A further trend across the grid is the role of clouds in setting a detectability--diagnostic trade-off. In cloudier cases (smaller $f_{\rm sed}$), the planet is generally brighter in the visible, so more bands reach ${\rm SNR}\ge3$ under the same geometry and observing effort. However, the spectrum becomes flatter and band-to-band differences are reduced, which weakens the visibility of absorption-driven structure in sparse photometry. In clearer cases (larger $f_{\rm sed}$), the forward models show stronger wavelength dependence, but the overall contrast drops and non-detections become more frequent; once key bands turn into upper limits, the data constrain only broad colors and a brightness scale rather than detailed feature depths.

Metallicity introduces another systematic effect. Higher $[\mathrm{M/H}]$ increases molecular opacity and tends to deepen absorption features in the visible, strengthening the contrast between absorption-sensitive bands and nearby continuum anchors in the forward spectra. This trend is most informative when multiple diagnostic bands are detected, so that band ratios directly probe the relative depths. When the reddest band(s) are upper-limit
dominated, the metallicity sensitivity is largely lost because the limits mainly exclude overly bright models instead of measuring the feature depth.

\subsubsection{Scenario B: Methane-abundance sweep}

\begin{figure}
  \centering
  \includegraphics[width=\linewidth]{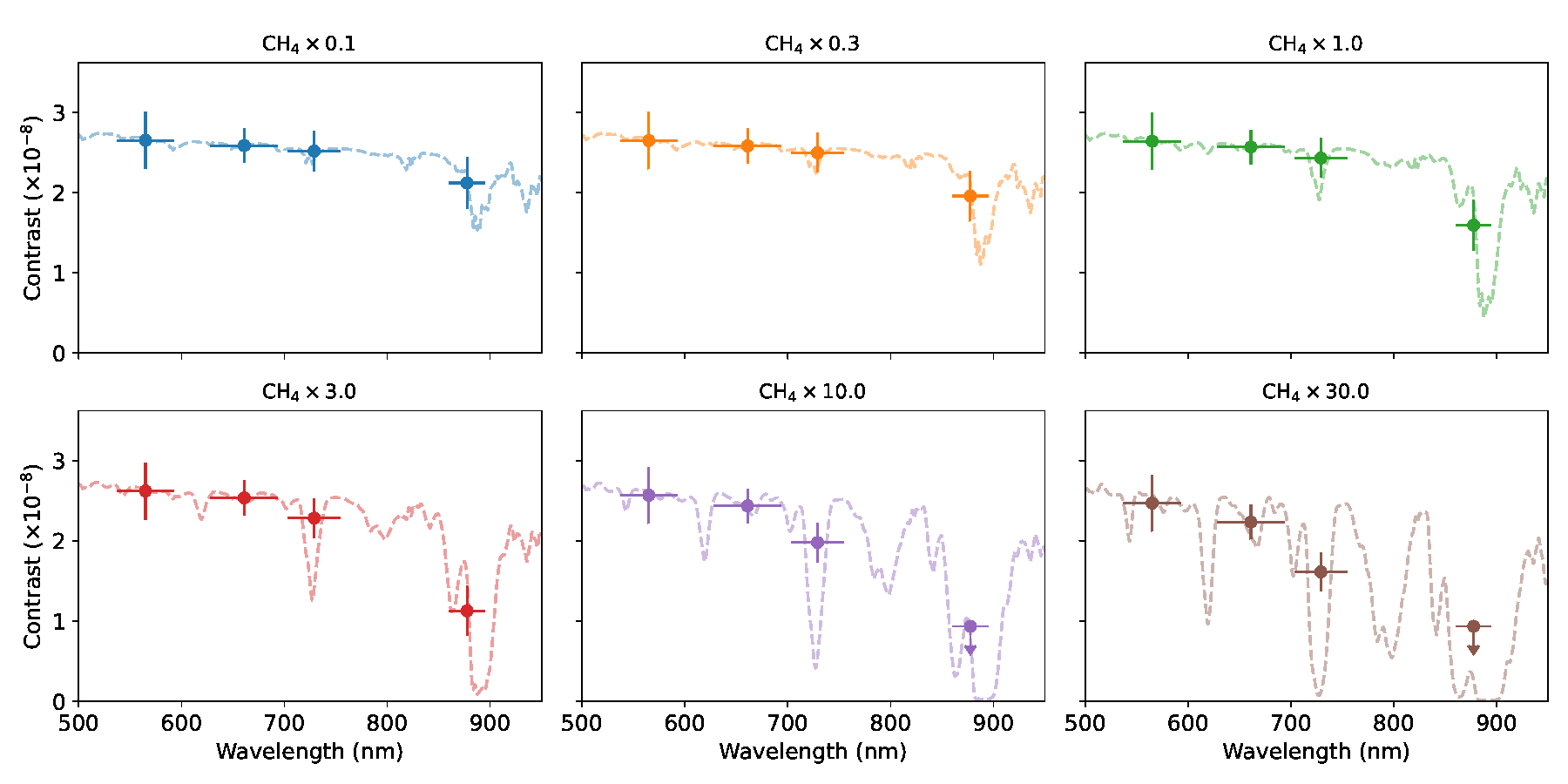}
  \caption{Visible-band scenario B: VIS4 contrasts for a controlled methane-abundance sweep under the $\pi^3$ Ori baseline geometry and observing setup. Panels show CH$_4$ abundance scaling factors from 0.1 to 30 relative to the baseline case. Symbols and arrows follow the same convention as Figure~\ref{fig:model_pi3_ori}.}
  \label{fig:ch4sweep}
\end{figure}

To further isolate the role of methane in the VIS4 diagnostics, we next consider a controlled methane-abundance experiment under the same baseline observing assumptions. Figure~\ref{fig:ch4sweep} shows the VIS4 response to a controlled methane-abundance sweep under the same baseline geometry and observing assumptions adopted above. The panels correspond to CH$_4$ abundance scaling factors of $0.1$, $0.3$, $1.0$, $3.0$, $10.0$, and $30.0$ relative to the baseline case, while the other assumptions are kept fixed. In each panel, the dashed curve represents the continuous reflected-light contrast spectrum and the points show the corresponding VIS4 band-integrated measurements. Bands with SNR $\geq 3$ are plotted with symmetric error bars, whereas bands with SNR $< 3$ are shown as $3\sigma$ upper limits.

A clear trend is that the shorter-wavelength bands, F565 and F661, change only modestly across the methane sweep. In this setup, these two bands mainly trace the overall reflected-light level and provide a local continuum reference. By contrast, the redder bands, F729 and especially F877, become progressively weaker as the methane abundance increases. This behavior is consistent with the band-placement logic introduced in Section~\ref{subsec:optical_reflected}.

This sweep also reveals how the extracted information changes with methane abundance. At low to moderate methane levels, all four VIS4 bands are detected, and their relative fluxes clearly define the overall spectral shape. At higher methane abundance, strong absorption can drive the reddest band (F877) into a low-SNR or upper-limit regime. In that case, the non-detection still provides useful information by excluding models that predict too much flux in the F877 band, but the constraint becomes one-sided rather than a direct symmetric measurement of the band depth.

\begin{figure}
  \centering
  \includegraphics[width=\linewidth]{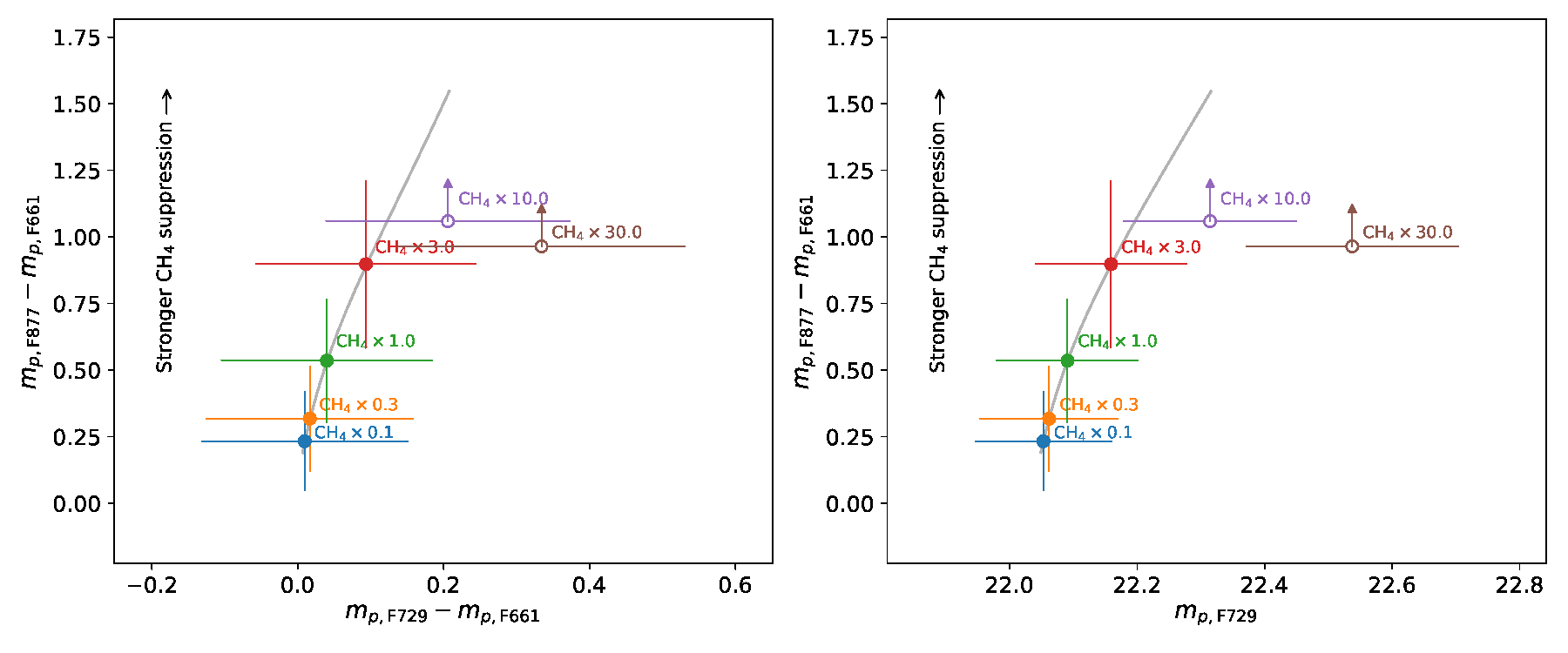}
  \caption{Methane-sensitive photometric diagnostics for the controlled methane-abundance sweep. Left: planet F877-F661 color versus F729-F661 color. Right: planet F877-F661 color versus apparent F729 AB magnitude. Symmetric error bars indicate measurements derived from detected bands, while open symbols with arrows denote one-sided limits when F877 is below the adopted detection threshold. The gray curves show the theoretical loci as the methane scaling factor changes.}
  \label{fig:colorcolor}
\end{figure}

Figure~\ref{fig:colorcolor} compares the methane-sensitive color--color and color--magnitude representations. In the left panel, increasing methane abundance produces progressively redder F729-F661 and F877-F661 colors because F729 and F877 are suppressed more strongly than the F661 continuum reference. The small changes in F565 and F661 across the sweep also imply weak methane sensitivity in F661-F565. In the right panel, the apparent F729 AB magnitude differs by only 0.038~mag between the CH$_4\times0.1$ and CH$_4\times1$ cases, compared with $1\sigma$ uncertainties of 0.108 and 0.112~mag. The color--magnitude representation therefore provides little additional leverage for distinguishing this one-order-of-magnitude abundance difference.

For the baseline integration of 3600~s per band, the conservative pairwise Mahalanobis separation is $D=1.02$ between CH$_4\times0.1$ and CH$_4\times1$, and $D=1.81$ between CH$_4\times0.1$ and CH$_4\times3$ in the color--color plane. The corresponding color--magnitude separations are 1.04 and 1.93. The similar separations in the two representations indicate that including the apparent F729 magnitude does not substantially improve the methane-abundance discrimination. We use D = 1 and D = 2 only as benchmarks for the pairwise separation metric, rather than as formal confidence levels. Under the baseline observing conditions, VIS4 is mainly sensitive to broad differences between low and high methane levels and provides only limited discrimination between CH$_4\times0.1$ and CH$_4\times1$.

At the adopted residual-speckle level of $f_{\rm pp}=0.05$, the CH$_4\times0.1$--CH$_4\times1$ color--color separation approaches $D\simeq1.22$ at long integration times and does not reach $D=2$ through increased exposure alone. Reducing the residual-speckle amplitude to half of the baseline value, $f_{\rm pp}=0.025$, allows the pairwise separation to reach $D=2$ after approximately 3.60~hr per band. In comparison, CH$_4\times0.1$ and CH$_4\times3$ reach $D=2$ after approximately 2.56~hr per band at the baseline residual-speckle level.

\subsubsection{Scenario C: Phase-angle sweep}
\begin{figure}
  \centering
  \includegraphics[width=\linewidth]{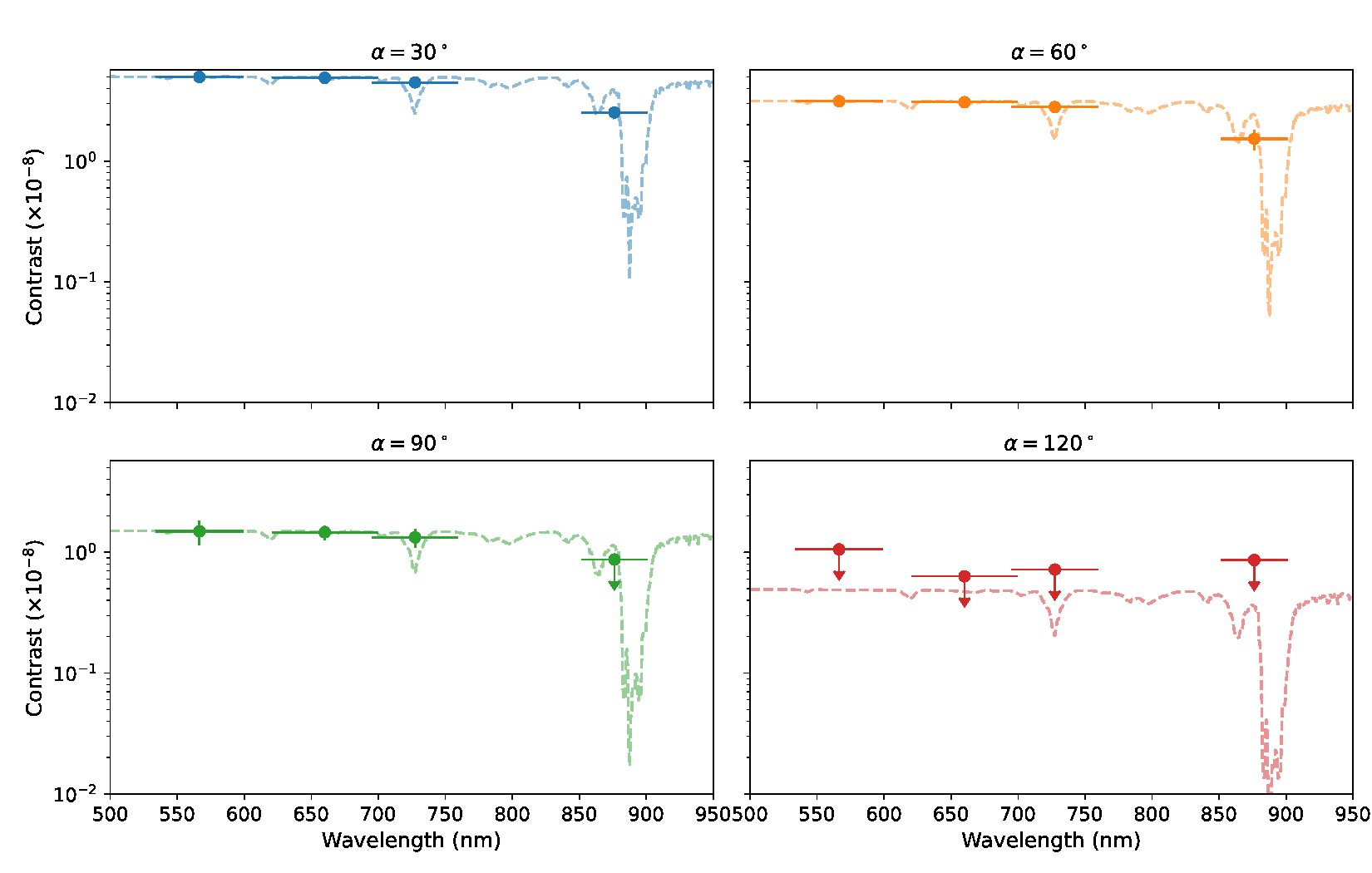}
  \caption{Visible-band scenario C: fixed-atmosphere VIS4 contrasts at four phase angles. Symbols and arrows
  follow the same convention as Figure~\ref{fig:model_pi3_ori}.}
  \label{fig:phase_spectra_pi3_ori}
\end{figure}

Figure~\ref{fig:phase_spectra_pi3_ori} isolates the phase-angle
dependence by fixing the atmospheric parameters and varying $\alpha$ over
$\{30^\circ,60^\circ,90^\circ,120^\circ\}$.  In the forward model, phase enters through the phase-dependent reflected-light spectra provided by the B18 atmospheric grid. The plotted points demonstrate the practical consequence for band photometry: as the overall contrast drops with phase, one or more bands can move from the detection regime into an upper-limit regime, reducing the amount of spectral-shape information available from VIS4.  For visual consistency across the phase-angle panels, the upper-limit arrows use the $3\sigma$ contrast limits calculated for the same separation and observing setup and therefore serve as a common reference threshold for display.  This phase-only experiment therefore provides a clear statement: even when the intrinsic albedo spectrum is fixed, phase can control whether the visible spectrum is constrained by multi-band detections or is dominated by upper limits in the reddest band, which directly changes the diagnostic content of the observation \citep[e.g.,][]{2016AJ....152..217L}.

\subsubsection{Scenario D: Post-processing-level sweep}
\begin{figure}
  \centering
  \includegraphics[width=\linewidth]{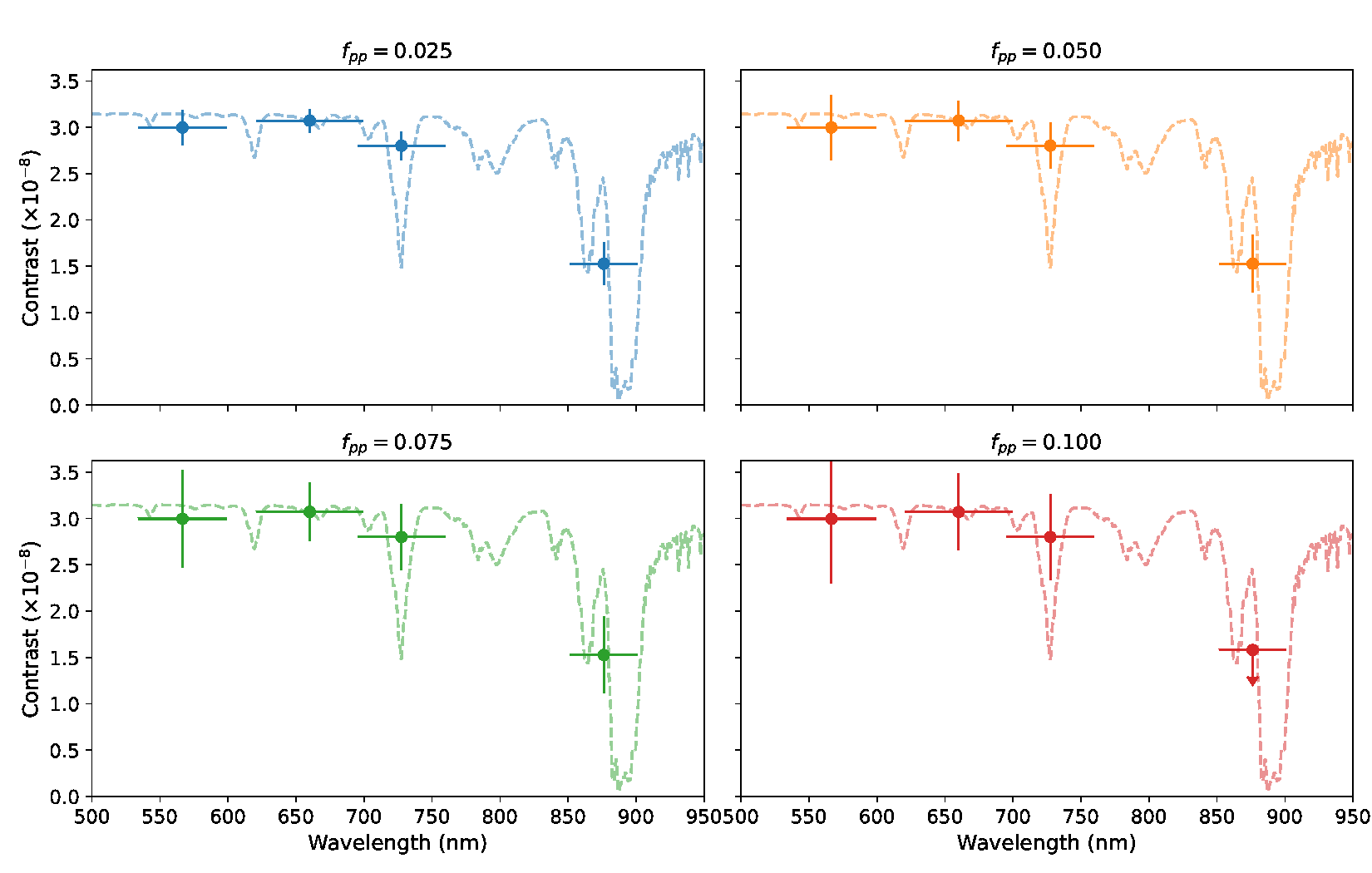}
  \caption{Visible-band scenario D: sensitivity of VIS4 constraints to the post-processing factor $f_{\rm pp}$.
  Panels show the same atmosphere and phase but different residual-speckle levels.}
  \label{fig:fpp_compare_pi3_ori}
\end{figure}

Figure~\ref{fig:fpp_compare_pi3_ori} explores the impact of the
post-processing factor $f_{\rm pp}$ that sets the residual-speckle amplitude in
Equation~\ref{eq:speck_sigma}.  Holding the atmosphere and phase fixed, increasing $f_{\rm pp}$ increases the speckle-floor variance term and therefore inflates the photometric uncertainties and promotes more bands into the upper-limit regime.  Conversely, smaller $f_{\rm pp}$ produces tighter error bars and preserves more
detectable bands, strengthening the constraint on the coarse spectral shape.
This behavior is a direct consequence of the scenario parameterization, demonstrating how $f_{\rm pp}$ parameterizes the net effectiveness of PSF subtraction (e.g., RDI/SVD-style processing) and is treated as an explicit input rather than a predicted quantity \citep[e.g.,][]{2016AJ....152..217L}.

\subsection{Atmospheric-parameter retrieval with the VIS4 filter set}
Using the synthetic VIS4 photometry generated for Scenario~A, we examine the atmospheric-parameter constraints supported by four-band reflected-light measurements. This analysis treats the simulated photometry as input to a parameter-inference calculation and evaluates the resulting posterior constraints and degeneracies.

\subsubsection{MCMC setup}
To assess what a small set of visible-band photometric points can constrain under the adopted assumptions, we performed an MCMC analysis of the synthetic VIS4 photometry from Scenario~A. Each fit corresponds to one input atmospheric grid case and uses a four-band data vector consisting of the simulated band-integrated contrasts in F565, F661, F729, and F877 and their SNR-derived uncertainties. At each MCMC step, the model contrast vector was computed from the B18 reflected-light grid at the fixed phase angle $\alpha=60^\circ$ by interpolating in $[\mathrm{M/H}]$ and $\log_{10}(f_{\rm sed})$, scaling the spectrum by $(R_p/r)^2$, and integrating it through the same VIS4 throughput curves.

The sampled parameters were $[\mathrm{M/H}]$, $\log_{10}(f_{\rm sed})$, and $R_p/r$. Uniform priors were adopted over $0\leq[\mathrm{M/H}]\leq2$, $-2\leq\log_{10}(f_{\rm sed})\leq1$, and $0<R_p/r\times10^4<20$. The upper cloud prior corresponds to $f_{\rm sed}=10$. As described in Section~\ref{sub:atmospheric_model}, the interval $6<f_{\rm sed}\leq10$ is retained as a limited interpolation range to prevent the fit from being artificially truncated at the native $f_{\rm sed}=6$ boundary, rather than as an extension of the native B18 cloudy-grid nodes. Bands with ${\rm SNR}\geq3$ were fitted with a Gaussian likelihood using the simulated contrast uncertainties, while bands with ${\rm SNR}<3$ were included as one-sided upper limits through the Gaussian cumulative probability below the adopted $3\sigma$ contrast limit.

The posterior distributions were sampled using the affine-invariant ensemble sampler implemented in \texttt{emcee} \citep{2013PASP..125..306F}. For each of the nine atmospheric input cases, we used 16 walkers and 5000 steps per walker. The first 2000 steps of each chain were discarded as burn-in, and the remaining samples were retained without thinning, giving 48,000 retained posterior samples for each fit. Parallel tempering was not used.

\subsubsection{Posterior constraints and degeneracies}

\begin{figure}
  \centering
  \includegraphics[width=\linewidth]{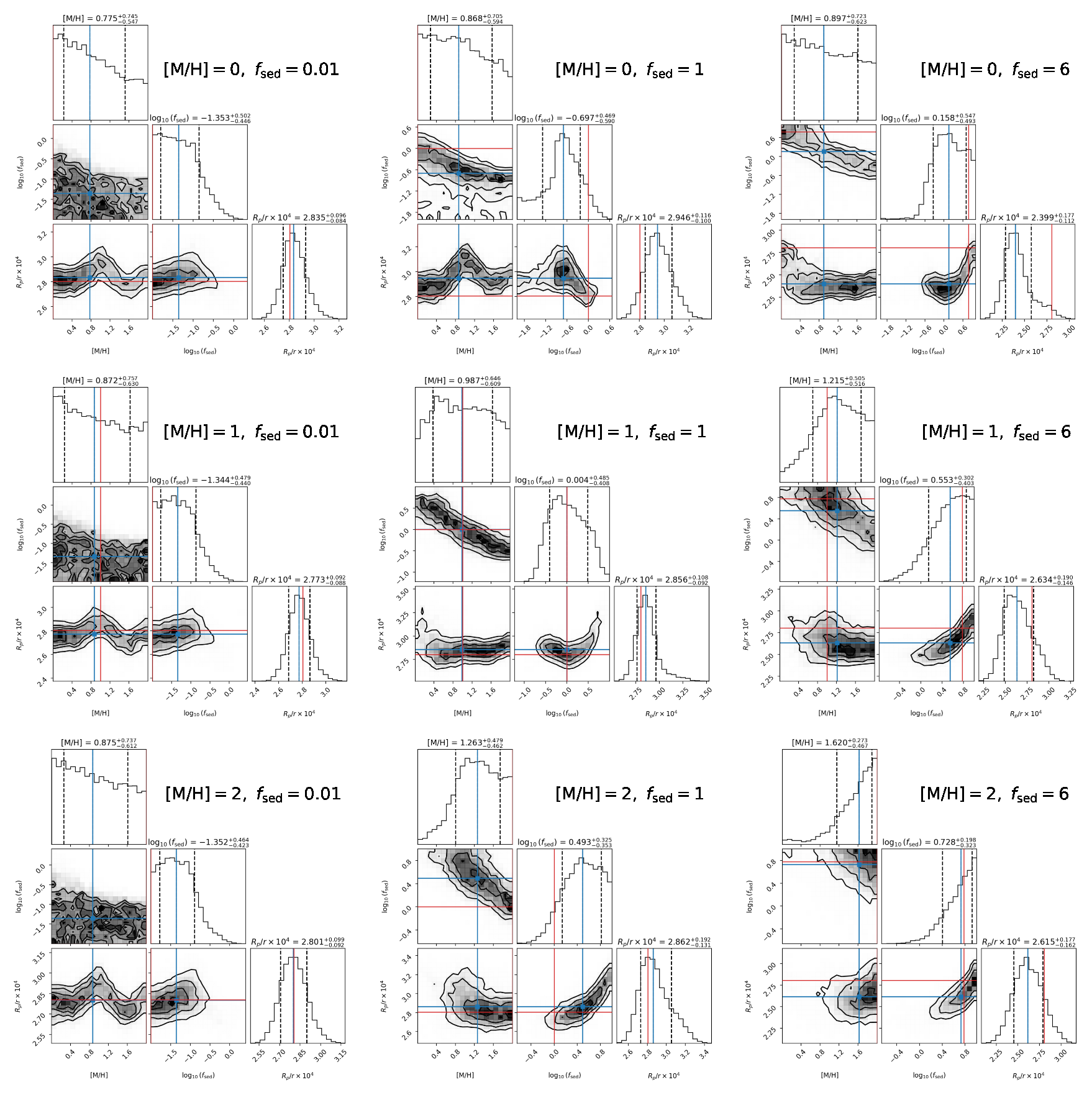}
  \caption{Posterior distributions from MCMC fits to the synthetic VIS4 photometry generated in Scenario~A. Each panel corresponds to one atmospheric input case. The marginal titles report the posterior median and the 16th and 84th percentiles. Blue lines indicate the posterior medians, and red lines indicate the input parameter values. The contours illustrate the degeneracies between the spectral-shape parameters and the reflected-light amplitude and show how upper-limit--dominated bands broaden the
  posterior volume.}
  \label{fig:corner_all}
\end{figure}

Figure~\ref{fig:corner_all} shows the posterior distributions from the independent fits to the Scenario~A atmospheric grid cases. Two common features are evident.
First, parameters that primarily modify the spectral shape, such as metallicity and cloud sedimentation efficiency, exhibit degeneracies with an overall amplitude degree of freedom, here represented by $R_p/r$ through the scaling in Eq.~\ref{eq:reflected_contrast_general}. This indicates that changes in spectral shape can be partially compensated by a global brightness rescaling when only a few bands are available. Second, when one or more bands fall into an upper-limit--dominated regime, the allowed region in parameter space broadens and the apparent correlations become stronger, because upper limits mainly exclude overly bright models rather than providing symmetric constraints around a measured value.

Across these independent fits, the posteriors generally show that the amplitude parameter is more robustly constrained when multiple bands are detected, whereas metallicity- and cloud-related parameters tend to remain weakly identified and/or degenerate under sparse visible-band sampling. Therefore, robustly characterizing exoplanetary atmospheres from sparse photometry requires mapping the full posterior distribution to capture these degeneracies, rather than relying on a single best-fit model \citep[e.g.,][]{2016AJ....152..217L}.

\section{Thermal Emission regime for warm self-luminous companions}	\label{sec:near_infrared_regime}

Having explored the optical bands as reflected-light diagnostics for cool giant planets, we now examine thermal-emission measurements for warm self-luminous companions across the CPI-C optical–NIR wavelength range. While reflected-light observations are naturally optimized for cool ($T_{\rm eff} \lesssim 1000$~K) planets, NIR photometry can extend the accessible parameter space to moderately warm ($T_{\rm eff} \sim 1000$--2000~K) companions whose self-luminosity provides the dominant flux contribution in this wavelength range. This thermal-emission regime is well represented by benchmark directly imaged planets such as $\beta$ Pictoris b and HR 8799 bcde, whose near-infrared spectral energy distributions have been extensively characterized by ground-based and space-based observations~\citep{2008Sci...322.1348M,2010Natur.468.1080M,2009A&A...493L..21L,2010Sci...329...57L,2025AJ....169..209B,2026ApJ..1000...27X}. 

For the benchmark warm companions considered in this section, the reflected-light contribution is small over the CPI-C wavelength range used in the thermal-emission calculation. Using the reflected-light scaling in Equation~\ref{eq:reflected_contrast_general},
and adopting $R_p \simeq 1.7\,R_{\rm J}$ and $r \simeq 9$--10~au for $\beta$ Pic b, we obtain
$
\left({R_p}/{r}\right)^2
\simeq (6.6\text{--}8.2)\times10^{-9}.
$
Thus, even in the limiting case $A_g(\lambda)\Phi(\alpha)=1$, the reflected-light contrast is bounded at $C_{\rm ref}\lesssim10^{-8}$, with smaller values expected at near-infrared wavelengths and non-zero phase angles. We therefore neglect the reflected-light contribution in the following thermal-emission analysis.

In Sections~\ref{sub:betapicb} and \ref{sub:synthetic_warm}, we assess CPI-C thermal-emission diagnostics using the \texttt{Sonora Diamondback} atmosphere grid~\citep{2024ApJ...975...59M}. We separate two related but different questions. First, we use $\beta$ Pic b as a benchmark object to check the \texttt{Sonora}-based fitting procedure with synthetic CPI-C VIS4+NIR4 photometry and extensive existing data. Second, because benchmark planets such as $\beta$ Pic b already have rich multi-instrument spectrophotometric coverage, we use a controlled synthetic warm-giant experiment to quantify the information content of the CPI-C NIR filters for a discovery-like target with sparse prior photometry. Thus, the $\beta$ Pic b calculation below is intended as a benchmark consistency check, while the parameter-gain assessment is carried out with the controlled synthetic experiment. The CPI-C synthetic photometry is generated with the channel-specific throughput curves and noise prescriptions used in the corresponding VIS4 and NIR4 simulations. Section~\ref{subsec:hr8799_optical_thermal} then uses atmosphere-model spectra of the HR~8799 planets to examine how thermal emission is distributed across the full CPI-C VIS4+NIR4 wavelength range.

\subsection{Sonora-based eight-band benchmark fit for beta Pic b} \label{sub:betapicb}

We first apply the Sonora-based framework to $\beta$ Pic b, a warm directly imaged planet with extensive high-contrast spectrophotometry. For the benchmark fit, we combine the GPI Y/J/H spectrum from \citet{2017AJ....153..182C}, the GRAVITY K-band spectrum from \citet{2020A&A...633A.110G}, broadband photometry from \citet{2013ApJ...776...15C}, and eight synthetic CPI-C photometric points covering VIS4 and NIR4. The GPI and GRAVITY spectra are treated as separate spectroscopic data sets, while the literature and CPI-C measurements are included as broadband photometric constraints.

The model spectra are computed with the \texttt{Sonora Diamondback} grid. We fit the data with $T_{\rm eff}$, $\log g$, $R_{\rm p}$, [Fe/H], and $f_{\rm sed}$ as free model parameters. We also include independent scaling factors for the GPI Y, J, H and GRAVITY K spectra to account for relative spectrophotometric calibration uncertainties between data sets. The distance is fixed to 19.63~pc. The posterior distributions are sampled with the \texttt{species} framework \citep{2023ascl.soft07057S}, using the reactive nested-sampling algorithm implemented in \texttt{UltraNest} \citep{2021JOSS....6.3001B}. We use a minimum of 300 live points and an evidence-tolerance criterion of $\Delta\ln Z=0.5$. Uniform priors are adopted over $1200<T_{\rm eff}<2200$~K, $3.5<\log g<5.5$, $0.8<R_{\rm p}/R_{\rm J}<2.2$, $-0.5<[\mathrm{Fe/H}]<0.5$, and $1<f_{\rm sed}<8$. The independent GPI Y, J, H and the GRAVITY K scaling factors have uniform priors between 0.5 and 1.5.

The simulated CPI-C photometry is generated with the actual F565, F661, F729, F877, F1040, F1265, F1425, and F1532 throughput curves. The VIS4 points sample the optical thermal tail of $\beta$~Pic~b. The NIR4 points sample the 1.0--1.6~$\mu$m continuum structure and the broad 1.4~$\mu$m molecular absorption region (Table~\ref{tab:betapicb_cpic_photometry}). For the NIR detector we use a CCD noise model with read noise of 70~$e^-$~pix$^{-1}$~frame$^{-1}$, dark current of 10~$e^-$~s$^{-1}$~pix$^{-1}$, plate scale of 0.025~arcsec~pix$^{-1}$, and sky background of 21~AB~mag~arcsec$^{-2}$. Residual speckle noise is included through a post-processing factor $f_{\rm pp}=0.15$. Each filter was simulated with 30 frames of 120 s each, corresponding to a total on-source integration time of 3600 s per band. The same exposure setup was adopted for all eight bands.

\begin{deluxetable}{lcccc}
\tablecaption{Simulated CPI-C VIS4+NIR4 photometry for $\beta$ Pic b.
\label{tab:betapicb_cpic_photometry}}
\tablehead{
\colhead{Band} &
\colhead{exposure time} &
\colhead{$m_{\rm AB}$} &
\colhead{$\sigma_m$} &
\colhead{Contrast} \\
\colhead{} &
\colhead{(seconds)} &
\colhead{(mag)} &
\colhead{(mag)} &
\colhead{}
}
\startdata
F565  & $120\times 30$ & 22.7211 & 0.1380 & $2.87 \times 10^{-8}$ \\
F661  & $120\times 30$ & 19.9426 & 0.0098 & $3.91 \times 10^{-7}$ \\
F729  & $120\times 30$ & 19.0196 & 0.0071 & $9.63 \times 10^{-7}$ \\
F877  & $120\times 30$ & 17.1668 & 0.0046 & $6.10 \times 10^{-6}$ \\
F1040 & $120\times 30$ & 15.9080 & 0.3201 & $2.19 \times 10^{-5}$ \\
F1265 & $120\times 30$ & 15.0229 & 0.0587 & $5.28 \times 10^{-5}$ \\
F1425 & $120\times 30$ & 15.1717 & 0.0421 & $5.15 \times 10^{-5}$ \\
F1532 & $120\times 30$ & 14.9163 & 0.0265 & $7.18 \times 10^{-5}$ \\
\enddata
\tablecomments{The CPI-C points are generated by integrating the Sonora model spectrum over the corresponding CPI-C VIS4 and NIR4 throughput curves and applying the channel-specific noise prescriptions.}
\end{deluxetable}

\begin{figure*}
\centering
\includegraphics[width=\linewidth]{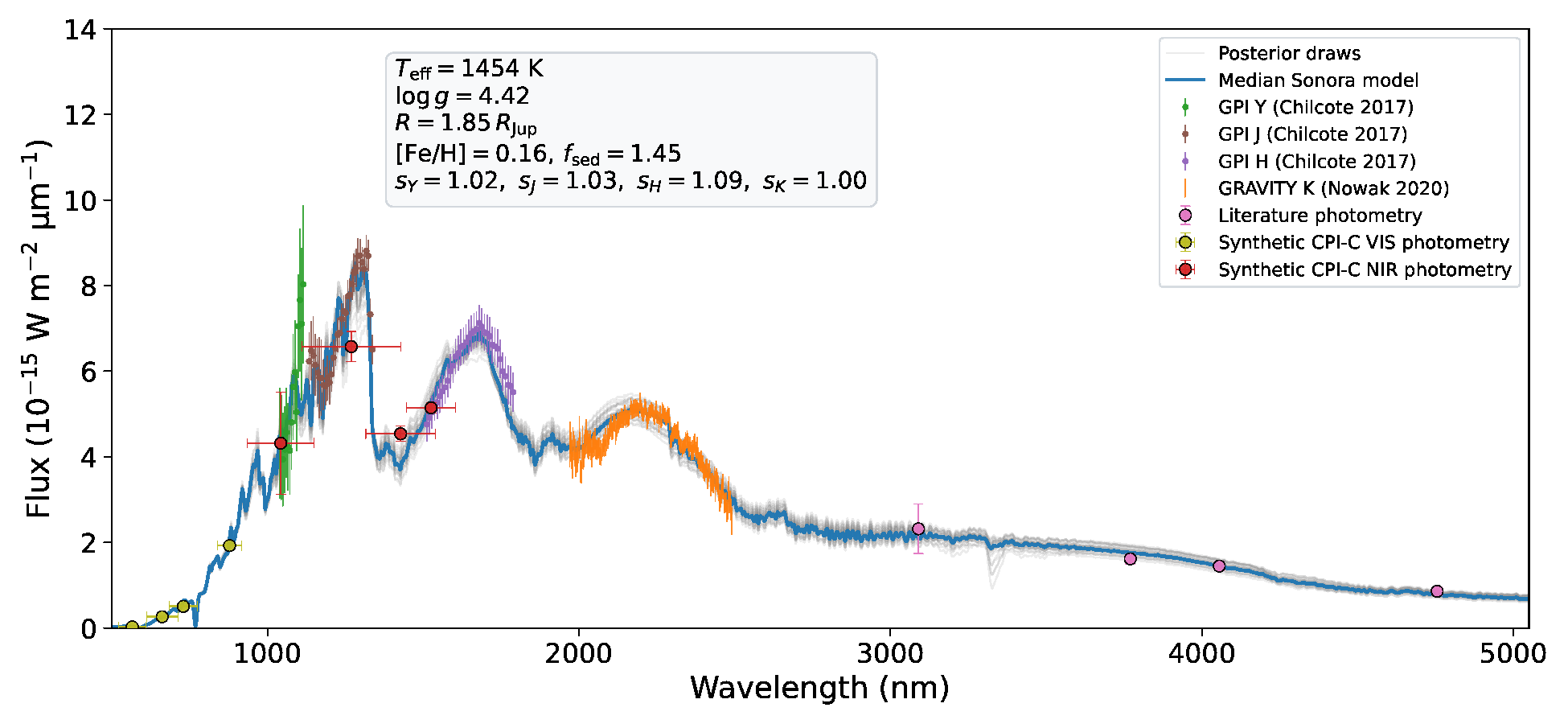}
\caption{Benchmark \texttt{Sonora Diamondback} fit to $\beta$ Pic b. The fit combines the GPI Y/J/H spectroscopy, GRAVITY K-band spectroscopy, broadband literature photometry, and synthetic CPI-C VIS4+NIR4 photometry. The blue curve is evaluated at the marginal posterior medians, and the gray curves show posterior draws. The yellow and red points mark the synthetic CPI-C VIS4 and NIR4 measurements.}
\label{fig:betapicb_benchmark}
\end{figure*}

Figure~\ref{fig:betapicb_benchmark} shows the resulting fit. The \texttt{Sonora} model reproduces the broad 0.5--5~$\mu$m spectral energy distribution traced by the GPI spectrum, the GRAVITY K-band spectrum, broadband photometry, and the synthetic CPI-C VIS4+NIR4 points. The synthetic CPI-C points extend the benchmark fit from the optical thermal tail to the short-wavelength near-infrared region. The marginal posterior medians and 16th--84th percentile credible intervals are $T_{\rm eff}=1454^{+34}_{-35}$~K, $\log g=4.42^{+0.16}_{-0.13}$, $R_{\rm p}=1.85^{+0.10}_{-0.10}\,R_{\rm J}$, [Fe/H]$=0.16^{+0.20}_{-0.20}$, and $f_{\rm sed}=1.45^{+0.52}_{-0.31}$. These values summarize the marginalized posterior distributions and are not the parameters of a single best-fitting grid spectrum.

The retrieved $T_{\rm eff}$, $\log g$, and $R_{\rm p}$ fall within the range of published estimates for $\beta$~Pic~b. Using \texttt{Exo-REM}, \citet{2015A&A...582A..83B} obtained $T_{\rm eff}=1550\pm150$~K, $\log g=3.5\pm1.0$, and $R_{\rm p}=1.76\pm0.24\,R_{\rm J}$, with the quoted intervals corresponding to $2\sigma$. Using the \texttt{Sonora} grid with GRAVITY, GPI, and literature photometry, \citet{2025A&A...704A.325R} obtained $T_{\rm eff}=1521.11^{+6.61}_{-6.36}$~K and $\log g=4.24\pm0.04$, which are consistent with our marginalized constraints within approximately $2\sigma$. From the bolometric luminosity and evolutionary models, \citet{2017AJ....153..182C} inferred a higher temperature and smaller radius, $T_{\rm eff}=1724\pm15$~K and $R_{\rm p}=1.46\pm0.01\,R_{\rm J}$. These differences reflect the sensitivity of the inferred parameters to the adopted atmosphere grid, wavelength coverage, data weighting, priors, and cloud treatment. In this study, the fit serves as a \texttt{Sonora}-based benchmark consistency test rather than a definitive determination of the atmospheric properties of $\beta$~Pic~b.

The physical visibility of the synthetic VIS4 points is strongly wavelength dependent. The published MagAO/VisAO detection of $\beta$ Pic b at $Y_S=0.985~\mu{\rm m}$ gives a contrast of $(1.63\pm0.49)\times10^{-5}$ at a projected separation of $0.470\pm0.010^{\prime\prime}$, providing an empirical anchor for the short-wavelength optical–NIR thermal tail of this planet \citep{2014ApJ...786...32M}. Our simulated F877 contrast, $6.10\times10^{-6}$, lies below the $Y_S$-band value and follows the expected decline of the thermal tail toward shorter wavelengths. The bluer VIS4 bands decrease further to $9.63\times10^{-7}$, $3.91\times10^{-7}$, and $2.87\times10^{-8}$ in F729, F661, and F565, respectively, making their physical detectability more sensitive to the adopted atmospheric temperature, cloud opacity, and radius. The nearly edge-on $\beta$ Pic debris disk adds an additional astrophysical component in optical scattered light. HST/ACS and HST/STIS imaging shows a warped inner disk and measurable disk surface brightness across the projected region where $\beta$ Pic b orbits \citep{2006AJ....131.3109G,2015ApJ...800..136A}. This disk emission can raise the local optical background and introduce band-dependent contamination to broadband point-source photometry. We use the VIS4 points in this benchmark calculation as synthetic thermal-tail photometry from the \texttt{Sonora} model, with the reddest visible bands providing the most physically plausible optical detections and the bluer bands providing model-sensitive constraints.

The agreement between the simulated CPI-C points and the existing SED indicates that the CPI-C eight-band set provides a consistent thermal-emission extension for a realistic warm giant planet spectrum. $\beta$ Pic b already has rich ground-based and interferometric spectrophotometry, so its benchmark fit mainly tests consistency across wavelength coverage. The next subsection uses a controlled synthetic case with sparse pre-existing data to quantify the diagnostic role of the CPI-C NIR4 bands.

\subsection{Controlled information-content experiment for a synthetic warm giant}	\label{sub:synthetic_warm}

To quantify the information content of these bands in a simpler and more controlled setting, we construct a synthetic warm giant with sparse pre-existing photometry. This experiment is designed to test how the CPI-C NIR4 bands improve the recovery of $T_{\rm eff}$, $\log g$, and $R_{\rm p}$ when only a small number of broad NIR photometric points are otherwise available.

The input planet is generated with the Sonora Diamondback grid. We adopt $T_{\rm eff}=1500$~K, $\log g=4.8$, $R_{\rm p}=1.8\,R_{\rm J}$, [Fe/H]$=0.0$, and $f_{\rm sed}=2.0$. The system is placed at 20~pc around a bright A-type host star with $V=4.0$~mag. These parameters place the synthetic planet in the warm self-luminous regime and yield detectable CPI-C NIR photometry under the adopted observing assumptions.

We generate two sets of synthetic observations from the same input Sonora spectrum. The first set represents sparse pre-existing follow-up photometry. It consists of three direct photometric measurements at the H-, K-, and L-band central wavelengths, 1.65, 2.20, and 3.80~$\mu$m. These points are obtained by directly sampling the model flux at the corresponding wavelengths, without convolving with filter profiles. The resulting AB magnitudes are 14.37, 14.13, and 14.18~mag in the H, K, and L bands, respectively. To define a transparent and band-independent photometric precision for this controlled experiment, we adopt ${\rm S/N}=10$ for each H/K/L measurement. This corresponds to a magnitude uncertainty of $\sigma_m=0.1086$~mag in all three bands.

The second set consists of the CPI-C NIR4 measurements. These points are generated by integrating the input Sonora spectrum through the actual CPI-C F1040, F1265, F1425, and F1532 throughput curves and applying the NIR noise model described above. The resulting CPI-C measurements are $15.73 \pm 0.26$, $14.96 \pm 0.05$, $15.08 \pm 0.03$, and $14.80 \pm 0.02$~mag in F1040, F1265, F1425, and F1532, respectively.

We fit the synthetic photometry under three data combinations: H/K/L only, CPI-C NIR4 only, and H/K/L + CPI-C NIR4. Model spectra are interpolated from the \texttt{Sonora Diamondback} grid with the \texttt{species} framework \citep{2023ascl.soft07057S}. For each trial parameter set, the model spectrum is scaled using the trial planet radius and the fixed distance of 20~pc.
In all three cases, the free parameters are $T_{\rm eff}$, $\log g$, and $R_{\rm p}$. The distance, [Fe/H], and $f_{\rm sed}$ are fixed at 20 pc, 0.0, and 2.0, respectively. We use broad uniform priors of $900<T_{\rm eff}<2200$~K, $3.5<\log g<5.5$, and $0.5<R_{\rm p}/R_{\rm J}<2.5$. The posterior distributions are sampled with \texttt{UltraNest}, using a minimum of 300 live points and an evidence-tolerance criterion of $\Delta\ln Z=0.5$.

\begin{figure*}
\centering
\includegraphics[width=\linewidth]{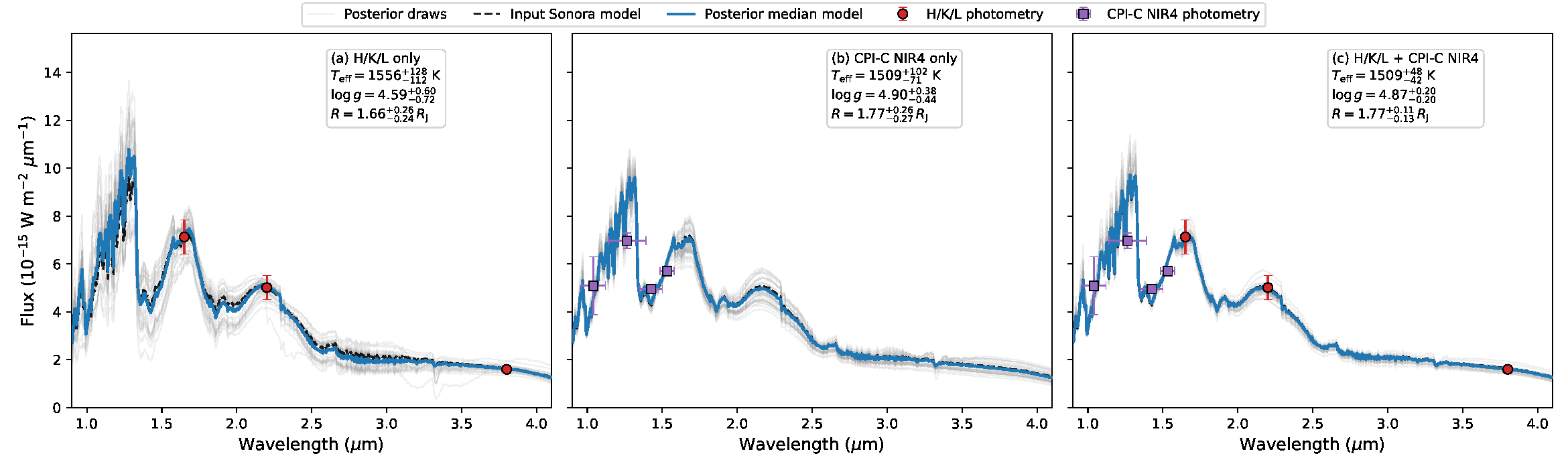}
\caption{Model fits to the synthetic warm giant for three photometric data combinations. The panels show fits to (a) H/K/L direct photometry only, (b) CPI-C NIR4 photometry only, and (c) the combined H/K/L + CPI-C NIR4 data set. The dashed black curve is the input model spectrum used to generate the synthetic data. Gray curves show posterior draws, and the blue curve shows the posterior median model. The H/K/L points are direct samples of the model flux at the adopted central wavelengths and are assigned ${\rm S/N}=10$ in each band. The CPI-C points are generated by integrating the model spectrum through the actual CPI-C NIR throughput curves and applying the adopted noise model.}
\label{fig:synthetic_sed_three_cases}
\end{figure*}

Figure~\ref{fig:synthetic_sed_three_cases} shows the fitted SEDs for the three data combinations. With H/K/L only, the overall flux level is constrained at longer NIR wavelengths, but the model remains weakly constrained between 1.0 and 1.6~$\mu$m. The CPI-C-only fit constrains the short-wavelength NIR spectral shape more directly. The combined fit uses both the H/K/L baseline and the CPI-C NIR4 shape information, producing a tighter set of posterior spectra around the input Sonora model.

\begin{figure*}
\centering
\includegraphics[width=\linewidth]{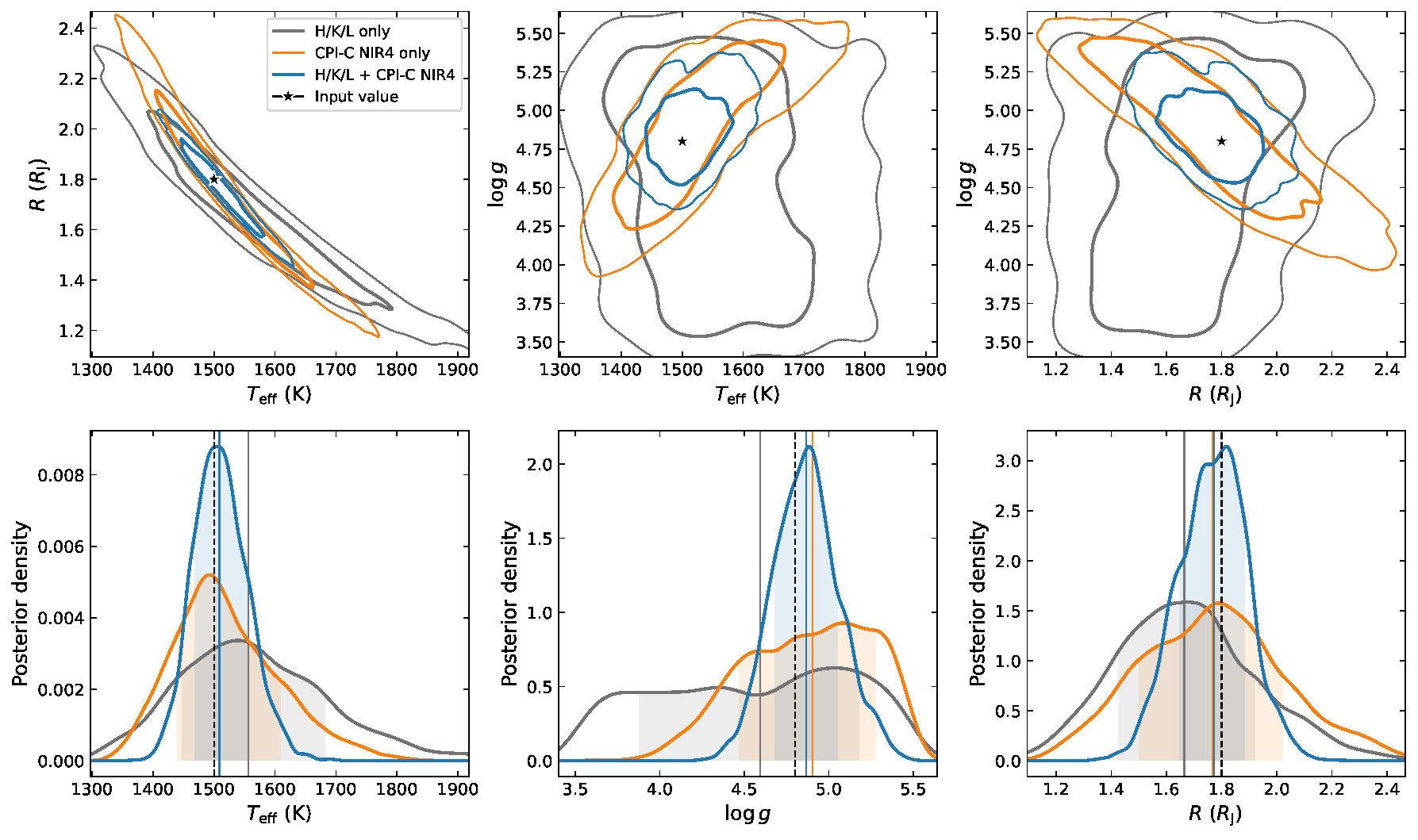}
\caption{Posterior comparison for the synthetic warm giant. The upper panels show Gaussian kernel-density estimates of the joint posterior distributions in the $T_{\rm eff}$--$R_{\rm p}$, $T_{\rm eff}$--$\log g$, and $R_{\rm p}$--$\log g$ planes. The contours enclose 68\% and 95\% of the posterior probability. The lower panels show the corresponding one-dimensional marginal distributions of $T_{\rm eff}$, $\log g$, and $R_{\rm p}$. Gray, orange, and blue denote the H/K/L-only, CPI-C NIR4-only, and combined fits, respectively. Colored vertical lines mark the posterior medians, and the shaded regions indicate the central 68\% credible intervals. The black stars and vertical dashed lines mark the input model parameters.}
\label{fig:synthetic_posterior_comparison}
\end{figure*}

The same behavior is seen in the posterior distributions. Figure~\ref{fig:synthetic_posterior_comparison} compares the three fits using kernel-density estimates of the joint and marginal posterior distributions. The upper panels show the $T_{\rm eff}$--$R_{\rm p}$, $T_{\rm eff}$--$\log g$, and $R_{\rm p}$--$\log g$ distributions, while the lower panels show the corresponding one-dimensional marginal distributions. The H/K/L-only case leaves a broad allowed region, mainly because the three points do not sample the 1.0--1.6~$\mu$m spectral structure. The CPI-C-only case provides stronger leverage on $T_{\rm eff}$ and $\log g$ through the short-wavelength NIR bands. The combined case gives the most compact joint and marginal posterior distributions among the three tests.

\begin{deluxetable*}{lccc}
\tablecaption{Posterior constraints for the synthetic warm giant}
\label{tab:synthetic_retrieval_results}
\tablehead{
\colhead{Case} &
\colhead{$T_{\rm eff}$ (K)} &
\colhead{$\log g$} &
\colhead{$R_{\rm p}$ ($R_{\rm J}$)}
}
\startdata
Input model &
1500 &
4.80 &
1.80 \\
H/K/L only &
$1556^{+128}_{-112}$ &
$4.59^{+0.60}_{-0.72}$ &
$1.66^{+0.26}_{-0.24}$ \\
CPI-C NIR4 only &
$1509^{+102}_{-71}$ &
$4.90^{+0.38}_{-0.44}$ &
$1.77^{+0.26}_{-0.27}$ \\
H/K/L + CPI-C NIR4 &
$1509^{+48}_{-42}$ &
$4.87^{+0.20}_{-0.20}$ &
$1.77^{+0.11}_{-0.13}$ \\
\enddata
\tablecomments{The fitted values are posterior medians with uncertainties defined by the 16th and 84th percentiles. The input-model values are listed for comparison.}
\end{deluxetable*}

Table~\ref{tab:synthetic_retrieval_results} summarizes the marginalized posterior constraints for the three data combinations. The H/K/L-only fit gives the broadest constraints because the three measurements provide limited information on the spectral shape between 1.0 and 1.6~$\mu$m. The CPI-C NIR4-only fit constrains the short-wavelength NIR shape more directly. The combined fit gives the narrowest marginal credible intervals for all three free parameters. The input values lie within the marginal 68\% credible intervals of the combined fit.

We quantify this improvement using the full width of the marginal $68\%$ credible interval, defined as $\Delta x_{68}=x_{84}-x_{16}$. Relative to the H/K/L-only fit, adding the CPI-C NIR4 points reduces the $68\%$ interval width of $T_{\rm eff}$ from approximately 240~K to 90~K, corresponding to a factor of 2.7 improvement, or a 62\% reduction. The $\log g$ interval decreases from 1.32~dex to 0.40~dex, a factor of 3.3 improvement, or a 70\% reduction. The radius interval decreases from 0.50~$R_{\rm J}$ to 0.25~$R_{\rm J}$, a factor of 2.0 improvement, or a 51\% reduction. As a simple proxy for the overall constraint volume, the product of the three marginal $68\%$ interval widths is reduced by approximately a factor of 18. This quantifies the visual trend in the posterior distributions: the CPI-C NIR4 bands provide additional leverage by sampling the 1.0--1.6~$\mu$m spectral shape that is absent from the H/K/L-only data set.

This controlled experiment shows that the CPI-C NIR4 filters are most useful when they complement sparse longer-wavelength NIR photometry. The H/K/L points set the broad thermal-emission level, while the CPI-C bands add short-wavelength shape information across 1.0--1.6~$\mu$m. Their combination reduces the main degeneracies among $T_{\rm eff}$, $\log g$, and $R_{\rm p}$ in this synthetic warm-giant case.

\subsection{Atmosphere-model thermal-emission estimates for HR 8799 planets}
\label{subsec:hr8799_optical_thermal}

We estimate the thermal-emission contribution of the HR 8799 planets across the full CPI-C optical--NIR wavelength range using atmosphere-model spectra. HR 8799 b/c/d/e are young, self-luminous giant planets with strong near-infrared thermal emission, and they provide a useful benchmark for evaluating how the CPI-C bands sample the short-wavelength tail and the near-infrared rise of planetary thermal spectra. For the HR~8799 system, the planets orbit at large separations (tens of au), so the reflected-light contribution scales as $(R_p/a)^2$ and is expected to be far below the thermal component at these wavelengths; we therefore neglect reflected light in the following.

For each planet, we adopt the single-best \texttt{Exo-REM} model parameters from recent atmospheric characterization of the HR 8799 system \citep{2024A&A...687A.298N}. The adopted parameters are $T_{\rm eff}=850$ K, $\log g=3.5$, $[{\rm M/H}]=+0.5$, C/O = 0.55, and $R_p=1.05\,R_{\rm J}$ for HR 8799 b; $T_{\rm eff}=1100$ K, $\log g=3.5$, $[{\rm M/H}]=+1.0$, C/O = 0.80, and $R_p=1.22\,R_{\rm J}$ for HR 8799 c; $T_{\rm eff}=1200$ K, $\log g=3.0$, $[{\rm M/H}]=+1.0$, C/O = 0.55, and $R_p=1.12\,R_{\rm J}$ for HR 8799 d; and $T_{\rm eff}=1100$ K, $\log g=3.5$, $[{\rm M/H}]=+1.0$, C/O = 0.80, and $R_p=1.15\,R_{\rm J}$ for HR 8799 e. The Exo-REM surface-flux spectra are scaled by $(R_p/d)^2$ and integrated through the CPI-C VIS4 and NIR4 throughput curves to obtain band-integrated AB magnitudes and planet--star contrasts.

\begin{figure*}
\centering
\includegraphics[width=\textwidth]{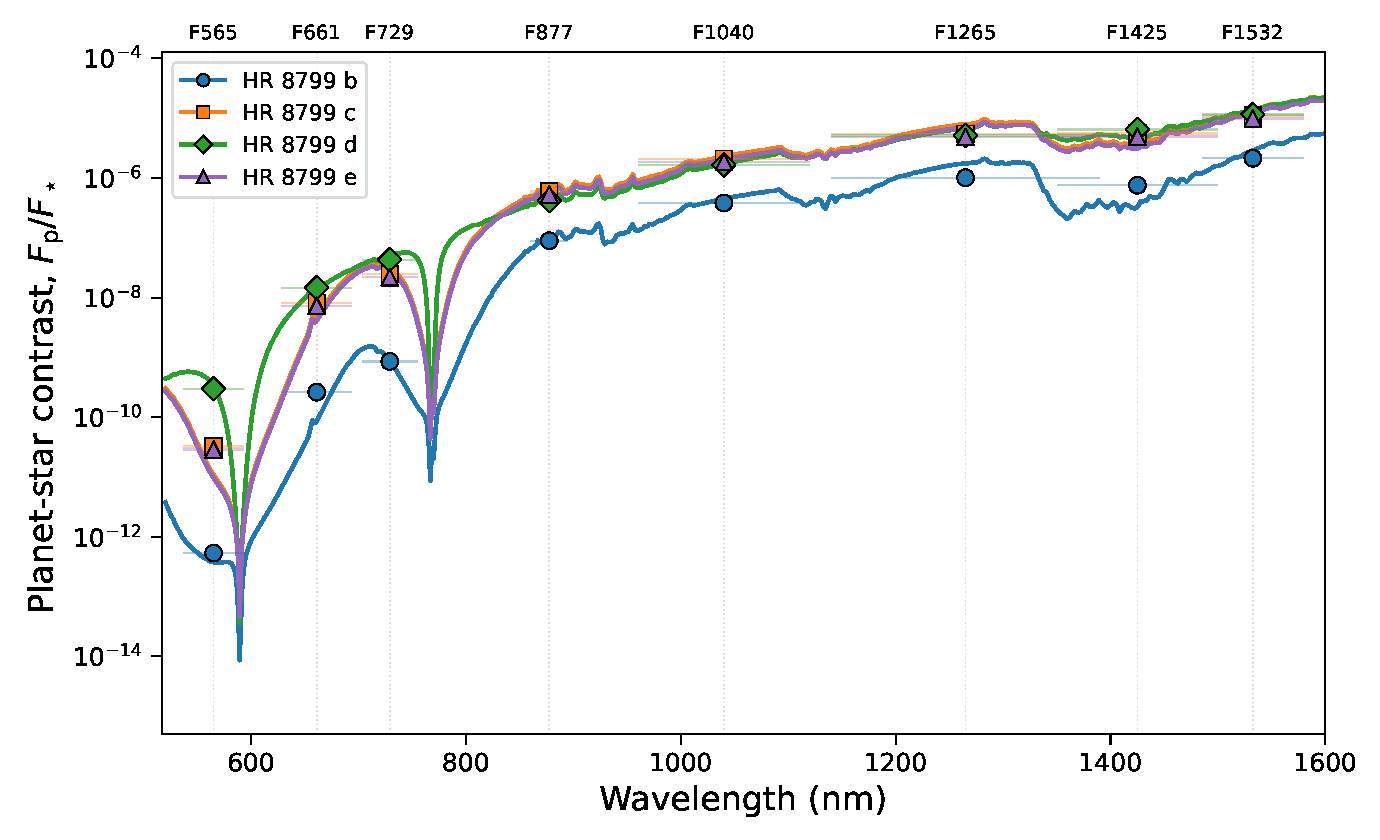}
\caption{Thermal-emission planet--star contrasts predicted for HR~8799 b, c, d, and e from the adopted \texttt{Exo-REM} models. The solid curves show the monochromatic model contrast spectra, and the symbols show the band-integrated contrasts listed in Table~\ref{tab:hr8799_exorem_cpic8}. The dotted vertical lines and upper labels mark the effective wavelengths of the eight CPI-C filters. Reflected light is not included.}
\label{fig:hr8799_cpic_contrasts}
\end{figure*}

Table~\ref{tab:hr8799_exorem_cpic8} lists the band-integrated AB magnitudes and planet--star contrasts, and Figure~\ref{fig:hr8799_cpic_contrasts} shows the corresponding monochromatic model contrast spectra and CPI-C photometric points. Together, the table and figure provide the numerical values and their wavelength dependence across the VIS4 and NIR4 bands.

\begin{deluxetable*}{l| cc| cc| cc| cc}
\tabletypesize{\scriptsize}
\tablecaption{Exo-REM thermal-emission synthetic photometry for HR~8799 b/c/d/e in the eight CPI-C bands.}
\label{tab:hr8799_exorem_cpic8}
\tablewidth{0pt}
\tablehead{
\multicolumn{1}{c|}{Band} &
\multicolumn{2}{c|}{Planet b} &
\multicolumn{2}{c|}{Planet c} &
\multicolumn{2}{c|}{Planet d} &
\multicolumn{2}{c}{Planet e} \\
\multicolumn{1}{c|}{} &
\multicolumn{1}{c}{$m_{b,\mathrm{AB}}$} & \multicolumn{1}{c|}{$C_b$} &
\multicolumn{1}{c}{$m_{c,\mathrm{AB}}$} & \multicolumn{1}{c|}{$C_c$} &
\multicolumn{1}{c}{$m_{d,\mathrm{AB}}$} & \multicolumn{1}{c|}{$C_d$} &
\multicolumn{1}{c}{$m_{e,\mathrm{AB}}$} & \multicolumn{1}{c}{$C_e$}
}
\startdata
F565  & 36.65 & $5.36\times10^{-13}$ & 32.23 & $3.24\times10^{-11}$ & 29.80 & $2.99\times10^{-10}$ & 32.36 & $2.88\times10^{-11}$ \\
F661  & 29.97 & $2.63\times10^{-10}$ & 26.24 & $8.20\times10^{-9}$  & 25.62 & $1.47\times10^{-8}$  & 26.37 & $7.29\times10^{-9}$  \\
F729  & 28.80 & $8.56\times10^{-10}$ & 25.15 & $2.47\times10^{-8}$  & 24.53 & $4.31\times10^{-8}$  & 25.28 & $2.19\times10^{-8}$  \\
F877  & 23.89 & $8.95\times10^{-8}$  & 21.84 & $5.97\times10^{-7}$  & 22.23 & $4.17\times10^{-7}$  & 21.96 & $5.30\times10^{-7}$  \\
F1040 & 22.40 & $3.81\times10^{-7}$  & 20.55 & $2.10\times10^{-6}$  & 20.80 & $1.66\times10^{-6}$  & 20.67 & $1.86\times10^{-6}$  \\
F1265 & 21.58 & $1.01\times10^{-6}$  & 19.74 & $5.47\times10^{-6}$  & 19.80 & $5.14\times10^{-6}$  & 19.87 & $4.86\times10^{-6}$  \\
F1425 & 22.03 & $7.55\times10^{-7}$  & 19.87 & $5.57\times10^{-6}$  & 19.71 & $6.49\times10^{-6}$  & 20.00 & $4.95\times10^{-6}$  \\
F1532 & 21.00 & $2.14\times10^{-6}$  & 19.21 & $1.11\times10^{-5}$  & 19.17 & $1.16\times10^{-5}$  & 19.34 & $9.90\times10^{-6}$  \\
\enddata
\tablecomments{
The values are band-integrated AB magnitudes and planet--star contrasts obtained by convolving \texttt{Exo-REM} thermal-emission spectra with the CPI-C throughput curves. The contrast is defined as $C=F_p/F_\star$. Reflected starlight is omitted.
}
\end{deluxetable*}

The thermal-emission contrasts rise by several orders of magnitude from the visible Wien tail toward the NIR. F565--F729 sample the faint short-wavelength tail, particularly for the cooler HR~8799 b model, whereas F877 marks the beginning of the rapid increase in thermal emission. The NIR4 bands trace the brighter part of the thermal spectral energy distribution, and F1532 gives the largest band-integrated contrast for all four planets. Among the VIS4 filters, F877 has the largest predicted thermal-emission contrast in these models.

\section{Joint Reflected-light and Thermal-emission Analysis}	\label{sec:joint__analysis}

\subsection{Synthetic Giant in a Mixed Reflected-light and Thermal-emission Regime}
\label{subsec:joint_synthetic_target}

The analyses in Sections~\ref{sec:optical_param} and~\ref{sec:near_infrared_regime} treated the optical and near-infrared regimes separately. The optical bands were used to diagnose reflected light from cool giant planets, while the near-infrared bands were used to characterize thermal emission from warmer self-luminous companions. In this section, we consider an intermediate case in which both components contribute to the same planet spectrum. Such a target must satisfy a narrow set of conditions. The planet must be close enough to the host star for reflected starlight to be measurable, and it must retain enough internal heat to produce detectable near-infrared thermal emission. This overlap is most likely for young or moderately young giant planets at small-to-moderate orbital separations \citep{1999ApJ...513..879M,2020ApJ...892..151L}.

We construct a synthetic giant planet in this mixed regime to test the CPI-C VIS4+NIR4 filters. The purpose of this example is to provide a controlled same-target experiment. We do not model a known planet. Instead, we choose a favorable but physically motivated set of parameters that places the planet in the regime where reflected light and thermal emission can both be present in the optical--NIR spectral energy distribution. 

The reference system adopts the same F-type stellar properties as the optical simulation setup used in Section~\ref{sec:optical_param}, providing a consistent stellar baseline for the present calculation. The stellar parameters are $T_\star=6441$~K, $R_\star=1.3223\,R_\odot$, $M_\star=1.262\,M_\odot$, and solar metallicity. We assume a system distance of 3~pc to represent a favorable nearby target and a planetary orbital distance of 1.6~au to maintain an appreciable reflected-light contribution while keeping stellar irradiation subdominant to the planet's internal heat. At this orbital distance, the resulting equilibrium temperature is about 282~K, so the near-infrared thermal flux in this model is determined mainly by the planet's internal heat rather than by stellar irradiation. We assume a phase angle of $\alpha=0^\circ$ when computing the reflected-light component. This geometry corresponds to full phase, where the illuminated hemisphere is fully visible to the observer and the reflected-light signal reaches its maximum value. As discussed by \citet{2020ApJ...892..151L}, adopting $\alpha=0^\circ$ provides a convenient reference case for evaluating instrument performance and atmospheric retrieval capability because it removes uncertainties associated with orbital phase and yields the highest reflected-light contrast. The assumption is therefore intended as an optimistic but standardized phase assumption for the spectral calculation, rather than a prediction for a specific planetary system.

The planet parameters are chosen to represent a young, self-luminous giant that has not yet cooled to a mature Jupiter-like state. We adopt a mass of $5\,M_{\rm J}$, a radius of $1.3\,R_{\rm J}$, and an intrinsic temperature of $T_{\rm int}=1000$~K. These values give $\log g \simeq 3.86$ in cgs units. Similar mass, radius, temperature, and gravity values occur in young giant-planet atmosphere calculations in which residual formation heat and reflected starlight can both affect the optical spectrum \citep{2020ApJ...892..151L}. 

We compute the optical--NIR planet spectrum with \texttt{PICASO} \citep{2019ApJ...878...70B,2023ApJ...942...71M}. The calculation uses a single atmosphere structure and a single set of stellar, orbital, and planetary parameters. The pressure grid spans $10^{-6}$--$10^2$ bar with 91 layers. The pressure--temperature profile is obtained with the irradiated radiative--convective calculation and then used to compute the emergent spectrum. Cloud opacity is calculated with \texttt{Virga} \citep{2026AJ....171...98B} using H$_2$O, MnS, Mg$_2$SiO$_4$, and Al$_2$O$_3$ condensates. We adopt $f_{\rm sed}=1.0$ and $K_{zz}=10^9~{\rm cm^2~s^{-1}}$ as the cloudy baseline.

The model output covers 0.35--1.80~$\mu$m and is saved as total, thermal, and reflected components. The total planet spectrum is used to generate the synthetic CPI-C photometry. The thermal and reflected components are retained to show which physical source dominates each filter. This decomposition is useful for interpreting the VIS4 and NIR4 bands, but the simulated measurements are derived from the same total spectrum. The VIS4 filters mainly sample the reflected component and its wavelength dependence, while the NIR4 filters sample the thermal continuum and broad molecular-band structure. The reddest optical band and the bluest near-infrared band provide the closest sampling of the transition between the two components.

\subsection{Simulated CPI-C VIS4+NIR4 photometry}
\label{subsec:joint_photometry}

We generate synthetic CPI-C photometry by integrating the total model contrast spectrum over the four visible bands and four near-infrared bands. The visible-channel calculation follows the same prescription used in Section~\ref{sec:optical_param}, and the near-infrared calculation follows the same prescription used in Section~\ref{sec:near_infrared_regime}. The resulting sensitivities should be regarded as representative performance estimates. The reflected and thermal components are integrated over the same bandpasses for interpretation, while the simulated measurements are based on the total planet flux. This gives a direct comparison between the two physical components in each CPI-C band.

\begin{figure*}
\centering
\includegraphics[width=\textwidth]
{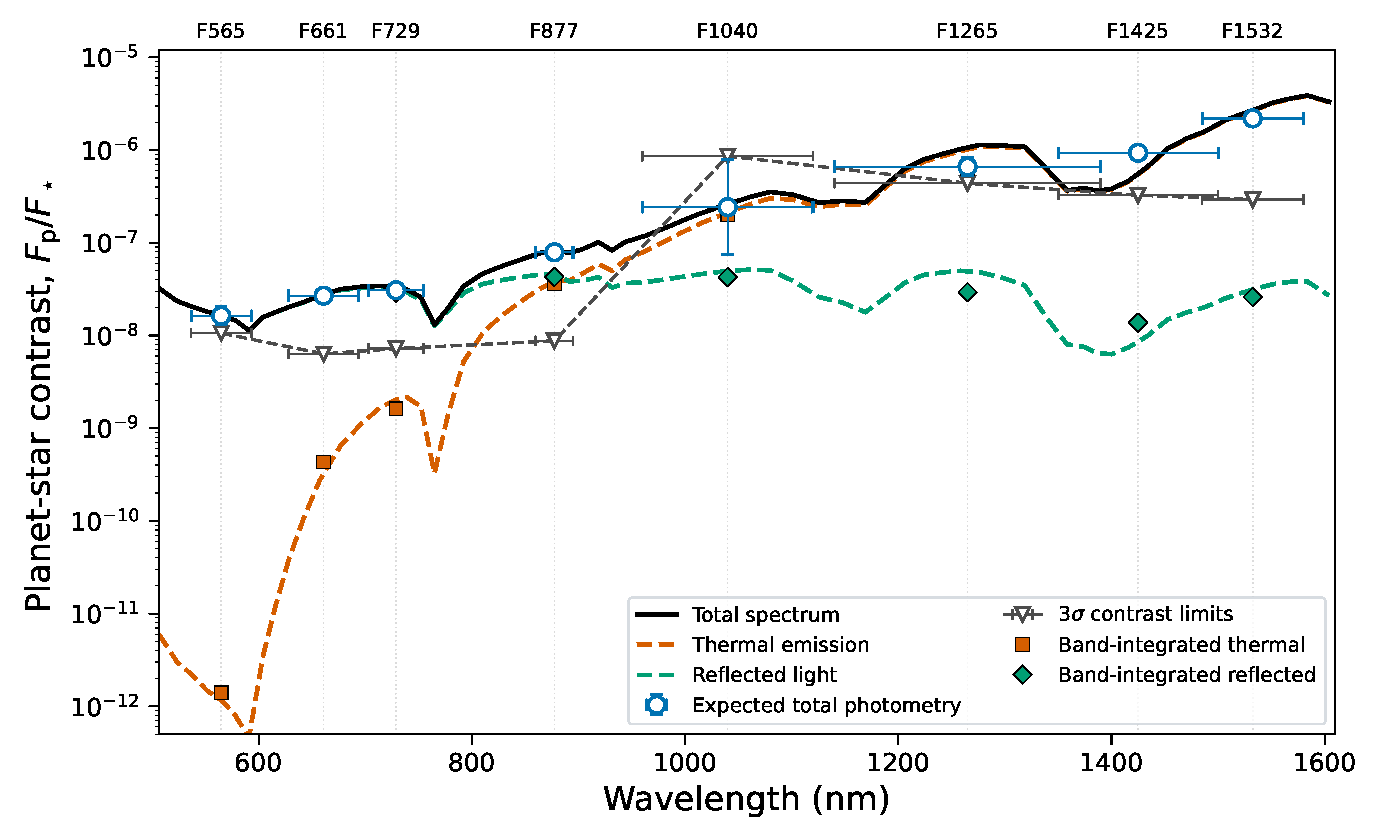}
\caption{Spectral decomposition and simulated CPI-C photometry for the synthetic giant planet. The black solid curve shows the total planet--star contrast spectrum, and the orange and green dashed curves show the thermal-emission and reflected-light components, respectively. Open blue circles show the expected band-integrated total contrasts. Their vertical error bars represent the propagated photometric uncertainties, and the horizontal bars show the nominal filter widths centered on the adopted band wavelengths. Orange squares and green diamonds show the band-integrated thermal and reflected components. Gray inverted triangles connected by a dashed line mark the per-band $3\sigma$ sensitivity thresholds. The NIR4 thresholds are first determined as apparent AB-magnitude limits from the adopted noise model and are then expressed in contrast units using the band-integrated PICASO stellar spectrum.}
\label{fig:joint_component_photometry}
\end{figure*}

Table~\ref{tab:joint_photometry} lists the resulting band-integrated contrasts and apparent AB photometry, and Figure~\ref{fig:joint_component_photometry} shows the same measurements together with the continuous spectral decomposition and the per-band sensitivity thresholds. The optical contrasts are between $1.6\times10^{-8}$ and $7.9\times10^{-8}$. The reflected component dominates F565, F661, and F729, with reflected-light fractions of 99.99\%, 98.4\%, and 94.7\%, respectively. F877 is the first strongly mixed band in this setup. Its total contrast is $7.90\times10^{-8}$, with 54.5\% from reflected light and 45.5\% from thermal emission.

The near-infrared bands trace the thermal component more directly. F1040 has a model contrast of $2.43\times10^{-7}$, but it is below the adopted detection threshold in this simulation and is treated as a $3\sigma$ upper limit. The longer-wavelength NIR bands are detected. F1265, F1425, and F1532 have signal-to-noise ratios of 4.5, 8.6, and 22.5, respectively. Their fluxes are thermal-emission dominated, with reflected-light fractions of only 4.4\%, 1.5\%, and 1.2\%.

These results show that the eight CPI-C bands sample three useful wavelength regimes for this synthetic planet. F565, F661, and F729 are dominated by reflected light. F877 samples a mixed regime where reflected light and thermal emission have comparable contributions. The NIR4 bands are thermal-emission dominated, although F1040 is below the adopted detection threshold in this simulation. The sharp increase in contrast from F729 to F877 marks the onset of the thermal continuum in the red optical and provides the main link between the visible and near-infrared measurements used in the retrieval tests below.

\begin{deluxetable*}{lcccc}
\tablecaption{Simulated CPI-C VIS4+NIR4 photometry for the synthetic giant planet.
\label{tab:joint_photometry}}
\tablehead{
\colhead{Band} &
\colhead{$C_{\rm total}$} &
\colhead{$C_{\rm therm}$} &
\colhead{$C_{\rm refl}$} &
\colhead{$m_{\rm AB}$}
\\
\colhead{} &
\colhead{} &
\colhead{} &
\colhead{} &
\colhead{(mag)}
}
\startdata
F565  & $1.633\times10^{-8}$ & $1.397\times10^{-12}$ & $1.633\times10^{-8}$ & $20.56\pm0.24$ \\
F661  & $2.687\times10^{-8}$ & $4.332\times10^{-10}$ & $2.644\times10^{-8}$ & $19.88\pm0.09$ \\
F729  & $3.072\times10^{-8}$ & $1.625\times10^{-9}$  & $2.909\times10^{-8}$ & $19.71\pm0.09$ \\
F877  & $7.903\times10^{-8}$ & $3.598\times10^{-8}$  & $4.305\times10^{-8}$ & $18.68\pm0.05$ \\
F1040 & $2.428\times10^{-7}$ & $2.002\times10^{-7}$  & $4.263\times10^{-8}$ & $>16.16$ \\
F1265 & $6.618\times10^{-7}$ & $6.325\times10^{-7}$  & $2.925\times10^{-8}$ & $16.60\pm0.24$ \\
F1425 & $9.357\times10^{-7}$ & $9.219\times10^{-7}$  & $1.379\times10^{-8}$ & $16.33\pm0.13$ \\
F1532 & $2.198\times10^{-6}$ & $2.172\times10^{-6}$  & $2.603\times10^{-8}$ & $15.48\pm0.05$ \\
\enddata
\tablecomments{
The total contrast is used as the synthetic measurement. The thermal and reflected contrasts are obtained by integrating the corresponding model components over the same filters and are used to diagnose the origin of the band-integrated signal.
The AB magnitudes refer to the total planet flux in each band. For the F1040 band, the planet signal is not significantly detected, and the reported value ($>16.16$) represents the $3\sigma$ upper limit.
}
\end{deluxetable*}

\subsection{Retrieval comparison}
\label{subsec:joint_retrieval}

We fit the synthetic photometry with the same forward-model framework used to generate the synthetic giant-planet spectrum. Three data sets are considered: VIS4-only, NIR4-only, and the combined VIS4+NIR4 set. The fitted parameters are the intrinsic temperature $T_{\rm int}$, planet radius $R_{\rm p}$, surface gravity $\log g$, cloud sedimentation parameter $f_{\rm sed}$, and cloud metallicity scaling $M_{\rm cloud}$. The planet mass is calculated as a derived parameter from $M_{\rm p}=gR_{\rm p}^2/G$. The stellar parameters, orbital distance, system distance, and phase angle are kept fixed to the values adopted in Section~\ref{subsec:joint_synthetic_target}. This comparison isolates the effect of wavelength coverage on the recovered parameters.

\begin{figure*}
\centering
\includegraphics[width=0.6\textwidth]{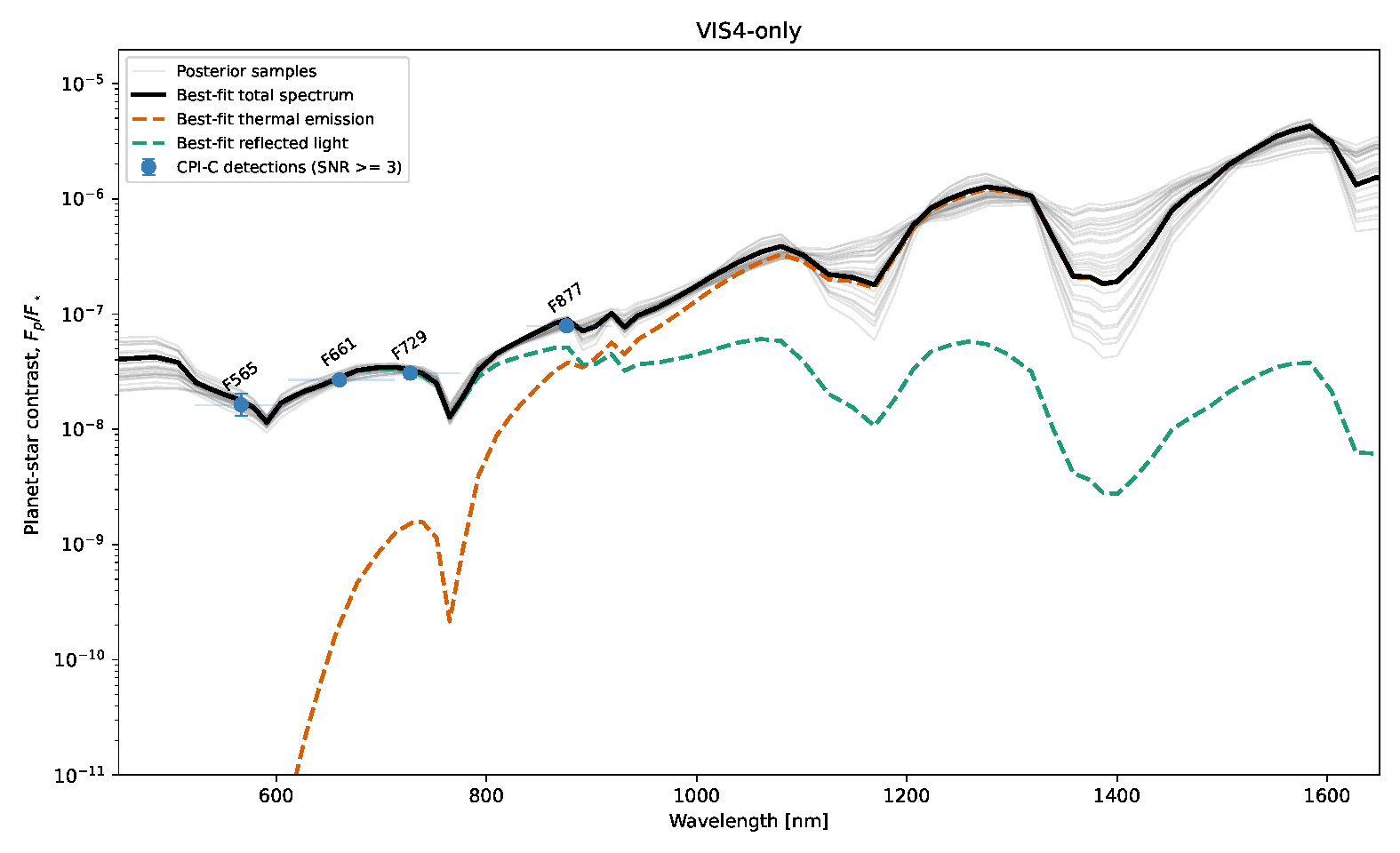}
\includegraphics[width=0.35\textwidth]{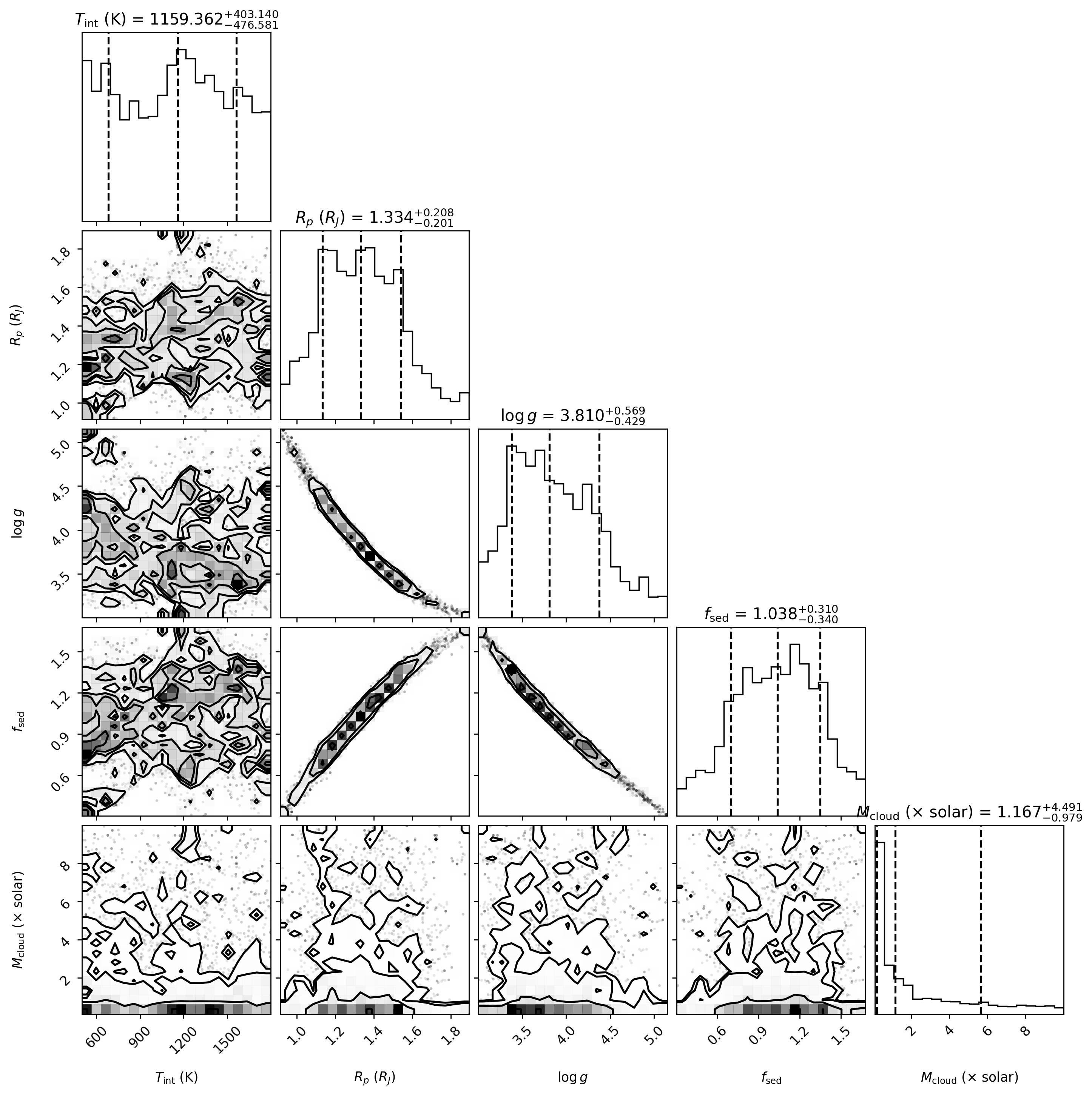}\\
\includegraphics[width=0.6\textwidth]{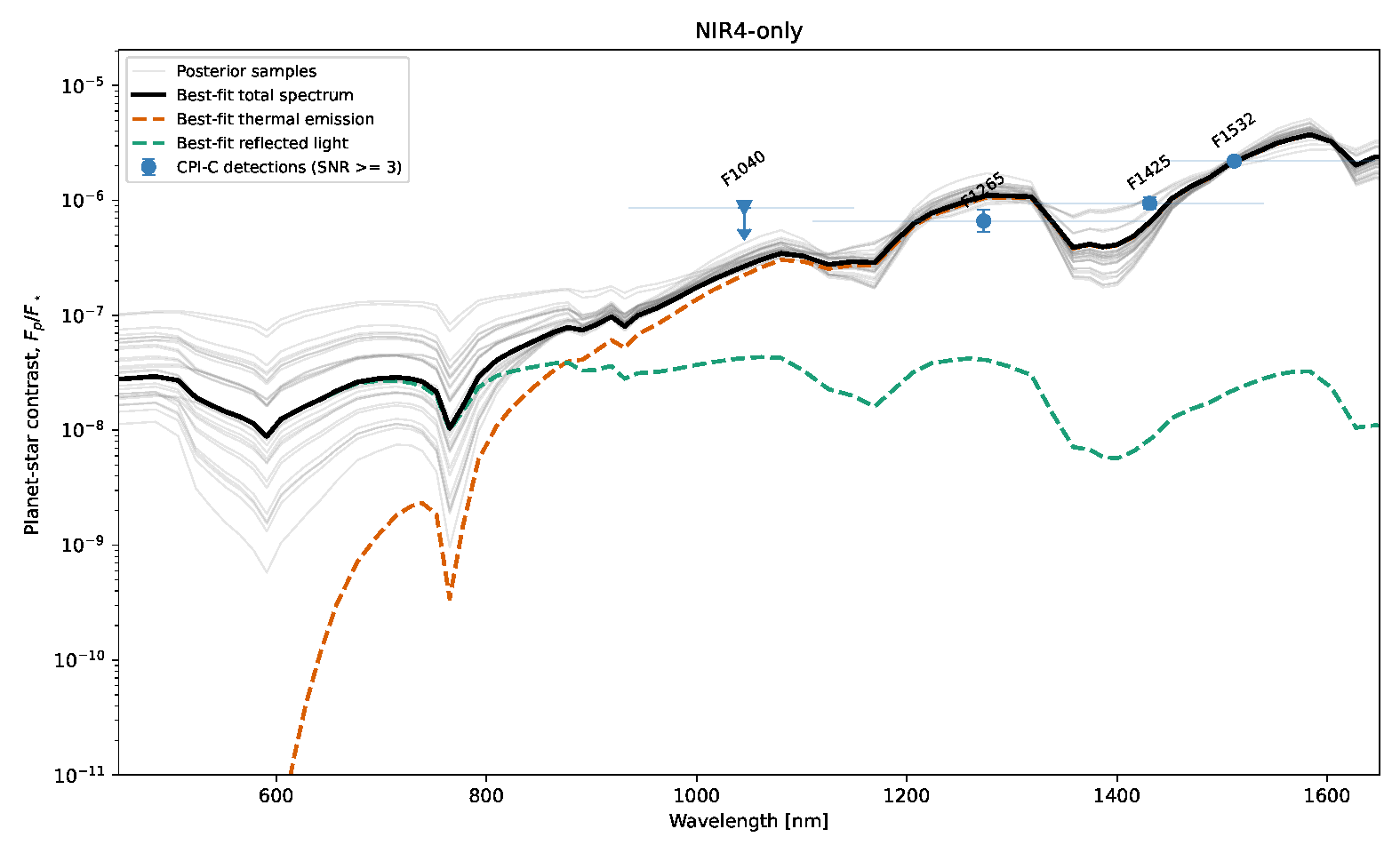}
\includegraphics[width=0.35\textwidth]{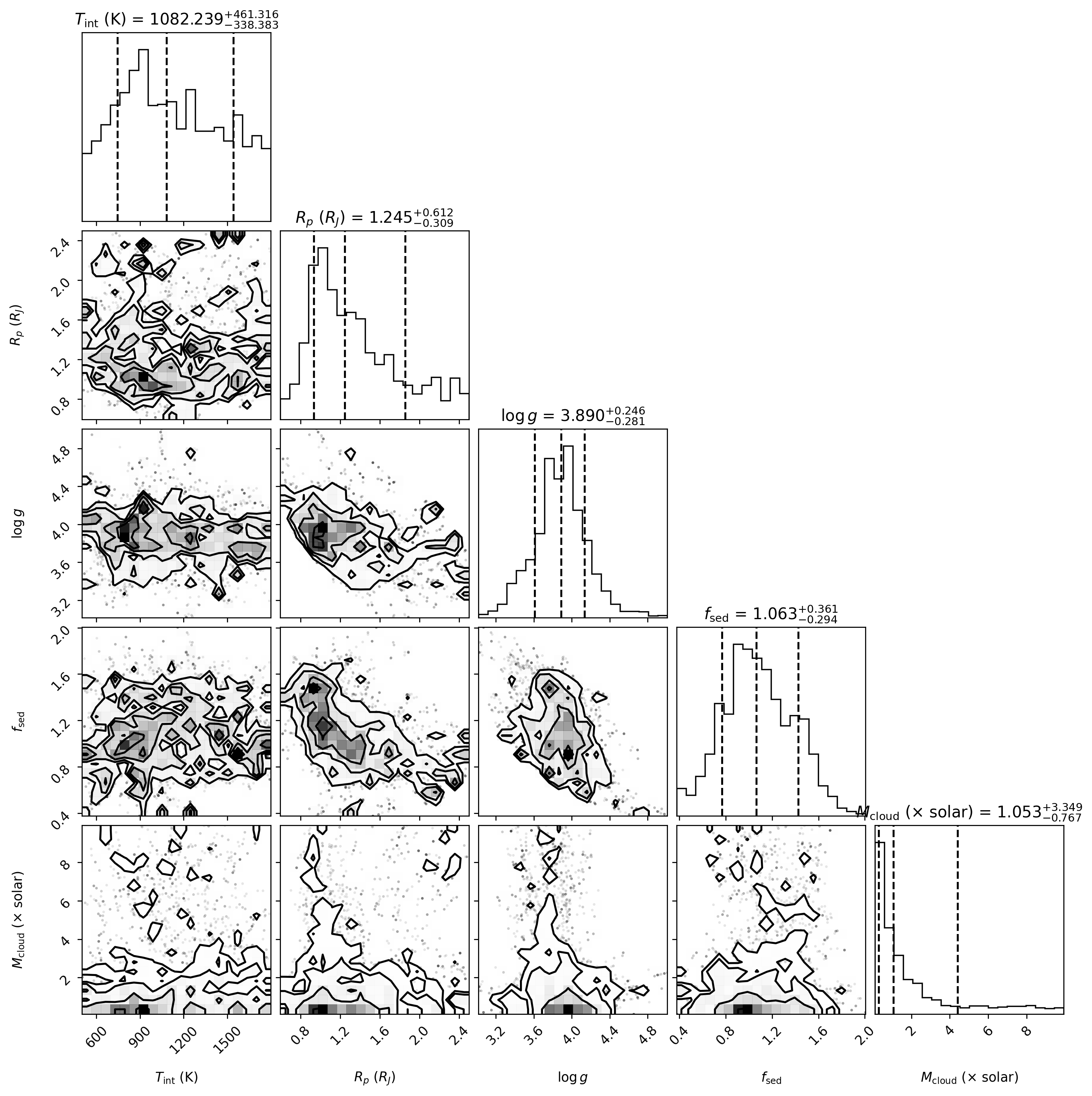}\\
\includegraphics[width=0.6\textwidth]{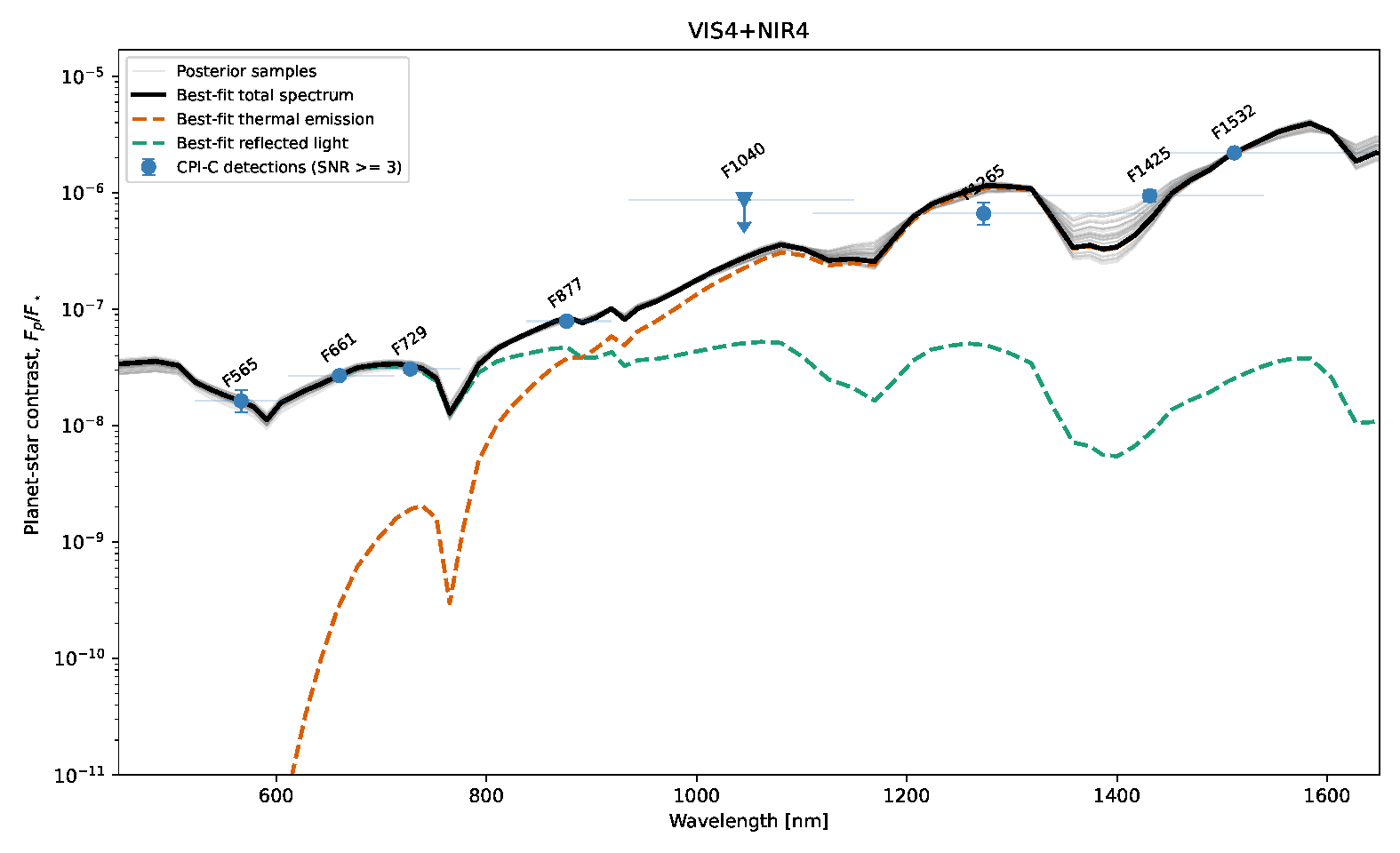}
\includegraphics[width=0.35\textwidth]{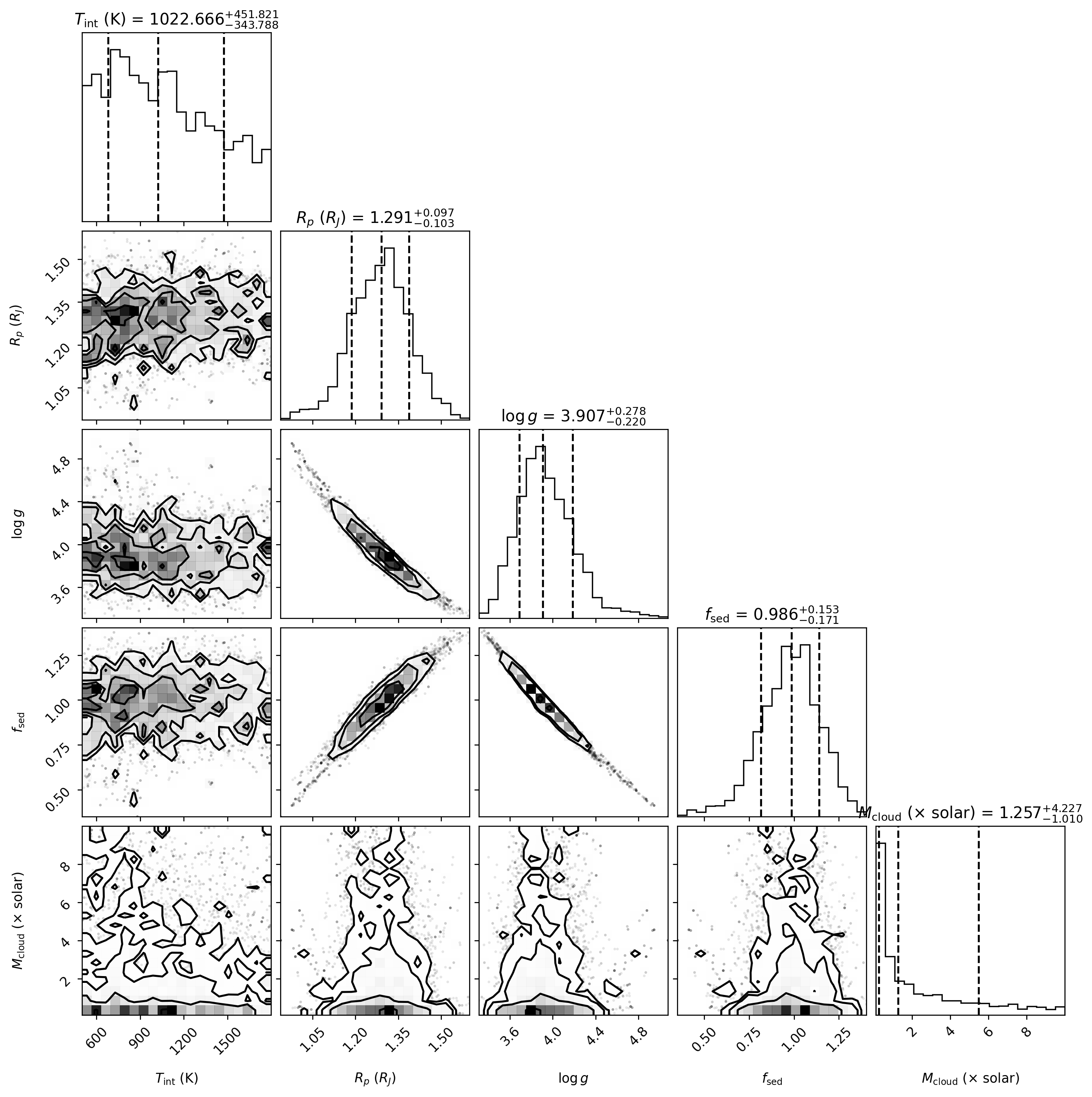}\\
\caption{Comparison of the VIS4-only, NIR4-only, and VIS4+NIR4 retrievals for the synthetic giant planet. Each row shows one data set. The left panels show the best-fit planet--star contrast spectra and posterior spectral samples. The black curves show the total model spectrum, the orange dashed curves show the thermal-emission component, and the green dashed curves show the reflected-light component. Blue points mark CPI-C detections with ${\rm S/N}\geq3$, and the F1040 measurement is shown as a $3\sigma$ upper limit when it is not significantly detected. The right panels show the corresponding posterior distributions for $T_{\rm int}$, $R_{\rm p}$, $\log g$, $f_{\rm sed}$, and $M_{\rm cloud}$.}
\label{fig:joint_retrieval_comparison}
\end{figure*}

Figure~\ref{fig:joint_retrieval_comparison} compares the spectral fits and posterior distributions for the three data sets. The VIS4-only fit is driven by the four optical detections. It recovers $R_{\rm p}=1.334^{+0.208}_{-0.201}\,R_{\rm J}$ and $f_{\rm sed}=1.038^{+0.310}_{-0.340}$, showing that the reflected-light amplitude and optical spectral shape provide useful constraints on the radius and cloud sedimentation parameter. The intrinsic temperature and surface gravity remain broad, with $T_{\rm int}=1159^{+403}_{-477}$~K and $\log g=3.810^{+0.569}_{-0.429}$. The posterior samples in the spectral panel show that VIS4-only data allow a wide range of near-infrared thermal spectra.

The NIR4-only fit is controlled by the thermal-emission spectrum. It gives $T_{\rm int}=1082^{+461}_{-338}$~K and $\log g=3.890^{+0.246}_{-0.281}$. The radius posterior is broader, with $R_{\rm p}=1.245^{+0.612}_{-0.309}\,R_{\rm J}$. This reflects the limited number of near-infrared constraints in this simulation: F1265, F1425, and F1532 are detected, while F1040 is treated as an upper limit. The NIR4 points trace the thermal continuum at longer wavelengths, but they provide limited leverage on the reflected-light level and on the radius--temperature scaling by themselves.

The combined VIS4+NIR4 fit gives the most stable constraints on the parameters shared by the reflected and thermal components. It recovers $T_{\rm int}=1023^{+452}_{-344}$~K, $R_{\rm p}=1.291^{+0.097}_{-0.103}\,R_{\rm J}$, $\log g=3.907^{+0.278}_{-0.220}$, and $f_{\rm sed}=0.986^{+0.153}_{-0.171}$. The derived mass is $M_{\rm p}=5.19^{+3.13}_{-1.63}\,M_{\rm J}$, consistent with the input value of $5\,M_{\rm J}$. The cloud metallicity scaling is only weakly constrained, with $M_{\rm cloud}=1.257^{+4.227}_{-1.010}$ times solar. The posterior allows a wide range of cloud metallicities, indicating that the eight-band photometry is sensitive to cloud-driven changes in the spectrum but does not precisely determine the cloud abundance scaling.

The gain from the joint fit is most evident for $R_{\rm p}$ and $f_{\rm sed}$. The 68\% credible interval of $R_{\rm p}$ decreases from 0.409~$R_{\rm J}$ in the VIS4-only fit and 0.921~$R_{\rm J}$ in the NIR4-only fit to 0.200~$R_{\rm J}$ in the VIS4+NIR4 fit. The interval for $f_{\rm sed}$ decreases from 0.650 and 0.655 to 0.324. The derived mass is also better constrained, mainly because the joint fit gives a tighter radius estimate and a more limited range of $\log g$. The constraint on $T_{\rm int}$ improves less, as the NIR4 bands sample only the short-wavelength side of the thermal-emission spectrum and F1040 is used as an upper limit.

The joint fit uses complementary information from the two wavelength ranges. The visible detections set the reflected-light level and constrain the cloud-dependent optical slope. The near-infrared detections constrain the thermal continuum at longer wavelengths. F877 samples the mixed regime where reflected light and thermal emission have comparable contributions, and the F1040 upper limit helps restrict the rise of the near-infrared thermal component. For this synthetic planet, the combined VIS4+NIR4 data give tighter and more consistent constraints on the same target than either filter set alone.

\section{Discussion and Conclusions\label{sec:dis}}

This work develops and evaluates an eight-band photometric design (VIS4+NIR4) for characterizing giant exoplanets across reflected-light and thermal-emission regimes. The central outcome of this study is a science-driven mapping between band placement and the atmospheric and physical information that can be extracted from sparse, multi-band photometry. Although we adopt filter curves based on the CPI-C design, the diagnostic logic relies primarily on the relative placement of bands with respect to broad molecular absorption features and continuum windows, and is therefore transferable to similar high-contrast, multi-band imaging platforms.

A key conclusion from the reflected-light experiments is that the information content of VIS4 is dominated by coarse spectral shape, rather than absolute flux normalization. In practice, the absolute photometric level is strongly coupled to geometry and scale factors (e.g., orbital phase and $(R_{\rm p}/r)^2$), so the most robust constraints arise from relative inter-band behavior: the depth and slope changes across the optical bands that respond to methane absorption, cloud/haze effects, and continuum curvature. A controlled methane-abundance sweep further shows that this spectral-shape leverage is expressed mainly through the relative suppression of F729 and F877, while F565 and F661 act primarily as local continuum anchors. The methane-sensitive colors trace a systematic sequence across the full abundance sweep, but the baseline measurements give a conservative pairwise separation of only $D=1.02$ between CH$_4\times0.1$ and CH$_4\times1$. VIS4 therefore provides useful sensitivity to broadly low and high methane regimes under favorable SNR conditions, but does not provide a precise methane-abundance measurement from a single baseline observation.

Orbital phase fundamentally impacts the retrievability of atmospheric properties, beyond simply altering the planet's overall brightness. Even for an identical intrinsic reflectance spectrum, changing phase angle can move a target between a regime where multiple bands yield detections and a regime where one-sided constraints dominate. Our simulations show that this transition directly alters which atmospheric properties remain identifiable: when several bands are measured, band-to-band ratios retain sensitivity to absorption depth and spectral curvature; when the redder points are replaced by upper limits, the diagnostic leverage on parameters that primarily affect those bands is substantially reduced. This highlights a practical implication for survey and follow-up design: multi-epoch measurements spanning different orbital phases can break degeneracies and improve overall constraints on planetary parameters without requiring that any single epoch be maximally informative.

Upper limits are retained in our likelihood framework as one-sided constraints rather than discarded as non-detections. They provide critical leverage for rejecting overly bright or spectrally inconsistent models. However, when multiple bands collapse to upper limits, the posterior inherently reflects the assumed model families more than the measured spectral structure, marking a fundamental boundary for sparse photometric inference.

By parameterizing residual speckle noise and post-processing efficacy using an explicit scenario factor, $f_{\rm pp}$, we can systematically isolate their effects on data identifiability. This factor controls the degree to which systematic residuals contribute to the uncertainty budget, and thus controls whether the inference is driven by photon-limited statistics or by a speckle floor. Importantly, the same band design operates in different information regimes depending on the residual noise level. This transition can be quantified by treating the post-processing efficacy as an explicit parameter.

For moderately warm and young self-luminous companions, the near-infrared bands extend the analysis beyond reflected-light diagnostics and provide constraints on fundamental parameters such as $T_{\rm eff}$, $\log g$, and $R_p$ from thermal emission. Using \texttt{Sonora Diamondback} thermal-emission models, the benchmark $\beta$~Pic~b fit verifies that the synthetic CPI-C NIR photometry is physically consistent with the observed spectral energy distribution of a well-studied warm giant planet. The controlled synthetic experiment, which compares H/K/L-only, CPI-C NIR4-only, and combined H/K/L+CPI-C NIR4 photometry, shows that the CPI-C NIR bands reduce degeneracies among effective temperature, surface gravity, and radius by adding short-wavelength SED-shape information across the 1.0--1.6~$\mu$m region. Given the retrieved $R_p$ and $\log g$, the planet mass can then be inferred as a derived quantity via $M_p = g R_p^2/G$, although tighter mass constraints require additional wavelength coverage or external priors. 

These results show that the same CPI-C band set can support both cool reflected-light targets and warmer self-luminous companions, with the useful diagnostic regime set by the target properties. The result also clarifies the role of the transition bands. F877 samples the mixed regime where reflected light and thermal emission have comparable contributions, and the F1040 upper limit restricts the onset of the near-infrared thermal component. Together, these bands connect the reflected-light and thermal-emission regimes in the CPI-C filter set.  

In summary, the eight-band design enables four main scientific capabilities. First, the optical bands provide compact reflected-light diagnostics that are most stable when interpreted through relative spectral shape rather than absolute normalization. Second, orbital phase and residual speckle noise strongly modulate the information content, shifting observations between detection- and upper-limit-dominated regimes with clear consequences for parameter identifiability. Third, the addition of near-infrared bands extends the framework to thermal-emission characterization and improves constraints on fundamental parameters for self-luminous companions. Fourth, for giant planets whose optical–NIR spectra contain measurable contributions from both reflected light and thermal emission, the combined VIS4+NIR4 data provide stronger same-target constraints than either filter set alone. These conclusions, derived from explicit noise and modeling scenarios, provide transferable guidance for science-driven band selection. CPI-C serves as a concrete implementation of this design philosophy for future high-contrast multi-band imaging of giant exoplanets.

\begin{acknowledgments}
We acknowledge the science research grants CMS-CSST-201906 and CMS-CSST-2025-A18 from the China Manned Space Project, as well as National Natural Science Foundation of China (NSFC) under grant nos U2031210 and 11827804. This research is also funded by the ``Jiangsu Funding Program for Excellent Postdoctoral Talent'' (grant 2024ZB178). We thank Prof Di-Chang Chen (Sun Yat-sen University) for helpful discussions on estimating the visible-band contrasts of the HR~8799 planets.
\end{acknowledgments}

%

\vspace{5mm}







\bibliography{bibtex}{}

@ARTICLE{2018AJ....156..158B,
       author = {{Batalha}, Natasha E. and {Smith}, Adam J.~R.~W. and {Lewis}, Nikole K. and {Marley}, Mark S. and {Fortney}, Jonathan J. and {Macintosh}, Bruce},
        title = "{Color Classification of Extrasolar Giant Planets: Prospects and Cautions}",
      journal = {AJ},
         year = 2018,
        month = oct,
       volume = {156},
       number = {4},
          eid = {158},
        pages = {158},
          doi = {10.3847/1538-3881/aad59d},
archivePrefix = {arXiv},
       eprint = {1807.08453},
 primaryClass = {astro-ph.EP},
       adsurl = {https://ui.adsabs.harvard.edu/abs/2018AJ....156..158B}
}

@ARTICLE{2010RAA....10..189D,
       author = {{Dou}, Jiangpei and {Ren}, Deqing and {Zhu}, Yongtian},
        title = "{High-contrast coronagraph for ground-based imaging of Jupiter-like planets}",
      journal = {Research in Astronomy and Astrophysics},
         year = 2010,
        month = feb,
       volume = {10},
       number = {2},
        pages = {189-198},
          doi = {10.1088/1674-4527/10/2/010},
archivePrefix = {arXiv},
       eprint = {0910.5355},
 primaryClass = {astro-ph.IM},
       adsurl = {https://ui.adsabs.harvard.edu/abs/2010RAA....10..189D}
}

@ARTICLE{2010PASP..122..590R,
       author = {{Ren}, Deqing and {Dou}, Jiangpei and {Zhu}, Yongtian},
        title = "{A Transmission-Filter Coronagraph: Design and Test}",
      journal = {PASP},
         year = 2010,
        month = may,
       volume = {122},
       number = {891},
        pages = {590},
          doi = {10.1086/652958},
archivePrefix = {arXiv},
       eprint = {1510.03796},
 primaryClass = {astro-ph.IM},
       adsurl = {https://ui.adsabs.harvard.edu/abs/2010PASP..122..590R}
}

@ARTICLE{2016ApJ...832...84D,
       author = {{Dou}, Jiangpei and {Ren}, Deqing},
        title = "{Phase Quantization Study of Spatial Light Modulator for Extreme High-contrast Imaging}",
      journal = {ApJ},
         year = 2016,
        month = nov,
       volume = {832},
       number = {1},
          eid = {84},
        pages = {84},
          doi = {10.3847/0004-637X/832/1/84},
archivePrefix = {arXiv},
       eprint = {1609.04870},
 primaryClass = {astro-ph.IM},
       adsurl = {https://ui.adsabs.harvard.edu/abs/2016ApJ...832...84D}
}

@ARTICLE{2006ApJS..167...81G,
       author = {{Guyon}, O. and {Pluzhnik}, E.~A. and {Kuchner}, M.~J. and {Collins}, B. and {Ridgway}, S.~T.},
        title = "{Theoretical Limits on Extrasolar Terrestrial Planet Detection with Coronagraphs}",
      journal = {ApJS},
         year = 2006,
        month = nov,
       volume = {167},
       number = {1},
        pages = {81-99},
          doi = {10.1086/507630},
archivePrefix = {arXiv},
       eprint = {astro-ph/0608506},
 primaryClass = {astro-ph},
       adsurl = {https://ui.adsabs.harvard.edu/abs/2006ApJS..167...81G}
}

@ARTICLE{2009A&A...495..363M,
       author = {{Martinez}, P. and {Dorrer}, C. and {Aller Carpentier}, E. and {Kasper}, M. and {Boccaletti}, A. and {Dohlen}, K. and {Yaitskova}, N.},
        title = "{Design, analysis, and testing of a microdot apodizer for the Apodized Pupil Lyot Coronagraph}",
      journal = {A\&A},
         year = 2009,
        month = feb,
       volume = {495},
       number = {1},
        pages = {363-370},
          doi = {10.1051/0004-6361:200810918},
archivePrefix = {arXiv},
       eprint = {0810.5678},
 primaryClass = {astro-ph},
       adsurl = {https://ui.adsabs.harvard.edu/abs/2009A&A...495..363M}
}

@ARTICLE{2007Natur.446..771T,
       author = {{Trauger}, John T. and {Traub}, Wesley A.},
        title = "{A laboratory demonstration of the capability to image an Earth-like extrasolar planet}",
      journal = {Nature},
         year = 2007,
        month = apr,
       volume = {446},
       number = {7137},
        pages = {771-773},
          doi = {10.1038/nature05729},
       adsurl = {https://ui.adsabs.harvard.edu/abs/2007Natur.446..771T}
}

@ARTICLE{2018JQSRT.217...86V,
       author = {{Villanueva}, G.~L. and {Smith}, M.~D. and {Protopapa}, S. and {Faggi}, S. and {Mandell}, A.~M.},
        title = "{Planetary Spectrum Generator: An accurate online radiative transfer suite for atmospheres, comets, small bodies and exoplanets}",
      journal = {J. Quant. Spec. Radiat. Transf.},
         year = 2018,
        month = sep,
       volume = {217},
        pages = {86-104},
          doi = {10.1016/j.jqsrt.2018.05.023},
archivePrefix = {arXiv},
       eprint = {1803.02008},
 primaryClass = {astro-ph.EP},
       adsurl = {https://ui.adsabs.harvard.edu/abs/2018JQSRT.217...86V}
}

@BOOK{2022fpsg.book.....V,
       author = {{Villanueva}, Geronimo Luis and {Liuzzi}, Giuliano and {Faggi}, Sara and {Protopapa}, Silvia and {Kofman}, Vincent and {Fauchez}, Thomas and {Stone}, Shane Wesley and {Mandell}, Avi Max},
        title = "{Fundamentals of the Planetary Spectrum Generator}",
         year = 2022,
       adsurl = {https://ui.adsabs.harvard.edu/abs/2022fpsg.book.....V}
}

@ARTICLE{2013PASP..125..306F,
       author = {{Foreman-Mackey}, Daniel and {Hogg}, David W. and {Lang}, Dustin and {Goodman}, Jonathan},
        title = "{emcee: The MCMC Hammer}",
      journal = {PASP},
         year = 2013,
        month = mar,
       volume = {125},
       number = {925},
        pages = {306},
          doi = {10.1086/670067},
archivePrefix = {arXiv},
       eprint = {1202.3665},
 primaryClass = {astro-ph.IM},
       adsurl = {https://ui.adsabs.harvard.edu/abs/2013PASP..125..306F}
}

@ARTICLE{2012ApJ...747...25M,
       author = {{Madhusudhan}, Nikku and {Burrows}, Adam},
        title = "{Analytic Models for Albedos, Phase Curves, and Polarization of Reflected Light from Exoplanets}",
      journal = {ApJ},
         year = 2012,
        month = mar,
       volume = {747},
       number = {1},
          eid = {25},
        pages = {25},
          doi = {10.1088/0004-637X/747/1/25},
archivePrefix = {arXiv},
       eprint = {1112.4476},
 primaryClass = {astro-ph.EP},
       adsurl = {https://ui.adsabs.harvard.edu/abs/2012ApJ...747...25M}
}

@ARTICLE{2010ApJ...724..189C,
       author = {{Cahoy}, Kerri L. and {Marley}, Mark S. and {Fortney}, Jonathan J.},
        title = "{Exoplanet Albedo Spectra and Colors as a Function of Planet Phase, Separation, and Metallicity}",
      journal = {ApJ},
         year = 2010,
        month = nov,
       volume = {724},
       number = {1},
        pages = {189-214},
          doi = {10.1088/0004-637X/724/1/189},
archivePrefix = {arXiv},
       eprint = {1009.3071},
 primaryClass = {astro-ph.EP},
       adsurl = {https://ui.adsabs.harvard.edu/abs/2010ApJ...724..189C}
}

@INPROCEEDINGS{2003SPIE.4796..164D,
  author    = {Denvir, D. J. and Conroy, E.},
  year      = {2003},
  booktitle = {Proc. SPIE},
  volume    = {4796},
  pages     = {164--174},
  doi       = {10.1117/12.457779}
}

@ARTICLE{2008Sci...322.1348M,
       author = {{Marois}, Christian and {Macintosh}, Bruce and {Barman}, Travis and {Zuckerman}, B. and {Song}, Inseok and {Patience}, Jennifer and {Lafreni{\`e}re}, David and {Doyon}, Ren{\'e}},
        title = "{Direct Imaging of Multiple Planets Orbiting the Star HR 8799}",
      journal = {Science},
         year = 2008,
        month = nov,
       volume = {322},
       number = {5906},
        pages = {1348},
          doi = {10.1126/science.1166585},
archivePrefix = {arXiv},
       eprint = {0811.2606},
 primaryClass = {astro-ph},
       adsurl = {https://ui.adsabs.harvard.edu/abs/2008Sci...322.1348M}
}

@ARTICLE{2003ITED...50.1227R,
       author = {{Robbins}, M.~S. and {Hadwen}, B.~J.},
        title = "{The noise performance of electron multiplying charge-coupled devices}",
      journal = {IEEE Transactions on Electron Devices},
         year = 2003,
        month = may,
       volume = {50},
       number = {5},
        pages = {1227-1232},
          doi = {10.1109/TED.2003.813462},
       adsurl = {https://ui.adsabs.harvard.edu/abs/2003ITED...50.1227R}
}

@ARTICLE{2011MNRAS.411..211T,
       author = {{Tulloch}, S.~M. and {Dhillon}, V.~S.},
        title = "{On the use of electron-multiplying CCDs for astronomical spectroscopy}",
      journal = {\mnras},
         year = 2011,
        month = feb,
       volume = {411},
       number = {1},
        pages = {211-225},
          doi = {10.1111/j.1365-2966.2010.17675.x},
archivePrefix = {arXiv},
       eprint = {1009.3403},
 primaryClass = {astro-ph.IM},
       adsurl = {https://ui.adsabs.harvard.edu/abs/2011MNRAS.411..211T}
}

@ARTICLE{2015ARA&A..53..279M,
       author = {{Marley}, M.~S. and {Robinson}, T.~D.},
        title = "{On the Cool Side: Modeling the Atmospheres of Brown Dwarfs and Giant Planets}",
      journal = {\araa},
         year = 2015,
        month = aug,
       volume = {53},
        pages = {279-323},
          doi = {10.1146/annurev-astro-082214-122522},
archivePrefix = {arXiv},
       eprint = {1410.6512},
 primaryClass = {astro-ph.EP},
       adsurl = {https://ui.adsabs.harvard.edu/abs/2015ARA&A..53..279M}
}

@ARTICLE{2013ApJ...776...15C,
       author = {{Currie}, Thayne and {Burrows}, Adam and {Madhusudhan}, Nikku and {Fukagawa}, Misato and {Girard}, Julien H. and {Dawson}, Rebekah and {Murray-Clay}, Ruth and {Kenyon}, Scott and {Kuchner}, Marc and {Matsumura}, Soko and {Jayawardhana}, Ray and {Chambers}, John and {Bromley}, Ben},
        title = "{A Combined Very Large Telescope and Gemini Study of the Atmosphere of the Directly Imaged Planet, {\ensuremath{\beta}} Pictoris b}",
      journal = {\apj},
         year = 2013,
        month = oct,
       volume = {776},
       number = {1},
          eid = {15},
        pages = {15},
          doi = {10.1088/0004-637X/776/1/15},
archivePrefix = {arXiv},
       eprint = {1306.0610},
 primaryClass = {astro-ph.EP},
       adsurl = {https://ui.adsabs.harvard.edu/abs/2013ApJ...776...15C}
}

@ARTICLE{2009ARA&A..47..253O,
       author = {{Oppenheimer}, Ben R. and {Hinkley}, Sasha},
        title = "{High-Contrast Observations in Optical and Infrared Astronomy}",
      journal = {\araa},
         year = 2009,
        month = sep,
       volume = {47},
       number = {1},
        pages = {253-289},
          doi = {10.1146/annurev-astro-082708-101717},
archivePrefix = {arXiv},
       eprint = {0903.4466},
 primaryClass = {astro-ph.IM},
       adsurl = {https://ui.adsabs.harvard.edu/abs/2009ARA&A..47..253O}
}

@INCOLLECTION{2010exop.book..111T,
       author = {{Traub}, W.~A. and {Oppenheimer}, B.~R.},
        title = "{Direct Imaging of Exoplanets}",
    booktitle = {Exoplanets},
         year = 2010,
       editor = {{Seager}, S.},
        pages = {111-156},
       adsurl = {https://ui.adsabs.harvard.edu/abs/2010exop.book..111T}
}

@ARTICLE{2026SCPMA..6939501C,
       author = {{CSST Collaboration} and {Gong}, Yan and {Miao}, Haitao and {Zhan}, Hu and {Li}, Zhao-Yu and {Shangguan}, Jinyi and {Li}, Haining and {Liu}, Chao and {Chen}, Xuefei and {Yuan}, Haibo and {Zhou}, Jilin and {Liu}, Hui-Gen and {Yu}, Cong and {Ji}, Jianghui and {Qi}, Zhaoxiang and {Liu}, Jiacheng and {Dai}, Zigao and {Wang}, Xiaofeng and {Zheng}, Zhenya and {Hao}, Lei and {Dou}, Jiangpei and {Ao}, Yiping and {Lin}, Zhenhui and {Zhang}, Kun and {Wang}, Wei and {Sun}, Guotong and {Li}, Ran and {Li}, Guoliang and {Xu}, Youhua and {Li}, Xinfeng and {Li}, Shengyang and {Wu}, Peng and {Zhang}, Jiuxing and {Wang}, Bo and {Bai}, Jinming and {Cai}, Yi-Fu and {Cai}, Zheng and {Cao}, Jie and {Chan}, Kwan Chuen and {Chang}, Jin and {Chen}, Xiaodian and {Chen}, Xuelei and {Chen}, Yuqin and {Chen}, Yun and {Cui}, Wei and {Dong}, Subo and {Du}, Pu and {Duan}, Wenying and {Fan}, Junhui and {Fan}, LuLu and {Fan}, Zhou and {Fan}, Zuhui and {Fang}, Taotao and {Fu}, Jianning and {Fu}, Liping and {Fu}, Zhensen and {Gao}, Jian and {Gu}, Shenghong and {Gu}, Yidong and {Guo}, Qi and {Han}, Zhanwen and {Hu}, Bin and {Huang}, Zhiqi and {Ho}, Luis C. and {Jiang}, Linhua and {Jiang}, Ning and {Jing}, Yipeng and {Kang}, Xi and {Kong}, Xu and {Li}, Cheng and {Li}, Chengyuan and {Li}, Di and {Li}, Jing and {Li}, Nan and {Li}, Yang A. and {Liao}, Shilong and {Lin}, Weipeng and {Liu}, Fengshan and {Liu}, Jifeng and {Liu}, Xiangkun and {Liu}, Zhuokai and {Mao}, Ruiqing and {Mao}, Shude and {Meng}, Xianmin and {Pang}, Xiaoying and {Peng}, Xiyan and {Peng}, Yingjie and {Shan}, Huanyuan and {Shen}, Juntai and {Shen}, Shiyin and {Shen}, Zhiqiang and {Shi}, Sheng-Cai and {Shi}, Yong and {Tan}, Siyuan and {Tian}, Hao and {Wang}, Jianmin and {Wang}, Jun-Xian and {Wang}, Xin and {Wang}, Yuting and {Wu}, Hong and {Wu}, Jingwen and {Wu}, Xuebing and {Xu}, Chun and {Xue}, Xiang-Xiang and {Xue}, Yongquan and {Yang}, Ji and {Yang}, Xiaohu and {Yao}, Qijun and {Yuan}, Fangting and {Yuan}, Zhen and {Zhang}, Jun and {Zhang}, Pengjie and {Zhang}, Tianmeng and {Zhang}, Wei and {Zhang}, Xin and {Zhao}, Gang and {Zhao}, Gongbo and {Zhong}, Hongen and {Zhong}, Jing and {Zhou}, Liyong and {Zhu}, Wei and {Zu}, Ying},
        title = "{Introduction to the Chinese Space Station Survey Telescope (CSST)}",
      journal = {Science China Physics, Mechanics, and Astronomy},
         year = 2026,
        month = jan,
       volume = {69},
       number = {3},
          eid = {239501},
        pages = {239501},
          doi = {10.1007/s11433-025-2809-0},
archivePrefix = {arXiv},
       eprint = {2507.04618},
 primaryClass = {astro-ph.IM},
       adsurl = {https://ui.adsabs.harvard.edu/abs/2026SCPMA..6939501C}
}

@ARTICLE{2021MNRAS.502.2158R,
       author = {{Ren}, Deqing and {Chen}, Yili},
        title = "{Global optimization-based reference star differential imaging for high-contrast exoplanet imaging survey}",
      journal = {\mnras},
         year = 2021,
        month = apr,
       volume = {502},
       number = {2},
        pages = {2158-2171},
          doi = {10.1093/mnras/stab022},
       adsurl = {https://ui.adsabs.harvard.edu/abs/2021MNRAS.502.2158R}
}

@ARTICLE{2026RAA....26b4011Z,
       author = {{Zhu}, YiMing and {Zhao}, Gang and {Dou}, JiangPei and {Lv}, ZhongHua and {Chen}, YiLi and {Ma}, Bo and {Yan}, ZhaoJun and {Tang}, Jing and {Li}, Ran},
        title = "{Mock Observations for the CSST Mission: CPI-C─Targets for High Contrast Imaging}",
      journal = {Research in Astronomy and Astrophysics},
         year = 2026,
        month = feb,
       volume = {26},
       number = {2},
          eid = {024011},
        pages = {024011},
          doi = {10.1088/1674-4527/ae2100},
archivePrefix = {arXiv},
       eprint = {2511.09862},
 primaryClass = {astro-ph.IM},
       adsurl = {https://ui.adsabs.harvard.edu/abs/2026RAA....26b4011Z}
}

@ARTICLE{2026RAA....26b4010Z,
       author = {{Zhao}, Gang and {Zhu}, Yiming and {Dou}, Jiangpei and {Zhang}, Xi and {Chen}, Yili and {Lv}, Zhonghua and {Niu}, Bingli and {Yan}, Zhaojun and {Ma}, Bo and {Li}, Ran},
        title = "{Mock Observations for the CSST Mission: CPI-C─Instrument Simulation}",
      journal = {Research in Astronomy and Astrophysics},
         year = 2026,
        month = feb,
       volume = {26},
       number = {2},
          eid = {024010},
        pages = {024010},
          doi = {10.1088/1674-4527/ae2102},
archivePrefix = {arXiv},
       eprint = {2511.08886},
 primaryClass = {astro-ph.IM},
       adsurl = {https://ui.adsabs.harvard.edu/abs/2026RAA....26b4010Z}
}

@ARTICLE{2010Natur.468.1080M,
       author = {{Marois}, Christian and {Zuckerman}, B. and {Konopacky}, Quinn M. and {Macintosh}, Bruce and {Barman}, Travis},
        title = "{Images of a fourth planet orbiting HR 8799}",
      journal = {\nat},
         year = 2010,
        month = dec,
       volume = {468},
       number = {7327},
        pages = {1080-1083},
          doi = {10.1038/nature09684},
archivePrefix = {arXiv},
       eprint = {1011.4918},
 primaryClass = {astro-ph.EP},
       adsurl = {https://ui.adsabs.harvard.edu/abs/2010Natur.468.1080M}
}

@ARTICLE{2009A&A...493L..21L,
       author = {{Lagrange}, A.-M. and {Gratadour}, D. and {Chauvin}, G. and {Fusco}, T. and {Ehrenreich}, D. and {Mouillet}, D. and {Rousset}, G. and {Rouan}, D. and {Allard}, F. and {Gendron}, {\'E}. and {Charton}, J. and {Mugnier}, L. and {Rabou}, P. and {Montri}, J. and {Lacombe}, F.},
        title = "{A probable giant planet imaged in the {\ensuremath{\beta}} Pictoris disk. VLT/NaCo deep L'-band imaging}",
      journal = {\aap},
         year = 2009,
        month = jan,
       volume = {493},
       number = {2},
        pages = {L21-L25},
          doi = {10.1051/0004-6361:200811325},
archivePrefix = {arXiv},
       eprint = {0811.3583},
 primaryClass = {astro-ph},
       adsurl = {https://ui.adsabs.harvard.edu/abs/2009A&A...493L..21L}
}

@ARTICLE{2010Sci...329...57L,
       author = {{Lagrange}, A.-M. and {Bonnefoy}, M. and {Chauvin}, G. and {Apai}, D. and {Ehrenreich}, D. and {Boccaletti}, A. and {Gratadour}, D. and {Rouan}, D. and {Mouillet}, D. and {Lacour}, S. and {Kasper}, M.},
        title = "{A Giant Planet Imaged in the Disk of the Young Star {\ensuremath{\beta}} Pictoris}",
      journal = {Science},
         year = 2010,
        month = jul,
       volume = {329},
       number = {5987},
        pages = {57},
          doi = {10.1126/science.1187187},
archivePrefix = {arXiv},
       eprint = {1006.3314},
 primaryClass = {astro-ph.EP},
       adsurl = {https://ui.adsabs.harvard.edu/abs/2010Sci...329...57L}
}

@software{2023ascl.soft07057S,
       author = {{Stolker}, Tomas},
        title = "{species: Atmospheric characterization of directly imaged exoplanets}",
 howpublished = {Astrophysics Source Code Library, record ascl:2307.057},
         year = 2023,
        month = jul,
          eid = {ascl:2307.057},
archivePrefix = {ascl},
       eprint = {2307.057},
       adsurl = {https://ui.adsabs.harvard.edu/abs/2023ascl.soft07057S}
}

@ARTICLE{2024A&A...687A.298N,
       author = {{Nasedkin}, E. and {Molli{\`e}re}, P. and {Lacour}, S. and {Nowak}, M. and {Kreidberg}, L. and {Stolker}, T. and {Wang}, J.~J. and {Balmer}, W.~O. and {Kammerer}, J. and {Shangguan}, J. and {Abuter}, R. and {Amorim}, A. and {Asensio-Torres}, R. and {Benisty}, M. and {Berger}, J.-P. and {Beust}, H. and {Blunt}, S. and {Boccaletti}, A. and {Bonnefoy}, M. and {Bonnet}, H. and {Bordoni}, M.~S. and {Bourdarot}, G. and {Brandner}, W. and {Cantalloube}, F. and {Caselli}, P. and {Charnay}, B. and {Chauvin}, G. and {Chavez}, A. and {Choquet}, E. and {Christiaens}, V. and {Cl{\'e}net}, Y. and {Coud{\'e} Du Foresto}, V. and {Cridland}, A. and {Davies}, R. and {Dembet}, R. and {Dexter}, J. and {Drescher}, A. and {Duvert}, G. and {Eckart}, A. and {Eisenhauer}, F. and {F{\"o}rster Schreiber}, N.~M. and {Garcia}, P. and {Garcia Lopez}, R. and {Gendron}, E. and {Genzel}, R. and {Gillessen}, S. and {Girard}, J.~H. and {Grant}, S. and {Haubois}, X. and {Hei{\ss}el}, G. and {Henning}, Th. and {Hinkley}, S. and {Hippler}, S. and {Houll{\'e}}, M. and {Hubert}, Z. and {Jocou}, L. and {Keppler}, M. and {Kervella}, P. and {Kurtovic}, N.~T. and {Lagrange}, A.-M. and {Lapeyr{\`e}re}, V. and {Le Bouquin}, J.-B. and {Lutz}, D. and {Maire}, A.-L. and {Mang}, F. and {Marleau}, G.-D. and {M{\'e}rand}, A. and {Monnier}, J.~D. and {Mordasini}, C. and {Ott}, T. and {Otten}, G.~P.~P.~L. and {Paladini}, C. and {Paumard}, T. and {Perraut}, K. and {Perrin}, G. and {Pfuhl}, O. and {Pourr{\'e}}, N. and {Pueyo}, L. and {Ribeiro}, D.~C. and {Rickman}, E. and {Ruffio}, J.~B. and {Rustamkulov}, Z. and {Shimizu}, T. and {Sing}, D. and {Stadler}, J. and {Straub}, O. and {Straubmeier}, C. and {Sturm}, E. and {Tacconi}, L.~J. and {van Dishoeck}, E.~F. and {Vigan}, A. and {Vincent}, F. and {von Fellenberg}, S.~D. and {Widmann}, F. and {Winterhalder}, T.~O. and {Woillez}, J. and {Yazici}, {\c{S}}. and {Gravity Collaboration}},
        title = "{Four-of-a-kind? Comprehensive atmospheric characterisation of the HR 8799 planets with VLTI/GRAVITY}",
      journal = {\aap},
         year = 2024,
        month = jul,
       volume = {687},
          eid = {A298},
        pages = {A298},
          doi = {10.1051/0004-6361/202449328},
archivePrefix = {arXiv},
       eprint = {2404.03776},
 primaryClass = {astro-ph.EP},
       adsurl = {https://ui.adsabs.harvard.edu/abs/2024A&A...687A.298N}
}

@ARTICLE{2025A&A...704A.325R,
       author = {{Ravet}, M. and {Bonnefoy}, M. and {Chauvin}, G. and {Lacour}, S. and {Nowak}, M. and {Charnay}, B. and {Tremblin}, P. and {Homeier}, D. and {Morley}, C. and {Fortney}, J. and {Denis}, A. and {Petrus}, S. and {Palma-Bifani}, P. and {Landman}, R. and {Parker}, L.~T. and {Houll{\'e}}, M. and {Chomez}, A. and {Worthen}, K. and {Kiefer}, F. and {Marleau}, G.-D. and {Zhang}, Z. and {Birkby}, J.~L. and {Millour}, F. and {Lagrange}, A.-M. and {Vigan}, A. and {Otten}, G.~P.~P.~L. and {Shangguan}, J.},
        title = "{Multimodal atmospheric characterization of {\ensuremath{\beta}} Pictoris b: Adding high-resolution continuum spectra from GRAVITY}",
      journal = {\aap},
         year = 2025,
        month = dec,
       volume = {704},
          eid = {A325},
        pages = {A325},
          doi = {10.1051/0004-6361/202553885},
archivePrefix = {arXiv},
       eprint = {2509.25338},
 primaryClass = {astro-ph.EP},
       adsurl = {https://ui.adsabs.harvard.edu/abs/2025A&A...704A.325R}
}

@ARTICLE{2015A&A...582A..83B,
       author = {{Baudino}, J.-L. and {B{\'e}zard}, B. and {Boccaletti}, A. and {Bonnefoy}, M. and {Lagrange}, A.-M. and {Galicher}, R.},
        title = "{Interpreting the photometry and spectroscopy of directly imaged planets: a new atmospheric model applied to {\ensuremath{\beta}} Pictoris b and SPHERE observations}",
      journal = {\aap},
         year = 2015,
        month = oct,
       volume = {582},
          eid = {A83},
        pages = {A83},
          doi = {10.1051/0004-6361/201526332},
archivePrefix = {arXiv},
       eprint = {1504.04876},
 primaryClass = {astro-ph.EP},
       adsurl = {https://ui.adsabs.harvard.edu/abs/2015A&A...582A..83B}
}

@ARTICLE{2016AJ....152..217L,
       author = {{Lupu}, Roxana E. and {Marley}, Mark S. and {Lewis}, Nikole and {Line}, Michael and {Traub}, Wesley A. and {Zahnle}, Kevin},
        title = "{Developing Atmospheric Retrieval Methods for Direct Imaging Spectroscopy of Gas Giants in Reflected Light. I. Methane Abundances and Basic Cloud Properties}",
      journal = {\aj},
         year = 2016,
        month = dec,
       volume = {152},
       number = {6},
          eid = {217},
        pages = {217},
          doi = {10.3847/0004-6256/152/6/217},
archivePrefix = {arXiv},
       eprint = {1604.05370},
 primaryClass = {astro-ph.IM},
       adsurl = {https://ui.adsabs.harvard.edu/abs/2016AJ....152..217L}
}

@BOOK{1975lpsa.book.....S,
       author = {{Sobolev}, V.~V.},
        title = "{Light scattering in planetary atmospheres}",
         year = 1975,
       adsurl = {https://ui.adsabs.harvard.edu/abs/1975lpsa.book.....S}
}

@ARTICLE{2010ARA&A..48..631S,
       author = {{Seager}, Sara and {Deming}, Drake},
        title = "{Exoplanet Atmospheres}",
      journal = {\araa},
         year = 2010,
        month = sep,
       volume = {48},
        pages = {631-672},
          doi = {10.1146/annurev-astro-081309-130837},
archivePrefix = {arXiv},
       eprint = {1005.4037},
 primaryClass = {astro-ph.EP},
       adsurl = {https://ui.adsabs.harvard.edu/abs/2010ARA&A..48..631S}
}

@ARTICLE{2016PASP..128b5003R,
       author = {{Robinson}, Tyler D. and {Stapelfeldt}, Karl R. and {Marley}, Mark S.},
        title = "{Characterizing Rocky and Gaseous Exoplanets with 2 m Class Space-based Coronagraphs}",
      journal = {\pasp},
         year = 2016,
        month = feb,
       volume = {128},
       number = {960},
        pages = {025003},
          doi = {10.1088/1538-3873/128/960/025003},
archivePrefix = {arXiv},
       eprint = {1507.00777},
 primaryClass = {astro-ph.EP},
       adsurl = {https://ui.adsabs.harvard.edu/abs/2016PASP..128b5003R}
}

@ARTICLE{2024A&A...682A.145S,
       author = {{Soubiran}, C. and {Creevey}, O.~L. and {Lagarde}, N. and {Brouillet}, N. and {Jofr{\'e}}, P. and {Casamiquela}, L. and {Heiter}, U. and {Aguilera-G{\'o}mez}, C. and {Vitali}, S. and {Worley}, C. and {de Brito Silva}, D.},
        title = "{Gaia FGK benchmark stars: Fundamental T$_{eff}$ and log g of the third version}",
      journal = {\aap},
         year = 2024,
        month = feb,
       volume = {682},
          eid = {A145},
        pages = {A145},
          doi = {10.1051/0004-6361/202347136},
archivePrefix = {arXiv},
       eprint = {2310.11302},
 primaryClass = {astro-ph.SR},
       adsurl = {https://ui.adsabs.harvard.edu/abs/2024A&A...682A.145S}
}

@ARTICLE{Dou2026CPIC,
  author       = {{Dou}, Jiangpei and {Zhang}, Xi and {Zhao}, Gang and {Xu}, Mingming and {Wu}, Zhen and {Wang}, Gang and {Yuan}, Baoning and {Kong}, Lingyi and {Zhu}, Yiming and {Niu}, Bingli and {Lv}, Zhonghua and {Qi}, Yongjun and {Jiang}, Shu and {Chen}, Bo and {Guo}, Wei and {Wang}, Di and {Lin}, Yinglu and {Zheng}, Liping and {Guo}, Jing and {Li}, Ruokun and {Xu}, Liyan and {Wu}, Huihai and {Wen}, Cheng and {Miao}, Shuwei and {Lv}, Boyang and {Li}, Weimiao},
  title        = "{CPI-C: Cool Planet Imaging Coronagraph on Chinese Space Station Survey Telescope}",
  journal      = {Research in Astronomy and Astrophysics},
  year         = {2026},
  doi          = {10.1088/1674-4527/ae4a04},
  note         = {online first},
  archivePrefix= {arXiv},
  eprint       = {2512.11292},
  primaryClass = {astro-ph.EP},
  adsurl       = {https://ui.adsabs.harvard.edu/abs/2025arXiv251211292D}
}

@ARTICLE{1999ApJ...513..879M,
       author = {{Marley}, Mark S. and {Gelino}, Christopher and {Stephens}, Denise and {Lunine}, Jonathan I. and {Freedman}, Richard},
        title = "{Reflected Spectra and Albedos of Extrasolar Giant Planets. I. Clear and Cloudy Atmospheres}",
      journal = {\apj},
         year = 1999,
        month = mar,
       volume = {513},
       number = {2},
        pages = {879-893},
          doi = {10.1086/306881},
archivePrefix = {arXiv},
       eprint = {astro-ph/9810073},
 primaryClass = {astro-ph},
       adsurl = {https://ui.adsabs.harvard.edu/abs/1999ApJ...513..879M}
}

@ARTICLE{2000ApJ...538..885S,
       author = {{Sudarsky}, David and {Burrows}, Adam and {Pinto}, Philip},
        title = "{Albedo and Reflection Spectra of Extrasolar Giant Planets}",
      journal = {\apj},
         year = 2000,
        month = aug,
       volume = {538},
       number = {2},
        pages = {885-903},
          doi = {10.1086/309160},
archivePrefix = {arXiv},
       eprint = {astro-ph/9910504},
 primaryClass = {astro-ph},
       adsurl = {https://ui.adsabs.harvard.edu/abs/2000ApJ...538..885S}
}

@ARTICLE{2004ApJ...609..407B,
       author = {{Burrows}, Adam and {Sudarsky}, David and {Hubeny}, Ivan},
        title = "{Spectra and Diagnostics for the Direct Detection of Wide-Separation Extrasolar Giant Planets}",
      journal = {\apj},
         year = 2004,
        month = jul,
       volume = {609},
       number = {1},
        pages = {407-416},
          doi = {10.1086/420974},
archivePrefix = {arXiv},
       eprint = {astro-ph/0401522},
 primaryClass = {astro-ph},
       adsurl = {https://ui.adsabs.harvard.edu/abs/2004ApJ...609..407B}
}

@ARTICLE{2013ApJ...775..137L,
       author = {{Line}, Michael R. and {Wolf}, Aaron S. and {Zhang}, Xi and {Knutson}, Heather and {Kammer}, Joshua A. and {Ellison}, Elias and {Deroo}, Pieter and {Crisp}, Dave and {Yung}, Yuk L.},
        title = "{A Systematic Retrieval Analysis of Secondary Eclipse Spectra. I. A Comparison of Atmospheric Retrieval Techniques}",
      journal = {\apj},
         year = 2013,
        month = oct,
       volume = {775},
       number = {2},
          eid = {137},
        pages = {137},
          doi = {10.1088/0004-637X/775/2/137},
archivePrefix = {arXiv},
       eprint = {1304.5561},
 primaryClass = {astro-ph.EP},
       adsurl = {https://ui.adsabs.harvard.edu/abs/2013ApJ...775..137L}
}

@ARTICLE{2014ApJ...786..154B,
       author = {{Barstow}, J.~K. and {Aigrain}, S. and {Irwin}, P.~G.~J. and {Hackler}, T. and {Fletcher}, L.~N. and {Lee}, J.~M. and {Gibson}, N.~P.},
        title = "{Clouds on the Hot Jupiter HD189733b: Constraints from the Reflection Spectrum}",
      journal = {\apj},
         year = 2014,
        month = may,
       volume = {786},
       number = {2},
          eid = {154},
        pages = {154},
          doi = {10.1088/0004-637X/786/2/154},
archivePrefix = {arXiv},
       eprint = {1403.6664},
 primaryClass = {astro-ph.EP},
       adsurl = {https://ui.adsabs.harvard.edu/abs/2014ApJ...786..154B}
}

@ARTICLE{2012ApJ...753...99R,
       author = {{Ren}, Deqing and {Dou}, Jiangpei and {Zhang}, Xi and {Zhu}, Yongtian},
        title = "{Speckle Noise Subtraction and Suppression with Adaptive Optics Coronagraphic Imaging}",
      journal = {\apj},
         year = 2012,
        month = jul,
       volume = {753},
       number = {2},
          eid = {99},
        pages = {99},
          doi = {10.1088/0004-637X/753/2/99},
       adsurl = {https://ui.adsabs.harvard.edu/abs/2012ApJ...753...99R}
}

@ARTICLE{2025Natur.642..905L,
       author = {{Lagrange}, A.-M. and {Wilkinson}, C. and {M{\^a}lin}, M. and {Boccaletti}, A. and {Perrot}, C. and {Matr{\`a}}, L. and {Combes}, F. and {Beust}, H. and {Rouan}, D. and {Chomez}, A. and {Milli}, J. and {Charnay}, B. and {Mazevet}, S. and {Flasseur}, O. and {Olofsson}, J. and {Bayo}, A. and {Kral}, Q. and {Carter}, A. and {Crotts}, K.~A. and {Delorme}, P. and {Chauvin}, G. and {Thebault}, P. and {Rubini}, P. and {Kiefer}, F. and {Radcliffe}, A. and {Mazoyer}, J. and {Bodrito}, T. and {Stasevic}, S. and {Langlois}, M.},
        title = "{Evidence for a sub-Jovian planet in the young TWA 7 disk}",
      journal = {\nat},
         year = 2025,
        month = jun,
       volume = {642},
       number = {8069},
        pages = {905-908},
          doi = {10.1038/s41586-025-09150-4},
archivePrefix = {arXiv},
       eprint = {2502.15081},
 primaryClass = {astro-ph.EP},
       adsurl = {https://ui.adsabs.harvard.edu/abs/2025Natur.642..905L}
}

@ARTICLE{2025Natur.643..938H,
       author = {{Hoch}, K.~K.~W. and {Rowland}, M. and {Petrus}, S. and {Nasedkin}, E. and {Ingebretsen}, C. and {Kammerer}, J. and {Perrin}, M. and {D'Orazi}, V. and {Balmer}, W.~O. and {Barman}, T. and {Bonnefoy}, M. and {Chauvin}, G. and {Chen}, C. and {De Rosa}, R.~J. and {Girard}, J. and {Gonzales}, E. and {Kenworthy}, M. and {Konopacky}, Q.~M. and {Macintosh}, B. and {Moran}, S.~E. and {Morley}, C.~V. and {Palma-Bifani}, P. and {Pueyo}, L. and {Ren}, B. and {Rickman}, E. and {Ruffio}, J.-B. and {Theissen}, C.~A. and {Ward-Duong}, K. and {Zhang}, Y.},
        title = "{Silicate clouds and a circumplanetary disk in the YSES-1 exoplanet system}",
      journal = {\nat},
         year = 2025,
        month = jul,
       volume = {643},
       number = {8073},
        pages = {938-942},
          doi = {10.1038/s41586-025-09174-w},
archivePrefix = {arXiv},
       eprint = {2507.18861},
 primaryClass = {astro-ph.EP},
       adsurl = {https://ui.adsabs.harvard.edu/abs/2025Natur.643..938H}
}

@ARTICLE{2024AstTI...1..166D,
       author = {{Dou}, Jiangpei and {Dong}, Huanyu},
        title = "{A high-contrast imaging coronagraph for segmented-mirror large aperture telescopes using a spatial light modulator}",
      journal = {Astronomical Techniques and Instruments},
         year = 2024,
        month = may,
       volume = {1},
       number = {3},
        pages = {166-170},
          doi = {10.61977/ati2024018},
       adsurl = {https://ui.adsabs.harvard.edu/abs/2024AstTI...1..166D}
}

@ARTICLE{2025ARA&A..63..179K,
       author = {{Kenworthy}, Matthew A. and {Haffert}, Sebastiaan Y.},
        title = "{High-Contrast Coronagraphy}",
      journal = {\araa},
         year = 2025,
        month = aug,
       volume = {63},
       number = {1},
        pages = {179-216},
          doi = {10.1146/annurev-astro-021225-022840},
archivePrefix = {arXiv},
       eprint = {2506.02907},
 primaryClass = {astro-ph.IM},
       adsurl = {https://ui.adsabs.harvard.edu/abs/2025ARA&A..63..179K}
}

@ARTICLE{2015ApJ...802...12D,
       author = {{Dou}, Jiangpei and {Ren}, Deqing and {Zhao}, Gang and {Zhang}, Xi and {Chen}, Rui and {Zhu}, Yongtian},
        title = "{A High-contrast Imaging Algorithm: Optimized Image Rotation and Subtraction}",
      journal = {\apj},
         year = 2015,
        month = mar,
       volume = {802},
       number = {1},
          eid = {12},
        pages = {12},
          doi = {10.1088/0004-637X/802/1/12},
archivePrefix = {arXiv},
       eprint = {1501.03893},
 primaryClass = {astro-ph.IM},
       adsurl = {https://ui.adsabs.harvard.edu/abs/2015ApJ...802...12D}
}

@ARTICLE{2019ApJ...878...70B,
       author = {{Batalha}, Natasha E. and {Marley}, Mark S. and {Lewis}, Nikole K. and {Fortney}, Jonathan J.},
        title = "{Exoplanet Reflected-light Spectroscopy with PICASO}",
      journal = {\apj},
         year = 2019,
        month = jun,
       volume = {878},
       number = {1},
          eid = {70},
        pages = {70},
          doi = {10.3847/1538-4357/ab1b51},
archivePrefix = {arXiv},
       eprint = {1904.09355},
 primaryClass = {astro-ph.EP},
       adsurl = {https://ui.adsabs.harvard.edu/abs/2019ApJ...878...70B}
}

@ARTICLE{2024Natur.633..789M,
       author = {{Matthews}, E.~C. and {Carter}, A.~L. and {Pathak}, P. and {Morley}, C.~V. and {Phillips}, M.~W. and {P.~M.}, S. Krishanth and {Feng}, F. and {Bonse}, M.~J. and {Boogaard}, L.~A. and {Burt}, J.~A. and {Crossfield}, I.~J.~M. and {Douglas}, E.~S. and {Henning}, Th. and {Hom}, J. and {Ko}, C.-L. and {Kasper}, M. and {Lagrange}, A.-M. and {Petit dit de la Roche}, D. and {Philipot}, F.},
        title = "{A temperate super-Jupiter imaged with JWST in the mid-infrared}",
      journal = {\nat},
         year = 2024,
        month = sep,
       volume = {633},
       number = {8031},
        pages = {789-792},
          doi = {10.1038/s41586-024-07837-8},
archivePrefix = {arXiv},
       eprint = {2503.01599},
 primaryClass = {astro-ph.EP},
       adsurl = {https://ui.adsabs.harvard.edu/abs/2024Natur.633..789M}
}

@ARTICLE{2025ApJ...988L..18B,
       author = {{Bardalez Gagliuffi}, Daniella C. and {Balmer}, William O. and {Pueyo}, Laurent and {Brandt}, Timothy D. and {Giovinazzi}, Mark R. and {Millholland}, Sarah and {Black}, Brennen and {Lu}, Tiger and {Rice}, Malena and {Mang}, James and {Morley}, Caroline and {Lacy}, Brianna and {Girard}, Julien H. and {Matthews}, Elisabeth C. and {Carter}, Aarynn L. and {Bowler}, Brendan P. and {Faherty}, Jacqueline K. and {Fontanive}, Clemence and {Rickman}, Emily},
        title = "{JWST Coronagraphic Images of 14 Her c: A Cold Giant Planet in a Dynamically Hot Multiplanet System}",
      journal = {\apjl},
         year = 2025,
        month = jul,
       volume = {988},
       number = {1},
          eid = {L18},
        pages = {L18},
          doi = {10.3847/2041-8213/ade30f},
archivePrefix = {arXiv},
       eprint = {2506.09201},
 primaryClass = {astro-ph.EP},
       adsurl = {https://ui.adsabs.harvard.edu/abs/2025ApJ...988L..18B}
}

@ARTICLE{2017PASP..129c4401N,
       author = {{Nayak}, Michael and {Lupu}, Roxana and {Marley}, Mark S. and {Fortney}, Jonathan J. and {Robinson}, Tyler and {Lewis}, Nikole},
        title = "{Atmospheric Retrieval for Direct Imaging Spectroscopy of Gas Giants in Reflected Light. II. Orbital Phase and Planetary Radius}",
      journal = {\pasp},
         year = 2017,
        month = mar,
       volume = {129},
       number = {973},
        pages = {034401},
          doi = {10.1088/1538-3873/129/973/034401},
archivePrefix = {arXiv},
       eprint = {1612.00342},
 primaryClass = {astro-ph.EP},
       adsurl = {https://ui.adsabs.harvard.edu/abs/2017PASP..129c4401N}
}

@ARTICLE{2024ApJ...969L..22S,
       author = {{Salvador}, Arnaud and {Robinson}, Tyler D. and {Fortney}, Jonathan J. and {Marley}, Mark S.},
        title = "{Influence of Orbit and Mass Constraints on Reflected Light Characterization of Directly Imaged Rocky Exoplanets}",
      journal = {\apjl},
         year = 2024,
        month = jul,
       volume = {969},
       number = {1},
          eid = {L22},
        pages = {L22},
          doi = {10.3847/2041-8213/ad54c5},
archivePrefix = {arXiv},
       eprint = {2406.07749},
 primaryClass = {astro-ph.EP},
       adsurl = {https://ui.adsabs.harvard.edu/abs/2024ApJ...969L..22S}
}

@ARTICLE{2025AJ....169...97D,
       author = {{Damiano}, Mario and {Burr}, Zachary and {Hu}, Renyu and {Burt}, Jennifer and {Kataria}, Tiffany},
        title = "{Effects of Planetary Mass Uncertainties on the Interpretation of the Reflectance Spectra of Earth-like Exoplanets}",
      journal = {\aj},
         year = 2025,
        month = feb,
       volume = {169},
       number = {2},
          eid = {97},
        pages = {97},
          doi = {10.3847/1538-3881/ada610},
archivePrefix = {arXiv},
       eprint = {2502.01513},
 primaryClass = {astro-ph.EP},
       adsurl = {https://ui.adsabs.harvard.edu/abs/2025AJ....169...97D}
}

@ARTICLE{2004A&A...420.1153A,
       author = {{Arnold}, L. and {Schneider}, J.},
        title = "{The detectability of extrasolar planet surroundings. I. Reflected-light photometry of unresolved rings}",
      journal = {\aap},
         year = 2004,
        month = jun,
       volume = {420},
        pages = {1153-1162},
          doi = {10.1051/0004-6361:20035720},
archivePrefix = {arXiv},
       eprint = {astro-ph/0403330},
 primaryClass = {astro-ph},
       adsurl = {https://ui.adsabs.harvard.edu/abs/2004A&A...420.1153A}
}

@ARTICLE{2018A&A...618A.162B,
       author = {{Berzosa Molina}, J. and {Rossi}, L. and {Stam}, D.~M.},
        title = "{Traces of exomoons in computed flux and polarization phase curves of starlight reflected by exoplanets}",
      journal = {\aap},
         year = 2018,
        month = oct,
       volume = {618},
          eid = {A162},
        pages = {A162},
          doi = {10.1051/0004-6361/201833320},
archivePrefix = {arXiv},
       eprint = {1807.10266},
 primaryClass = {astro-ph.EP},
       adsurl = {https://ui.adsabs.harvard.edu/abs/2018A&A...618A.162B}
}

@ARTICLE{2021Msngr.182...38K,
       author = {{Kasper}, M. and {Cerpa Urra}, N. and {Pathak}, P. and {Bonse}, M. and {Nousiainen}, J. and {Engler}, B. and {Heritier}, C.~T. and {Kammerer}, J. and {Leveratto}, S. and {Rajani}, C. and {Bristow}, P. and {Le Louarn}, M. and {Madec}, P.-Y. and {Str{\"o}bele}, S. and {Verinaud}, C. and {Glauser}, A. and {Quanz}, S.~P. and {Helin}, T. and {Keller}, C. and {Snik}, F. and {Boccaletti}, A. and {Chauvin}, G. and {Mouillet}, D. and {Kulcs{\'a}r}, C. and {Raynaud}, H.-F.},
        title = "{PCS {\textemdash} A Roadmap for Exoearth Imaging with the ELT}",
      journal = {The Messenger},
         year = 2021,
        month = mar,
       volume = {182},
        pages = {38-43},
          doi = {10.18727/0722-6691/5221},
archivePrefix = {arXiv},
       eprint = {2103.11196},
 primaryClass = {astro-ph.IM},
       adsurl = {https://ui.adsabs.harvard.edu/abs/2021Msngr.182...38K}
}

@INPROCEEDINGS{2024SPIE13096E..0YM,
       author = {{Males}, Jared R. and {Close}, Laird M. and {Haffert}, Sebastiaan Y. and {Kautz}, Maggie Y. and {Kelly}, Doug and {Fletcher}, Adam and {Salanski}, Thomas and {Durney}, Olivier and {Noenickx}, Jamison and {Ford}, John and {Gasho}, Victor and {Pearce}, Logan and {Kueny}, Jay and {Guyon}, Olivier and {Weinberger}, Alycia and {Bowler}, Brendan and {Kraus}, Adam and {Batalha}, Natasha},
        title = "{High-contrast imaging at first-light of the GMT: the preliminary design of GMagAO-X}",
    booktitle = {Ground-based and Airborne Instrumentation for Astronomy X},
         year = 2024,
       editor = {{Bryant}, Julia J. and {Motohara}, Kentaro and {Vernet}, Jo{\"e}l. R.~D.},
       series = {Society of Photo-Optical Instrumentation Engineers (SPIE) Conference Series},
       volume = {13096},
        month = jul,
          eid = {130960Y},
        pages = {130960Y},
          doi = {10.1117/12.3018157},
       adsurl = {https://ui.adsabs.harvard.edu/abs/2024SPIE13096E..0YM}
}

@INPROCEEDINGS{2022SPIE12180E..1WM,
       author = {{Mennesson}, B. and {Bailey}, V.~P. and {Zellem}, R. and {Hildebrandt}, S. and {Ygouf}, M. and {Rhodes}, J. and {Zimmerman}, N. and {Nemati}, B. and {Gonzalez}, G. and {Cady}, E. and {Kern}, B. and {Koch}, T. and {Krist}, J. and {Heydorff}, K. and {Luchik}, T. and {Mok}, F. and {Morrissey}, P. and {Poberezhskiy}, I. and {Riggs}, A.~J. and {Shi}, F. and {Zhao}, F. and {Akeson}, R. and {Armus}, L. and {Greenbaum}, A. and {Ingalls}, J. and {Lowrance}, P.},
        title = "{The Roman Space Telescope coronagraph technology demonstration: current status and relevance to future missions}",
    booktitle = {Space Telescopes and Instrumentation 2022: Optical, Infrared, and Millimeter Wave},
         year = 2022,
       editor = {{Coyle}, Laura E. and {Matsuura}, Shuji and {Perrin}, Marshall D.},
       series = {Society of Photo-Optical Instrumentation Engineers (SPIE) Conference Series},
       volume = {12180},
        month = aug,
          eid = {121801W},
        pages = {121801W},
          doi = {10.1117/12.2629176},
       adsurl = {https://ui.adsabs.harvard.edu/abs/2022SPIE12180E..1WM}
}

@misc{2026arXiv260106233B,
      title={Exoplanet characterization with NASA's Habitable Worlds Observatory}, 
      author={Joanna K. Barstow and Beth Biller and Mei Ting Mak and Sarah Rugheimer and Amaury Triaud and Hannah R. Wakeford},
      year={2026},
      eprint={2601.06233},
      archivePrefix={arXiv},
      primaryClass={astro-ph.IM},
      url={https://arxiv.org/abs/2601.06233}
}

@misc{2026arXiv260102556R,
      title={The Lazuli Space Observatory: Architecture \& Capabilities}, 
      author={Arpita Roy and Stuart Feldman and Pete Klupar and John DiPalma and Saul Perlmutter and Ewan S. Douglas and Greg Aldering and Gabor Furesz and Patrick Ingraham and Gudmundur Stefansson and others},
      year={2026},
      eprint={2601.02556},
      archivePrefix={arXiv},
      primaryClass={astro-ph.IM},
      url={https://arxiv.org/abs/2601.02556}
}

@ARTICLE{2018ApJ...858...69M,
       author = {{MacDonald}, Ryan J. and {Marley}, Mark S. and {Fortney}, Jonathan J. and {Lewis}, Nikole K.},
        title = "{Exploring H$_{2}$O Prominence in Reflection Spectra of Cool Giant Planets}",
      journal = {\apj},
         year = 2018,
        month = may,
       volume = {858},
       number = {2},
          eid = {69},
        pages = {69},
          doi = {10.3847/1538-4357/aabb05},
archivePrefix = {arXiv},
       eprint = {1804.00662},
 primaryClass = {astro-ph.EP},
       adsurl = {https://ui.adsabs.harvard.edu/abs/2018ApJ...858...69M}
}

@ARTICLE{2024ApJ...975...59M,
       author = {{Morley}, Caroline V. and {Mukherjee}, Sagnick and {Marley}, Mark S. and {Fortney}, Jonathan J. and {Visscher}, Channon and {Lupu}, Roxana and {Gharib-Nezhad}, Ehsan and {Thorngren}, Daniel and {Freedman}, Richard and {Batalha}, Natasha},
        title = "{The Sonora Substellar Atmosphere Models. III. Diamondback: Atmospheric Properties, Spectra, and Evolution for Warm Cloudy Substellar Objects}",
      journal = {\apj},
         year = 2024,
        month = nov,
       volume = {975},
       number = {1},
          eid = {59},
        pages = {59},
          doi = {10.3847/1538-4357/ad71d5},
archivePrefix = {arXiv},
       eprint = {2402.00758},
 primaryClass = {astro-ph.SR},
       adsurl = {https://ui.adsabs.harvard.edu/abs/2024ApJ...975...59M}
}

@ARTICLE{2020A&A...633A.110G,
       author = {{GRAVITY Collaboration} and {Nowak}, M. and {Lacour}, S. and {Molli{\`e}re}, P. and {Wang}, J. and {Charnay}, B. and {van Dishoeck}, E.~F. and {Abuter}, R. and {Amorim}, A. and {Berger}, J.~P. and {Beust}, H. and {Bonnefoy}, M. and {Bonnet}, H. and {Brandner}, W. and {Buron}, A. and {Cantalloube}, F. and {Collin}, C. and {Chapron}, F. and {Cl{\'e}net}, Y. and {Coud{\'e} Du Foresto}, V. and {de Zeeuw}, P.~T. and {Dembet}, R. and {Dexter}, J. and {Duvert}, G. and {Eckart}, A. and {Eisenhauer}, F. and {F{\"o}rster Schreiber}, N.~M. and {F{\'e}dou}, P. and {Garcia Lopez}, R. and {Gao}, F. and {Gendron}, E. and {Genzel}, R. and {Gillessen}, S. and {Hau{\ss}mann}, F. and {Henning}, T. and {Hippler}, S. and {Hubert}, Z. and {Jocou}, L. and {Kervella}, P. and {Lagrange}, A.-M. and {Lapeyr{\`e}re}, V. and {Le Bouquin}, J.-B. and {L{\'e}na}, P. and {Maire}, A.-L. and {Ott}, T. and {Paumard}, T. and {Paladini}, C. and {Perraut}, K. and {Perrin}, G. and {Pueyo}, L. and {Pfuhl}, O. and {Rabien}, S. and {Rau}, C. and {Rodr{\'\i}guez-Coira}, G. and {Rousset}, G. and {Scheithauer}, S. and {Shangguan}, J. and {Straub}, O. and {Straubmeier}, C. and {Sturm}, E. and {Tacconi}, L.~J. and {Vincent}, F. and {Widmann}, F. and {Wieprecht}, E. and {Wiezorrek}, E. and {Woillez}, J. and {Yazici}, S. and {Ziegler}, D.},
        title = "{Peering into the formation history of {\ensuremath{\beta}} Pictoris b with VLTI/GRAVITY long-baseline interferometry}",
      journal = {\aap},
         year = 2020,
        month = jan,
       volume = {633},
          eid = {A110},
        pages = {A110},
          doi = {10.1051/0004-6361/201936898},
archivePrefix = {arXiv},
       eprint = {1912.04651},
 primaryClass = {astro-ph.EP},
       adsurl = {https://ui.adsabs.harvard.edu/abs/2020A&A...633A.110G}
}

@ARTICLE{2017AJ....153..182C,
       author = {{Chilcote}, Jeffrey and {Pueyo}, Laurent and {De Rosa}, Robert J. and {Vargas}, Jeffrey and {Macintosh}, Bruce and {Bailey}, Vanessa P. and {Barman}, Travis and {Bauman}, Brian and {Bruzzone}, Sebastian and {Bulger}, Joanna and {Burrows}, Adam S. and {Cardwell}, Andrew and {Chen}, Christine H. and {Cotten}, Tara and {Dillon}, Daren and {Doyon}, Rene and {Draper}, Zachary H. and {Duch{\^e}ne}, Gaspard and {Dunn}, Jennifer and {Erikson}, Darren and {Fitzgerald}, Michael P. and {Follette}, Katherine B. and {Gavel}, Donald and {Goodsell}, Stephen J. and {Graham}, James R. and {Greenbaum}, Alexandra Z. and {Hartung}, Markus and {Hibon}, Pascale and {Hung}, Li-Wei and {Ingraham}, Patrick and {Kalas}, Paul and {Konopacky}, Quinn and {Larkin}, James E. and {Maire}, J{\'e}r{\^o}me and {Marchis}, Franck and {Marley}, Mark S. and {Marois}, Christian and {Metchev}, Stanimir and {Millar-Blanchaer}, Maxwell A. and {Morzinski}, Katie M. and {Nielsen}, Eric L. and {Norton}, Andrew and {Oppenheimer}, Rebecca and {Palmer}, David and {Patience}, Jennifer and {Perrin}, Marshall and {Poyneer}, Lisa and {Rajan}, Abhijith and {Rameau}, Julien and {Rantakyr{\"o}}, Fredrik T. and {Sadakuni}, Naru and {Saddlemyer}, Leslie and {Savransky}, Dmitry and {Schneider}, Adam C. and {Serio}, Andrew and {Sivaramakrishnan}, Anand and {Song}, Inseok and {Soummer}, Remi and {Thomas}, Sandrine and {Wallace}, J. Kent and {Wang}, Jason J. and {Ward-Duong}, Kimberly and {Wiktorowicz}, Sloane and {Wolff}, Schuyler},
        title = "{1-2.4 {\ensuremath{\mu}}m Near-IR Spectrum of the Giant Planet {\ensuremath{\beta}} Pictoris b Obtained with the Gemini Planet Imager}",
      journal = {\aj},
         year = 2017,
        month = apr,
       volume = {153},
       number = {4},
          eid = {182},
        pages = {182},
          doi = {10.3847/1538-3881/aa63e9},
archivePrefix = {arXiv},
       eprint = {1703.00011},
 primaryClass = {astro-ph.EP},
       adsurl = {https://ui.adsabs.harvard.edu/abs/2017AJ....153..182C}
}

@ARTICLE{2021ApJ...910..158M,
       author = {{Mukherjee}, Sagnick and {Batalha}, Natasha E. and {Marley}, Mark S.},
        title = "{Cloud Parameterizations and their Effect on Retrievals of Exoplanet Reflection Spectroscopy}",
      journal = {\apj},
         year = 2021,
        month = apr,
       volume = {910},
       number = {2},
          eid = {158},
        pages = {158},
          doi = {10.3847/1538-4357/abe53b},
archivePrefix = {arXiv},
       eprint = {2102.05305},
 primaryClass = {astro-ph.EP},
       adsurl = {https://ui.adsabs.harvard.edu/abs/2021ApJ...910..158M}
}

@ARTICLE{2023ApJ...942...71M,
       author = {{Mukherjee}, Sagnick and {Batalha}, Natasha E. and {Fortney}, Jonathan J. and {Marley}, Mark S.},
        title = "{PICASO 3.0: A One-dimensional Climate Model for Giant Planets and Brown Dwarfs}",
      journal = {\apj},
         year = 2023,
        month = jan,
       volume = {942},
       number = {2},
          eid = {71},
        pages = {71},
          doi = {10.3847/1538-4357/ac9f48},
archivePrefix = {arXiv},
       eprint = {2208.07836},
 primaryClass = {astro-ph.EP},
       adsurl = {https://ui.adsabs.harvard.edu/abs/2023ApJ...942...71M}
}

@ARTICLE{2026ApJ..1000...98M,
       author = {{Mang}, James and {Batalha}, Natasha E. and {Morley}, Caroline V. and {Wogan}, Nicholas F. and {Mukherjee}, Sagnick and {Visscher}, Channon and {Marley}, Mark S. and {Fortney}, Jonathan J. and {Chubb}, Katy L. and {Gao}, Peter and {Malsky}, Isaac},
        title = "{PICASO 4.0: Clouds and Photochemistry in Climate Models of Brown Dwarfs and Exoplanets}",
      journal = {\apj},
         year = 2026,
        month = mar,
       volume = {1000},
       number = {1},
          eid = {98},
        pages = {98},
          doi = {10.3847/1538-4357/ae47ff},
archivePrefix = {arXiv},
       eprint = {2602.22468},
 primaryClass = {astro-ph.EP},
       adsurl = {https://ui.adsabs.harvard.edu/abs/2026ApJ..1000...98M}
}

@ARTICLE{2016JATIS...2a1020T,
       author = {{Traub}, Wesley A. and {Breckinridge}, James and {Greene}, Thomas P. and {Guyon}, Olivier and {Jeremy Kasdin}, N. and {Macintosh}, Bruce},
        title = "{Science yield estimate with the Wide-Field Infrared Survey Telescope coronagraph}",
      journal = {Journal of Astronomical Telescopes, Instruments, and Systems},
         year = 2016,
        month = jan,
       volume = {2},
          eid = {011020},
        pages = {011020},
          doi = {10.1117/1.JATIS.2.1.011020},
       adsurl = {https://ui.adsabs.harvard.edu/abs/2016JATIS...2a1020T}
}

@ARTICLE{2026ApJ..1000...27X,
       author = {{Xuan}, Jerry W. and {Ruffio}, Jean-Baptiste and {Chachan}, Yayaati and {Ohno}, Kazumasa and {Kesseli}, Aurora and {Murray-Clay}, Ruth and {Lee}, Eve J. and {Moses}, Julianne I. and {Balmer}, William O. and {Baburaj}, Aneesh and {Blake}, Geoffrey A. and {Johnstone}, Doug and {Zhang}, Yapeng and {Knutson}, Heather A. and {Mawet}, Dimitri and {Beichman}, Charles and {Hodapp}, Klaus and {Perrin}, Marshall D. and {Konopacky}, Quinn and {Meyer}, Michael and {Bryden}, Geoffrey and {Greene}, Thomas P. and {Leisenring}, Jarron and {Ygouf}, Marie and {Benneke}, Bj{\"o}rn and {Inglis}, Julie and {Wallack}, Nicole L.},
        title = "{The Compositions of the HR 8799 Planets Reflect Accretion of Both Solids and Metal-enriched Gas}",
      journal = {\apj},
         year = 2026,
        month = mar,
       volume = {1000},
       number = {1},
          eid = {27},
        pages = {27},
          doi = {10.3847/1538-4357/ae448f},
archivePrefix = {arXiv},
       eprint = {2602.09422},
 primaryClass = {astro-ph.EP},
       adsurl = {https://ui.adsabs.harvard.edu/abs/2026ApJ..1000...27X}
}

@ARTICLE{2025AJ....169..209B,
       author = {{Balmer}, William O. and {Kammerer}, Jens and {Pueyo}, Laurent and {Perrin}, Marshall D. and {Girard}, Julien H. and {Leisenring}, Jarron M. and {Lawson}, Kellen and {Dennen}, Henry and {van der Marel}, Roeland P. and {Beichman}, Charles A. and {Bryden}, Geoffrey and {Llop-Sayson}, Jorge and {Valenti}, Jeff A. and {Lothringer}, Joshua D. and {Lewis}, Nikole K. and {M{\^a}lin}, Mathilde and {Rebollido}, Isabel and {Rickman}, Emily and {Hoch}, Kielan K.~W. and {Soummer}, R{\'e}mi and {Clampin}, Mark and {Mountain}, C. Matt},
        title = "{JWST-TST High Contrast: Living on the Wedge, or, NIRCam Bar Coronagraphy Reveals CO$_{2}$ in the HR 8799 and 51 Eri Exoplanets' Atmospheres}",
      journal = {\aj},
         year = 2025,
        month = apr,
       volume = {169},
       number = {4},
          eid = {209},
        pages = {209},
          doi = {10.3847/1538-3881/adb1c6},
archivePrefix = {arXiv},
       eprint = {2503.13608},
 primaryClass = {astro-ph.EP},
       adsurl = {https://ui.adsabs.harvard.edu/abs/2025AJ....169..209B}
}

@ARTICLE{2006AJ....131.3109G,
       author = {{Golimowski}, D.~A. and {Ardila}, D.~R. and {Krist}, J.~E. and {Clampin}, M. and {Ford}, H.~C. and {Illingworth}, G.~D. and {Bartko}, F. and {Ben{\'\i}tez}, N. and {Blakeslee}, J.~P. and {Bouwens}, R.~J. and {Bradley}, L.~D. and {Broadhurst}, T.~J. and {Brown}, R.~A. and {Burrows}, C.~J. and {Cheng}, E.~S. and {Cross}, N.~J.~G. and {Demarco}, R. and {Feldman}, P.~D. and {Franx}, M. and {Goto}, T. and {Gronwall}, C. and {Hartig}, G.~F. and {Holden}, B.~P. and {Homeier}, N.~L. and {Infante}, L. and {Jee}, M.~J. and {Kimble}, R.~A. and {Lesser}, M.~P. and {Martel}, A.~R. and {Mei}, S. and {Menanteau}, F. and {Meurer}, G.~R. and {Miley}, G.~K. and {Motta}, V. and {Postman}, M. and {Rosati}, P. and {Sirianni}, M. and {Sparks}, W.~B. and {Tran}, H.~D. and {Tsvetanov}, Z.~I. and {White}, R.~L. and {Zheng}, W. and {Zirm}, A.~W.},
        title = "{Hubble Space Telescope ACS Multiband Coronagraphic Imaging of the Debris Disk around {\ensuremath{\beta}} Pictoris}",
      journal = {\aj},
         year = 2006,
        month = jun,
       volume = {131},
       number = {6},
        pages = {3109-3130},
          doi = {10.1086/503801},
archivePrefix = {arXiv},
       eprint = {astro-ph/0602292},
 primaryClass = {astro-ph},
       adsurl = {https://ui.adsabs.harvard.edu/abs/2006AJ....131.3109G}
}

@ARTICLE{2015ApJ...800..136A,
       author = {{Apai}, D{\'a}niel and {Schneider}, Glenn and {Grady}, Carol A. and {Wyatt}, Mark C. and {Lagrange}, Anne-Marie and {Kuchner}, Marc J. and {Stark}, Christopher J. and {Lubow}, Stephen H.},
        title = "{The Inner Disk Structure, Disk-Planet Interactions, and Temporal Evolution in the {\ensuremath{\beta}} Pictoris System: A Two-epoch HST/STIS Coronagraphic Study}",
      journal = {\apj},
         year = 2015,
        month = feb,
       volume = {800},
       number = {2},
          eid = {136},
        pages = {136},
          doi = {10.1088/0004-637X/800/2/136},
archivePrefix = {arXiv},
       eprint = {1501.03181},
 primaryClass = {astro-ph.EP},
       adsurl = {https://ui.adsabs.harvard.edu/abs/2015ApJ...800..136A}
}

@ARTICLE{2014ApJ...786...32M,
       author = {{Males}, Jared R. and {Close}, Laird M. and {Morzinski}, Katie M. and {Wahhaj}, Zahed and {Liu}, Michael C. and {Skemer}, Andrew J. and {Kopon}, Derek and {Follette}, Katherine B. and {Puglisi}, Alfio and {Esposito}, Simone and {Riccardi}, Armando and {Pinna}, Enrico and {Xompero}, Marco and {Briguglio}, Runa and {Biller}, Beth A. and {Nielsen}, Eric L. and {Hinz}, Philip M. and {Rodigas}, Timothy J. and {Hayward}, Thomas L. and {Chun}, Mark and {Ftaclas}, Christ and {Toomey}, Douglas W. and {Wu}, Ya-Lin},
        title = "{Magellan Adaptive Optics First-light Observations of the Exoplanet {\ensuremath{\beta}} Pic B. I. Direct Imaging in the Far-red Optical with MagAO+VisAO and in the Near-ir with NICI}",
      journal = {\apj},
         year = 2014,
        month = may,
       volume = {786},
       number = {1},
          eid = {32},
        pages = {32},
          doi = {10.1088/0004-637X/786/1/32},
archivePrefix = {arXiv},
       eprint = {1403.0560},
 primaryClass = {astro-ph.EP},
       adsurl = {https://ui.adsabs.harvard.edu/abs/2014ApJ...786...32M}
}

@ARTICLE{2020ApJ...892..151L,
       author = {{Lacy}, Brianna and {Burrows}, Adam},
        title = "{Prospects for Directly Imaging Young Giant Planets at Optical Wavelengths}",
      journal = {\apj},
         year = 2020,
        month = apr,
       volume = {892},
       number = {2},
          eid = {151},
        pages = {151},
          doi = {10.3847/1538-4357/ab7017},
archivePrefix = {arXiv},
       eprint = {1911.10585},
 primaryClass = {astro-ph.EP},
       adsurl = {https://ui.adsabs.harvard.edu/abs/2020ApJ...892..151L}
}

@ARTICLE{2021JOSS....6.3001B,
       author = {{Buchner}, Johannes},
        title = "{UltraNest - a robust, general purpose Bayesian inference engine}",
      journal = {The Journal of Open Source Software},
         year = 2021,
        month = apr,
       volume = {6},
       number = {60},
          eid = {3001},
        pages = {3001},
          doi = {10.21105/joss.03001},
archivePrefix = {arXiv},
       eprint = {2101.09604},
 primaryClass = {stat.CO},
       adsurl = {https://ui.adsabs.harvard.edu/abs/2021JOSS....6.3001B}
}

@ARTICLE{2026AJ....171...98B,
       author = {{Batalha}, Natasha E. and {Rooney}, Caoimhe M. and {Visscher}, Channon and {Moran}, Sarah E. and {Marley}, Mark S. and {Sengupta}, Aditya R. and {Kiefer}, Sven and {Lodge}, Matt G. and {Mang}, James and {Morley}, Caroline V. and {Mukherjee}, Sagnick and {Fortney}, Jonathan J. and {Gao}, Peter and {Lewis}, Nikole K. and {Mayorga}, L.~C. and {Pearce}, Logan A. and {Wakeford}, Hannah R.},
        title = "{Condensation Clouds in Substellar Atmospheres with Virga}",
      journal = {\aj},
         year = 2026,
        month = feb,
       volume = {171},
       number = {2},
          eid = {98},
        pages = {98},
          doi = {10.3847/1538-3881/ae29e5},
archivePrefix = {arXiv},
       eprint = {2508.15102},
 primaryClass = {astro-ph.EP},
       adsurl = {https://ui.adsabs.harvard.edu/abs/2026AJ....171...98B}
}
\bibliographystyle{aasjournal}



\end{document}